\documentclass[12pt,letterpaper]{article}
\usepackage[usenames, dvipsnames]{xcolor}
\usepackage{jcapmod}

\usepackage{tocloft}
\usepackage[]{todonotes}
\usepackage{verbatim}
\usepackage{mathrsfs}
\usepackage{cleveref}
\usepackage[shortlabels]{enumitem}
\usepackage{pgf}
\usepackage{tikz-cd}
\usetikzlibrary{shapes,arrows}
\usetikzlibrary{calc}
\usepackage{pgfplots}
\pgfplotsset{compat=1.12}
\usepackage{tikz-3dplot}

\usepackage{amsthm}

\usepackage{tikz}
\usepackage{setspace,caption}
\usepackage{lipsum}
\usetikzlibrary{matrix,arrows,calc}
\makeatletter
\newsavebox\myboxA
\newsavebox\myboxB
\newlength\mylenA

\definecolor{cornellRed}{HTML}{B31B1B}
\definecolor{cornellBlue}{HTML}{0068AC}
\definecolor{cornellGreen}{HTML}{6EB43F}

\usepackage{bbm} 					
\usepackage{slashed} 				
\usepackage{graphicx}				
\usepackage{subcaption}			
\usepackage{psfrag}				
\usepackage{tensor}				
\usepackage{fouridx}				
\usepackage{bm}					
\usepackage{mdframed}				
\usepackage{multirow}				
\usepackage{soul}					
\usepackage{bbold}				
\usepackage{multicol}				
\usepackage{tikz-cd}
\usepackage{rotating}

\usetikzlibrary{arrows}

\tikzset{
commutative diagrams/.cd,
arrow style=tikz,
diagrams={>=latex}}

\usepackage{amsmath}
\usepackage{amssymb}
\usepackage{amsfonts}
\usepackage{mathtools}

\usepackage{feynmf}
\usepackage{marvosym}

\usepackage{import}

\newcommand*\xoverline[2][0.75]{%
    \sbox{\myboxA}{$\m@th#2$}%
    \setbox\myboxB\null
    \ht\myboxB=\ht\myboxA%
    \dp\myboxB=\dp\myboxA%
    \wd\myboxB=#1\wd\myboxA
    \sbox\myboxB{$\m@th\overline{\copy\myboxB}$}
    \setlength\mylenA{\the\wd\myboxA}
    \addtolength\mylenA{-\the\wd\myboxB}%
    \ifdim\wd\myboxB<\wd\myboxA%
       \rlap{\hskip 0.5\mylenA\usebox\myboxB}{\usebox\myboxA}%
    \else
        \hskip -0.5\mylenA\rlap{\usebox\myboxA}{\hskip 0.5\mylenA\usebox\myboxB}%
    \fi}
\makeatother

\definecolor{cobalt}{RGB}{44, 98, 120}
\definecolor{celadon}{rgb}{0.67, 0.88, 0.69}
\definecolor{dm}{cmyk}{.20, 0, .30, 0}
\definecolor{burgundy}{rgb}{0.5, 0.0, 0.13}
\definecolor{plotBlue}{RGB}{94, 130, 181}
\definecolor{bisque}{rgb}{1.0, 0.89, 0.77}

\hypersetup{
  colorlinks,
  citecolor=Violet,
  linkcolor=cobalt,
  urlcolor=Blue}

\DeclareSymbolFontAlphabet{\mathbb}{AMSb}
\NewDocumentCommand{\xrightarrows}{ O{}O{} }{%
\mathrel{%
\vcenter{\hbox{%
\begin{tikzpicture}
  \node[minimum width=1cm,minimum height=1ex,anchor=south,align=center] (a){\text{\vphantom{hg}#1}\\[0.5ex] \vphantom{hg}#2};
  \draw[<-] ([yshift=0.35ex]a.west) -- ([yshift=0.35ex]a.east);
  \draw[->] ([yshift=-0.35ex]a.west) -- ([yshift=-0.35ex]a.east);
\end{tikzpicture}
}}%
}%
}

\newif\iffastcompile

\fastcompilefalse

\iffastcompile
\newcommand{\mk}[1]{}
\else
\newcommand{\mk}[1]{\todo[color=burgundy!30, size=\scriptsize, bordercolor=burgundy!30]{MK: #1}}
\fi

\iffastcompile
\newcommand{\mc}[1]{}
\else
\newcommand{\mc}[1]{\todo[color=bisque!30, size=\scriptsize, bordercolor=bisque!30]{MC: #1}}
\fi

\ProvideTextCommandDefault{\Dbar}{%
\leavevmode\lower.5ex\rlap{\hskip-.07em\accent"16}D%
}

\usepackage{environ}
\usepackage{changepage}

\begin{document}
	\newcommand{\main}{.}
\begin{titlepage}

\setcounter{page}{1} \baselineskip=15.5pt \thispagestyle{empty}
\setcounter{tocdepth}{2}
\bigskip\

\vspace{1cm}
\begin{center}
{\large \bfseries String perturbation theory of Klebanov-Strassler throat}
\end{center}

\vspace{0.55cm}

\begin{center}
\scalebox{0.95}[0.95]{{\fontsize{14}{30}\selectfont  Manki Kim$^{a}$\vspace{0.25cm}}}

\end{center}

\begin{center}

\vspace{0.15 cm}
{\fontsize{11}{30}
\textsl{$^{a}$Stanford Institute for Theoretical Physics,
Stanford University, Stanford, CA 94305}}\\
\vspace{0.25cm}

\vskip .5cm
\end{center}

\vspace{0.8cm}
\noindent

\vspace{1.1cm}
In this work, we develop a tool to study string perturbation theory of the Klebanov-Strassler solution in the large radius approximation based on open-closed superstring field theory. Combining the large radius expansion and a double scaling limit, we find a perturbative background solution of open-closed superstring field theory that corresponds to the Klebanov-Strassler solution. To illustrate the utilities of this approach, we break supersymmetry of the background by placing a stack of anti-D3-branes at the tip of the throat. We then find a perturbative open string background solution to the third order in the large radius approximation, which agrees with the well-known supergravity analysis of Kachru-Pearson-Verlinde (KPV) on the stability of the anti-D3-brane supersymmetry breaking. The perturbative background solution to the open string field theory we found is expected to be dual to an NS5-brane probing the KS solution.

\vspace{3.1cm}

\noindent\today

\end{titlepage}
\tableofcontents\newpage

\section{Introduction}
The late-time cosmology is well approximated with the four-dimensional de Sitter space. However, many quantum gravitational properties of the de Sitter spacetime remain elusive. One of the many promising ways to advance our understanding of quantum gravitational aspects of de Sitter space is to construct de Sitter spaces in string theory and analyze them through the eyes of string theory.

In recent years, the fate of four-dimensional de Sitter solutions in string theory in the context of flux compactifications\footnote{For review on flux compactification, see, for example, \cite{Hebecker:2020aqr,Cicoli:2023opf,McAllister:2023vgy}.} welcomed extensive investigations \cite{Bena:2018fqc, Bena:2019sxm,Bena:2020xrh,Bena:2022ive,Lust:2022mhk,Lust:2022xoq,Lust:2024aeg,Randall:2019ent,Gao:2020xqh,Gao:2022fdi,Gao:2022uop,Junghans:2022exo,Junghans:2022kxg,Junghans:2023lpo,Schreyer:2022len,Schreyer:2024pml,Hebecker:2022zme,Blaback:2019ucp,Krippendorf:2023idy,Grana:2022nyp,Coudarchet:2022fcl,Alvarez-Garcia:2020pxd,Demirtas:2020ffz,Demirtas:2021nlu,McAllister:2024lnt}. Of the recent works, we would like to highlight one of the recent papers appeared in the literature that found candidate de Sitter vacua in the low energy approximation of type IIB string theory \cite{McAllister:2024lnt} as envisioned by Kachru-Kallosh-Linde-Trivedi \cite{Kachru:2003aw}.

Although the result of \cite{McAllister:2024lnt} is an impressive progress in search of de Sitter vacua of string theory, the verdict on the fate of de Sitter in string theory is not yet conclusive. As was stressed in \cite{McAllister:2024lnt}, layers of required flux tuning, given the modest size of the maximum D3-brane tadpole of order $\mathcal{O}(500)$ in the known landscape of the Calabi-Yau threefolds \cite{Kreuzer:2000xy}, pose a significant challenge towards constructing de Sitter solutions whose theoretical control is arbitrarily excellent in light of our insufficient understanding of $\alpha'$ and $g_s$ corrections in string theory. Among the possible sources of the unknown corrections, the worst offender appears to be $\alpha'$ corrections to the effect of anti-D3-brane supersymmetry breaking \'a la Kachru-Pearson-Verlinde \cite{Kachru:2002gs}. As was also emphasized in \cite{Hebecker:2022zme}, such $\alpha'$ corrections may be more important than what was naively thought before.\footnote{We thank Arthur Hebecker for emphasizing the importance of determining precise numerical factors of $\alpha'$ corrections.} Therefore, it is of high importance to understand the $\alpha'$ corrections to the anti-D3-brane supersymmetry breaking effects \cite{Junghans:2022exo,Junghans:2022kxg,Hebecker:2022zme,Schreyer:2022len,Schreyer:2024pml}.

In this work, we develop a systematic approach to study string perturbation theory of the KS solution based on the recently constructed open-closed superstring field theory \cite{FarooghMoosavian:2019yke}.\footnote{For review on string field theory, see, for example, \cite{Erler:2019loq,Erler:2019vhl,deLacroix:2017lif,Erbin:2021smf,Maccaferri:2023vns,Sen:2024npu}.} Superstring field theory recently attracted intense investigations partially due to its capability to study many questions that couldn't be addressed in the conventional string perturbation theory, e.g., D-instanton amplitudes \cite{Sen:2020cef,Sen:2020eck,Sen:2021qdk,Sen:2021tpp,Alexandrov:2021shf,Alexandrov:2021dyl,Alexandrov:2022mmy,Eniceicu:2022xvk,Chakravarty:2022cgj,Eniceicu:2022nay,Agmon:2022vdj,Kim:2023cbh}, study of Ramond-Ramond backgrounds \cite{Cho:2018nfn,Cho:2023mhw}. Although the computations in open-closed superstring field theory are much more cumbersome than the equivalent counterpart of the low-energy supergravity approximation, the string field theoretic approach appears to be a very promising in the long run as string field theory provides a systematic set up to formulate string perturbation theory in Ramond-Ramond backgrounds. 

We shall list a few reasons to study string field theoretic approaches. First, understanding of the D-brane action is incomplete, leading to an essentially incomplete treatment in the low-energy supergravity. The difficulty of the computation of the anti-D3-brane action stems from the fact that a D3-brane in a flat spacetime sources Ramond-Ramond fluxes, in addition to the NSNS profile it sources, and henceforth obscuring the computation of the amplitudes in the presence of a D3-brane in the RNS formalism. However, string field theory provides a systematic tool to study Ramond-Ramond backgrounds as was pioneered in \cite{Cho:2018nfn}. Second, extracting the off-shell supergravity action from the on-shell amplitude is notoriously complicated. However, as far as we are concerned with the well-defined on-shell quantities, there is no need to stick to the off-shell supergravity action. String field theory provides a gauge invariant off-shell action that is easier to compute, reducing some complexities. Third, some versions of open string field theory admit a compact form of the action \cite{Berkovits:1995ab,Kunitomo:2015usa}. Furthermore, it is known that the moduli space of the Riemann surfaces with boundaries can be completely covered by some of the fundamental string vertices \cite{Zwiebach:1992bw,Cho:2019anu}. Therefore, although it is futuristic at present, in principle, it is possible to study non-perturbative open string solutions for the anti-D3-branes, which will provide a definite answer for anti-D3-brane supersymmetry breaking. 

With this in mind, we will study the Klebanov-Strassler solution in the large radius limit in open-closed string field theory building on the earlier work of \cite{Cho:2023mhw}.\footnote{We shall use the $SL(2;R)$ vertices for practical reasons. For more elegant mathematical approaches to the construction of string vertices, see, for example, \cite{Zwiebach:1992ie,Cho:2019anu,Firat:2021ukc,Firat:2023gfn,Firat:2023suh,Firat:2024ajp,Costello:2019fuh,Pius:2018pqr,Moosavian:2017uso}. Also, the flat vertex developed in \cite{Mazel:2024alu} appears to be a very promising approach. } We shall explicitly solve the closed string field theory equations of motion to the second order in the large radius expansion. Then, we will study the open string background solution of a stack of anti-D3-branes to the third order in the large radius expansion. We find that our result agrees with the result of \cite{Kachru:2002gs}, modulo the difference that originates from the S-dual form of the action used in \cite{Kachru:2002gs}. This perturbative open string background solution is expected to be dual to an NS5-brane probing the KS solution. 

This paper is organized as follows. In \S\ref{sec:strategy}, we will explain the strategy we will employ with some comments on the limitations of the conventional supergravity approaches. In \S\ref{sec:convention CFT SFT}, we collect conventions for the worldsheet CFT and open-closed string field theory. In \S\ref{sec:review KS}, we review the Klebanov-Strassler solution and the anti-D3-brane supersymmetry breaking studied by \cite{Kachru:2002gs}. In \S\ref{sec:KS in SFT}, we study the KS solution in closed string field theory and solve the background to the second order in the large radius expansion. In \S\ref{sec:anti SFT}, we study the anti-D3-brane supersymmetry breaking with open-closed string field theory. We solve the background solution of the open string field theory to the third order in the large radius expansion and find an agreement with \cite{Kachru:2002gs}. In \S\ref{sec:conclusions}, we conclude. In \S\ref{sec:deformed conifold}, we summarize the metric of the deformed conifold. In \S\ref{sec:vertices}, we construct the string vertices that are relevant to our work using the $SL(2;R)$ vertices. In \S\ref{app:off diag}, we compute the source terms for the off-diagonal string modes in the third-order open string equations of motion. In \S\ref{app:diag} and \S\ref{app:S3+S4+S6}, we compute the source terms that are vanishing in the third-order open string equations of motion.

\newpage
\section{Strategy}\label{sec:strategy}
Searching for a classically stable string vacuum with broken supersymmetry and positive cosmological constant has been a major theoretical challenge. To the core of the theoretical challenge, there lies a lack of suitable theoretical tools to analyze the stability of a supersymmetry broken phase of string theory. In this section, we shall briefly summarize some of the challenges and how we shall overcome them concerning the anti-D3-brane supersymmetry braking of \cite{Kachru:2002gs}.

One of the most well-studied proposals for supersymmetry breaking by Kachru-Pearson-Verlinde (KPV) \cite{Kachru:2002gs} proceeds by placing a stack of anti-D3-branes at the tip of the Klebanov-Strassler(KS) throat \cite{Klebanov:2000hb}. Provided that the anti-D3-brane stack added to the closed string background is meta-stable and supersymmetry is broken, the setup of KPV can be used to engineer de Sitter vacua of string theory \cite{Kachru:2003aw}. However, the stack of anti-D3-branes backreacts to the geometry, and the brane stack itself can decay into the flux via nucleation. Hence, it is crucial to understand the stability of the open-string and closed-string backgrounds.

There are phenomenological and theoretical challenges that are interwoven to understanding the susy breaking \'a la KPV. 
\begin{itemize}
	\item For a successful uplift from $AdS_4$ to a $dS_4$ via anti-D3-brane supersymmetry breaking, one would like to maximize $g_sM,$ $g_sM^2,$ and the D3-brane tadpole anchored in the Klebanov-Strassler throat engineered in the Calabi-yau orientifold. However, combined with the fact that the maximum D3-brane tadpole for the weakly coupled type IIB orientifold vacua is somewhat limited $\sim 500,$ it does not seem very feasible to attain arbitrarily good control parameters, as was emphasized in the recent talk by Liam McAllister \cite{liamtalk}. 
 
    Therefore, to put the de Sitter solutions of string theory \'a la KKLT on a firmer footing, either one should seek a way to enlarge the D3-brane tadpole, or one must develop tools to compute the $\alpha'$ and $g_s$ corrections. As invoking a larger D3-brane tadpole will likely require one to consider genuine F-theory compactifications, whose theoretical control is much harder to achieve, the computation of $\alpha'$ and $g_s$ corrections in weakly coupled type IIB string theory appears to be a more realistic goal. 
	\item KS throat is a minimally supersymmetric and highly warped background with a large number of Ramond flux quanta. At the same time, one of the main motivations for studying KS throat is to understand low-energy supersymmetry breaking, after which all supercharges are broken. Therefore, exact techniques available for highly supersymmetric backgrounds cannot be applied. 
 
	\item In the absence of the worldsheet and exact techniques, one can resort to the low-energy supergravity approximation of string theory to understand the stability of anti-D3-brane supersymmetry breaking. This approach typically proceeds by taking $\alpha'$ corrected spacetime action of closed string fields with and without coupling to D-branes and computing relevant physical observables. This approach is not the most practical. Extracting the off-shell supergravity action from scattering amplitudes is plagued with ambiguities. 
 
    At higher order in the amplitudes, the field basis used in string perturbation theory does not necessarily agree with the field basis one naturally uses in low-energy supergravity. Related to this issue, at higher order in the derivative expansion, there is a huge redundancy in the off-shell action that complicates the computation of a usable off-shell action. 
    
    Furthermore, in order to compute the amplitudes involving D-branes, to extract the off-shell action involving D-branes, one needs to cancel the tadpole by shifting the background \cite{Fischler:1986ci,Hashimoto:1996bf}. Otherwise, the amplitudes one computes are plagued with divergences due to both the NSNS and RR tadpoles, which make even the computation of the D-brane scattering ambiguous. However, to shift the background properly, one needs to take into account that the D-branes source Ramond-Ramond profiles, for which RNS formulation fails to give a suitable formalism. This problem leads to an insufficient understanding of the D-brane actions in low-energy supergravity.

    \item The insufficient understanding of the D-brane action of low-energy supergravity then presents a few technical challenges.

    Due to the lack of proper understanding of $B_6$ coupling to anti-D3-branes, the S-dualized form of the anti-D3-brane action is commonly used to study the supersymmetry breaking. Strictly speaking, without the proof that the off-shell anti-D3-brane expanded around a flux background is invariant under the S-duality, usage of the S-dual action forces one to go beyond the regime of validity of parameters.\footnote{An interesting approach is to study the S-dual version of the KS solution and anti-D3-brane susy breaking therein \cite{Gautason:2016cyp,Blaback:2019ucp}.} Although it may turn out that even in flux backgrounds, the D3-brane action is invariant under the S-duality, developing more direct approaches to understand anti-D3-brane supersymmetry breaking is desirable.

    The other popular choice of attack in the supergravity approach is to use the worldvolume theory of NS5-branes to study the supersymmetry breaking. However, as was already noted in \cite{Kachru:2002gs}, the worldvolume theory of NS5-brane is expected to be strongly coupled, and therefore, it is doubtful that the cavalier treatment of the NS5-brane worldvolume theory is a controlled approximation.

	\item Furthermore, the number of Ramond flux quanta in the KS solution is gigantic. On the other hand, to compute the Ramond coupling to the D-brane action, one must treat the Ramond states perturbatively. Therefore, it is not immediately clear if it is even possible to compute the off-shell action of the anti-D3-brane to a sufficient precision. Note that this problem is also shared with the string field theoretic approach. We will explain how one can understand this problem later in this section.
 
\end{itemize}

To overcome the theoretical challenges mentioned above, we shall study the KS background with the recently developed superstring field theory based on RNS formulation \cite{deLacroix:2017lif,FarooghMoosavian:2019yke,Sen:2024npu}. Two main advantages of using string field theory are that computation of the off-shell action from the off-shell amplitude is immediate, and that one can perform unambiguous on-shell amplitude computations in the Ramond-Ramond background. The first advantage comes at the expense of making the identifications of the supergravity field basis obscure. For understanding general spacetime symmetries, usually supergravity basis is better to work with as the general covariance of spacetime is more manifest. But, as we summarized above, finding the off-shell action in the supergravity basis is not always the easiest. Also, physical quantities, such as the mass spectrum of string states, do not depend on which field basis one uses. As the stability of a supersymmetry broken background can be studied by computing the spectrum of the low-lying states, we can choose the most convenient field basis to work with. Also, as was recently studied in \cite{Cho:2018nfn,Cho:2023mhw}, computation of on-shell quantities in Ramond-Ramond background in string field theory does not require much more tool building than what is already understood in the RNS formalism, provided that a reasonable expansion scheme is identified. 

However, string field theory still comes with one major limitation at the current stage of the understanding. At best, one can only perform perturbative computations, with an exceptional case being bosonic open string field theory. Also, string field theory requires a well-defined string theory formulated around a string background as input. As the Klebanov-Strassler solution is obtained as a solution to the Einstein equation in the deformed conifold with large flux quanta, it is therefore of crucial importance to first understand if such a solution with strong warping can be understood as a small deformation away from a background for which we have access to the worldsheet degrees of freedom.

We shall use the fact that the KS solution admits an analog of the near horizon limit, which we shall call the near tip limit. As the near horizon limit of a D3-brane solution, which is $AdS_5\times S^5,$ can be understood as a small deformation from flat ten-dimensional Minkowski spacetime in the large radius limit, the KS solution also admits a similar access point. In the near tip limit, the metric of the KS background approximates to that of $\Bbb{R}^{1,3}\times X,$ where $X$ is the deformed conifold \cite{Klebanov:2000hb,Herzog:2001xk}
\begin{equation}\label{eqn:KS near tip metric}
ds^2= \frac{\epsilon^{4/3}}{2^{1/3}a_0^{1/2}g_s M} dx_{\Bbb{R}^{1,3}}^2 +\frac{2^{1/3}a_0^{1/2} g_sM}{ \epsilon^{4/3}}ds^2_{X}\,,
\end{equation}
where
\begin{equation}
ds_X^2=\epsilon^{4/3}\left[2^{-5/3}3^{-1/3}(d\tau^2+(g^5)^2)+((g^3)^2+(g^4)^2)+ 2^{-2}\tau^2 ( (g^1)^2+(g^2)^2)\right]+\mathcal{O}(\alpha'^2)\,,
\end{equation}
and the deformed conifold is defined as 
\begin{equation}
z_1^2+z_2^2+z_3^2+z_4^2=\epsilon^2\,.
\end{equation}

This is a very favorable situation. Because the deformed conifold admits an exact worldsheet description, this could imply that one can study the KS solution in string field theory provided that one can treat the Ramond fluxes as small deformations away from the deformed conifold of \eqref{eqn:KS near tip metric}. As we shall verify in \S\ref{sec:large volume limit KS}, the energy density contained in the three form fluxes that are used to generate the KS solution are suppressed in the radius of the $S^3$ of the deformed conifold $R=\mathcal{O}(\sqrt{g_s M}).$ 

Therefore, we conclude that when $g_sM\gg1,$ we can understand the KS solution as a perturbation from the deformed conifold in the large volume limit in string field theory.\footnote{For recent progress in understanding the large N limit in string field theory, see, for example, \cite{Maccaferri:2023gof,Firat:2023gfn}.}

We shall close this section by commenting on the large volume expansion. Although the exact worldsheet CFT probing Calabi-Yau backgrounds is formally well understood, computation of the off-shell amplitudes involving D-branes in such a background is far beyond the reach of current CFT capabilities. Therefore, for us to make progress, it will be essential to identify suitable approximation schemes to deal with the Calabi-Yau sector of the matter CFT. One very promising approach is to study the background through $\mathcal{N}=2$ Liouville theory that can be attained as a double scaling limit of the deformed conifold CFT \cite{Aharony:1998ub,Giveon:1999px,Giveon:1999tq}.\footnote{We thank Juan Maldacena for illuminating discussions.} Unfortunately, unlike its less supersymmetric cousins, $\mathcal{N}=2$ supersymmetric Liouville theory is relatively poorly understood \cite{Nakayama:2004vk}, and it does not seem feasible with the current technologies to compute required amplitudes in the Liouville theory. Also, the deformed conifold contains a shrinking $S^2$, which vanishes in size at the tip of the throat. Therefore, a na\"ive large volume approximation to the worldsheet computation breaks down. 

To alleviate the difficulty, we shall combine the idea of the Liouville theory with the large radius expansion. Let us recall that the metric of the deformed conifold around the tip is written as
\begin{equation}
ds_{CY}^2= g_s M  b_0^2\left( \frac{1}{2} d\tau^2 + \alpha' d\Omega_3^2+\tau^2 d\Omega_2^2\right)\,,
\end{equation}
where $\tau$ is the radial direction of the throat. We shall define a new coordinate $r$
\begin{equation}
r:= \sqrt{\frac{g_sM}{2}} b_0\tau\,,
\end{equation}
and we will take a double scaling limit
\begin{equation}
R=\mathcal{O}(\sqrt{g_sM}) \rightarrow\infty\,,\quad \tau\rightarrow 0\,,
\end{equation}
while fixing the ratio $r.$ When $r$ is a large parameter, the metric of the deformed conifold can be approximated to 
\begin{equation}
ds^2_{CY}= dr^2 +ds_{\Bbb{R}^3}^2 + ds^2_{\Bbb{R}^2}+\mathcal{O}(\alpha')\,,
\end{equation}
where $d\Omega^2_3$ is approximated to $ds^2_{\Bbb{R}^3}+\mathcal{O}(\alpha')$ and $r^2d\Omega_2^2$ is approximated to $ds^2_{\Bbb{R}^2}+\mathcal{O}(\alpha').$ Therefore, by combining the large radius limit with the double scaling limit near the tip, we can treat the deformed conifold sector of the matter CFT as a large radius limit of the non-linear sigma model. In the context of string field theory, this will correspond to turning on an additional background at order $\alpha'.$ Since the Ricci curvature of the Calabi-Yau is trivial, the addition of this $\alpha'$ corrected metric at the first order of the large radius expansion, in the absence of the flux, is BRST-trivial. This is expected as having different numerical factors of the $\alpha'$ correction to the background solution, in the absence of the flux, shall correspond to choosing a different point in the moduli space. Since any value of such a background solution in the large radius expansion is allowed, we shall carefully choose the numerical factor that corresponds to the appropriate supergravity background. 

\section{Worldsheet and SFT conventions}\label{sec:convention CFT SFT}
In this section, we shall collect conventions for the worldsheet CFT and open-closed string field theory.

\subsection{Worldsheet convention}
As the Klebanov-Strassler solution asymptotes to the rescaled deformed conifold compactification in the large radius limit, we shall, therefore, spell out the conventions for the Calabi-Yau compactification in the large radius expansion. Throughout the paper, we shall use $\alpha'=1$ unit. However, to estimate the order of corrections, we shall reinstate $\alpha'.$

The worldsheet CFT is $\mathcal{N}=1$ supersymmetric CFT, consisting of the matter CFT and the usual $b,~c,~\beta,~\gamma$ ghost CFT. The matter CFT is a direct sum of $\mathcal{N}=1$ CFT with central charge $c=6$ for the four ``non-compact" directions and $\mathcal{N}=2$ CFT with central charge $c=9$ for the deformed conifold. As we are working in the large radius limit, we can adopt the free field presentation for both the Minkowski and the deformed conifold sector of the matter CFT. In the NS sector, we have the usual worldsheet fields 
\begin{equation}
X^A\,,\psi^A\,.
\end{equation}
Note that captial Roman indices $A,~B,\dots,$ will range from $0$ to $9,$ lower Roman indices $a,~b,\dots,$ will range from $4~9,$ and lower Greek indices $\alpha,~\beta,\dots,$ will range from $0$ to $3.$ We shall bosonize the $\beta,~\gamma$ system \cite{Friedan:1985ge}
\begin{equation}
\beta =\partial\xi e^{-\phi}\,,\quad \gamma=\eta e^\phi\,,\quad \delta(\gamma)=e^{-\phi}\,,\quad \delta(\beta)=e^\phi\,.
\end{equation}
We shall define Vielbein
\begin{equation}
e_A^{~\tilde{B}}\,,\quad e^A_{~\tilde{B}}\,,
\end{equation}
such that
\begin{equation}
\eta_{\tilde{A}\tilde{B}}= G_{CD}e^C_{~\tilde{A}}e^D_{~\tilde{B}}\,,\quad \eta^{\tilde{A}\tilde{B}}= G^{CD} e_C^{~\tilde{A}}e_D^{~\tilde{B}}\,, 
\end{equation}
and
\begin{equation}
e_A^{~\tilde{B}} e^{A}_{~\tilde{C}}=\delta^{\tilde{B}}_{\tilde{C}}\,.
\end{equation}
We can then define renormalized worldsheet fields
\begin{equation}
\tilde{X}^{\tilde{A}}:= X^B e_B^{~\tilde{A}}\,,\quad \tilde{\psi}^{\tilde{A}}:=\psi^Be_B^{~\tilde{A}}\,,
\end{equation}
The OPEs of the worldsheet fields are given as
\begin{align}
&\tilde{X}^{\tilde{A}} (z,\bar{z}) \tilde{X}^{\tilde{B}} (0,0) \sim-\frac{1}{2}\eta^{\tilde{A}\tilde{B}} \log|z|^2\,,\quad\tilde{\psi}^{\tilde{A}}(z) \tilde{\psi}^{\tilde{B}}(0)\sim \frac{\eta^{\tilde{A}\tilde{B}}}{z}\,,\\
&c(z)b(0)\sim \frac{1}{z}\,,\quad \xi(z)\eta(0)\sim \frac{1}{z}\,,\\
&\partial\phi(z)\partial\phi(0)\sim-\frac{1}{z^2}\,,\quad e^{q_1\phi(z)}e^{q_2\phi(0)}\sim z^{-q_1q_2}e^{(q_1+q_2)\phi(0)}\,.
\end{align}

We shall now introduce the spin fields.  Let us start with constructing the spin fields in flat ten-dimensional Minkowski with the metric $\eta_{AB}$. In ten-dimensional Minkowski target spacetime, we have 16 components chiral spinors both in holomorphic and anti-holomorphic sectors. We denote the ten-dimensional chiral spin field by $\Sigma_\alpha$ and the anti-chiral spin field by $\Sigma^\alpha.$ We shall choose the GSO projection, which we will summarize later so that the following fields are chosen to be GSO even
\begin{equation}
e^{-\phi/2} \Sigma_\alpha\,,\quad e^{-3\phi/2}\Sigma^\alpha\,.
\end{equation}
Some useful OPEs involving the spin fields are 
\begin{align}
&\psi^A(z) e^{-3\phi/2}\Sigma^\alpha(0)\sim -\frac{(\Gamma^A)^{\alpha\beta}}{\sqrt{2}z} e^{-3\phi/2}\Sigma_\beta(0)\,,\\
&\psi^A(z) e^{-\phi/2}\Sigma_\alpha(0)\sim -\frac{(\Gamma^A)_{\alpha\beta}}{\sqrt{2}z} e^{-\phi/2}\Sigma^\beta(0)\,,\\
&e^{-\phi/2}\Sigma_\alpha(z)e^{-\phi/2}\Sigma_\beta(0)\sim \frac{(\Gamma_A)_{\alpha\beta}}{\sqrt{2}z} e^{-\phi}\psi^A(0)\,,\\
&e^{-3\phi/2}\Sigma^\alpha (z)e^{-\phi/2}\Sigma_\beta(0)\sim \frac{\delta^\alpha_\beta}{z^2} e^{-2\phi}(0) -\frac{3}{2z} e^{-2\phi}\partial\phi(0) -\frac{(\Gamma_{AB})^\alpha_{~\beta}}{2z} e^{-2\phi}\psi^A\psi^B(0)\,.
\end{align}
Note that we shall use the Gamma matrix representation of \cite{Grassi:2003cm}, that is
\begin{equation}
(\Gamma^A)_{\alpha\beta}=(\Gamma^A)_{\beta\alpha}\,,\quad (\Gamma^A)^{\alpha\beta}=(\Gamma^A)_{\alpha\beta}\,,
\end{equation}
for $A\neq0,$ and
\begin{equation}
(\Gamma^0)^{\alpha\beta}=\delta_{\alpha\beta}\,,\quad (\Gamma^0)_{\alpha\beta}=-\delta_{\alpha\beta}\,.
\end{equation}

We shall normalize the closed string ghost correlator as
\begin{equation}
\langle c_{-1}\bar{c}_{-1}c_0\bar{c}_0 c_1\bar{c}_1 e^{-2\phi} e^{-2\bar{\phi}}\rangle = - \int d^{10}X \sqrt{-G}\,,
\end{equation}
and the open string ghost correlator on a Dp-brane as
\begin{equation}
\langle c_{-1} c_0 c_1 e^{-2\phi}\rangle =- \int d^{p+1}X\sqrt{-G}\,.
\end{equation}
Therefore, with the tilde-ed spacetime coordinates, we have the usual normalization for the amplitudes. For the sphere diagrams, we shall explicitly include the factor of $C_{S^2},$ and correspondingly, for the disk diagrams, we shall include the factor of $C_{D^2}.$ The overall normalizations are related to the closed string and open string couplings as follows
\begin{equation}
C_{S^2} =\frac{8\pi}{g_c^2} \,,\quad C_{D^2}=\frac{1}{g_o^2}\,.
\end{equation}

We define the BRST current as
\begin{equation}
j_B=c\left(T_m-\frac{1}{2}(\partial\phi)^2-\partial^2\phi -\eta\partial\xi\right) +\eta e^\phi T_F+bc\partial c-\eta\partial\eta b e^{2\phi}\,,
\end{equation}
where $T_m$  is the energy-momentum tensor of the matter CFT, which is normalized as
\begin{equation}
T_m=-\partial \tilde{X}^A \partial \tilde{X}_A-\frac{1}{2} \tilde{\psi}_A\tilde{\partial}\psi^A\,,
\end{equation}
and $T_F$ is the worldsheet supercharge of the matter sector CFT
\begin{equation}
T_F=i\sqrt{2} \tilde{\psi}^A\partial \tilde{X}_A\,.
\end{equation}
We define the BRST charge as
\begin{equation}
Q_B:= \oint dz j_B+\oint d\bar{z} \bar{j}_B\,.
\end{equation}
Note that we defined $\oint dz:= \frac{1}{2\pi i} \int_\mathcal{C}dz $ and $\oint d\bar{z} :=-\frac{1}{2\pi i}\int_\mathcal{C}d\bar{z}$ for a closed contour $\mathcal{C}.$  We define the PCO as follows
\begin{equation}
\mathcal{X}:=\{Q_B,\xi\}=c\partial \xi  +e^\phi T_F-\partial\eta be^{2\phi}-\partial(\eta b e^{2\phi})\,.
\end{equation}
Anti-holomorphic PCO is defined similarly $\overline{\mathcal{X}}:=\{Q_B,\bar{\xi}\}.$ 

Finally, we shall summarize the convention for the doubling trick for the disk diagrams. We shall use a conformal map from the upper half plane to a unit disk to study the disk diagrams. In particular, we are primarily interested in D3-branes and anti-D3-branes that are spacetime filling. Let us introduce a matrix $S^{AB}$ 
\begin{equation}
    S^{AB}=\begin{cases}
        \delta^{AB}& \text{for} \quad 0\leq A,~B,\leq 3\\
        -\delta^{AB}&\text{for} \quad4\leq A,~B,\leq 9\\
        0&\text{else}
    \end{cases}\,.
\end{equation}
For the worldsheet fields, we impose the following boundary conditions
\begin{equation}
    \partial \tilde{X}^A (z)= S^{AB}\bar{\partial}\bar{\tilde{X}}_B(\bar{z})\,,\quad \tilde{\psi}^A(z)=S^{AB}\bar{\tilde{\psi}}_B(\bar{z})\,,\quad c(z)=\bar{c}(\bar{z})\,,
\end{equation}
\begin{equation}
    b(z)=\bar{b}(\bar{z})\,,\quad \xi(z)=\bar{\xi}(\bar{z})\,, \quad\eta(z)=\bar{\eta}(\bar{z})\,,\quad e^{q\phi}(z)=e^{q\bar{\phi}}(\bar{z})\,.
\end{equation}
For the spin fields, we choose the following boundary conditions
\begin{equation}
    \Sigma_\alpha(z)= M_\alpha^{~\beta} \overline{\Sigma}_\beta(\bar{z})\,,\quad \Sigma^\alpha(z)=N^\alpha_{~\beta}\overline{\Sigma}^\beta(\bar{z})\,,
\end{equation}
where
\begin{equation}
    M_\alpha^{~\beta} =(-1)^w (\Gamma^{0123})_{\alpha}^{~\beta}\,,\quad N^\alpha_{~\beta}=(-1)^{1+w} (\Gamma^{0123})^\alpha_{~\beta}\,.
\end{equation}
Note that $w=\pm1$ corresponds to a choice of the orientation of the D3-brane. Depending on the Ramond-Ramond flux, one choice of $w$ corresponds to a supersymmetric brane, and the other corresponds to an anti-brane. We shall determine the sign convention in \S\ref{sec:spacetime susy}.

We shall close this section with a comment on $\alpha'$ correction to the worldsheet CFT probing the Calabi-Yau geometry. As we explained in \S\ref{sec:strategy}, we shall treat the $\alpha'$ corrections to the worldsheet CFT as non-trivial background solutions in string field theory. This means, the Vielbeins we chose here shall completely trivialize the non-linear spacetime metric into a constant metric
\begin{equation}
\eta_{AB}=G_{CD} e_{~A}^{C}e_{~B}^{D}\,.
\end{equation}
This can be understood as choosing a Riemann normal coordinate such that the constant metric is that of flat space. This is not always a great approximation scheme for a generic Calabi-Yau with a vanishing cycle. But, as we explained in \S\ref{sec:strategy}, the large radius limit combined with the double scaling limit, this approximation scheme is well justified in the deformed conifold.

\subsection{Open-closed superstring field theory}
In this section, we will collect the conventions for open-closed superstring field theory constructed in \cite{FarooghMoosavian:2019yke}. For the orientation and the normalization of open-closed string field theory, see \cite{Sen:2024npu}.

We shall use the 1PI string field theory \cite{deLacroix:2017lif}, where each string vertex is obtained by integrating over the moduli space of puntured Riemann surfaces that correspond to 1PI diagrams. We shall define closed string state space with picture number $(p,q)$ to be $H_{p,q}.$ We further require that every physical state $|\Psi\rangle$ in $H_{p,q}$ satisfies
\begin{equation}
b_0^-|\Psi\rangle=L_0^-|\Psi\rangle=0\,.
\end{equation}
Note that the first requirement can be relaxed for the test field for the equations of motion for gauge symmetries, for example. Similarly, we define open string state space with picture number $p$ to be $H_p.$ We define the state spaces $H^c,~\tilde{H}^c,~H^o,~\tilde{H}^o$ as
\begin{align}
&H^c:= H_{-1,-1}\oplus H_{-1/2,-1}\oplus H_{-1,-1/2}\oplus H_{-1/2,-1/2}\,,\\
&\tilde{H}^c:=H_{-1,-1}\oplus H_{-3/2,-1}\oplus H_{-1,-3/2}\oplus H_{-3/2,-3/2}\,\\
&H^o:=H_{-1}\oplus H_{-1/2}\,,\\
&\tilde{H}^o:= H_{-1}\oplus H_{-3/2}\,.
\end{align}
We denote string fields living in $H^c,~\tilde{H}^c,~H^o,~\tilde{H}^o$ by $\Psi^c,~\tilde{\Psi}^c,~\Psi^o,~\tilde{\Psi}^o,$ respectively. We define operator $\mathcal{G}$ as
\begin{equation}
\mathcal{G}|s^o\rangle=\begin{cases}
|s^o\rangle& \text{if }|s^o\rangle\in H_{-1}\\
\frac{1}{2}(\mathcal{X}_0+\overline{\mathcal{X}}_0)|s^o\rangle& \text{if }|s^o\rangle\in H_{-3/2}
\end{cases}\,,
\end{equation}
\begin{equation}
\mathcal{G}|s^c\rangle=\begin{cases}
|s^c\rangle&\text{if }|s^c\rangle \in H_{-1,-1}\\
\mathcal{X}_0 |s^c\rangle &\text{if }|s^c\rangle\in H_{-3/2,-1}\\
\overline{\mathcal{X}}_0|s^c\rangle&\text{if }|s^c\rangle\in H_{-1,-3/2}\\
\mathcal{X}_0\overline{\mathcal{X}}_0|s^c\rangle&\text{if }|s^c\rangle\in H_{-3/2,-3/2}
\end{cases}\,,
\end{equation}
where $\mathcal{X}_0$ and $\overline{\mathcal{X}}_0$ are zero modes of the PCOs 
\begin{equation}
    \mathcal{X}_0:= \oint \frac{dz}{z} \mathcal{X}\,,\quad \overline{\mathcal{X}}_0:=\oint \frac{d\bar{z}}{\bar{z}} \overline{\mathcal{X}}\,.
\end{equation}

We now write the 1PI action of open-closed string field theory
\begin{align}
S=&-\frac{2}{g_c^2}\langle \tilde{\Psi}^c|c_0^-Q_B\mathcal{G}|\tilde{\Psi}^c\rangle+\frac{4}{g_c^2} \langle \tilde{\Psi}^c|c_0^-Q_B|\Psi^c\rangle -\frac{1}{2g_o^2} \langle \tilde{\Psi}^o|Q_B\mathcal{G}|\tilde{\Psi}^o\rangle+\frac{1}{g_o^2}\langle \tilde{\Psi}^o|Q_B|\Psi^o\rangle\nonumber\\
&+Z_{D^2}+\{\tilde{\Psi}^c\}_{D^2+}+\sum_{N,M}\frac{1}{N!M!}\{(\Psi^c)^N;(\Psi^o)^M\}\,,
\end{align}
where $\{\}$ is the 1PI string vertex \cite{Sen:2014dqa,Sen:2015hha,deLacroix:2017lif}. We shall determine the vertex regions and Feynman regions in \S\ref{sec:vertices}. We also define string brackets $[]^c$ and $[]^o$ by using the following equations
\begin{equation}
\langle A_0 |c_0^-|[ A_1\dots A_N; B_1\dots B_M]^c \rangle= \{A_0 A_1\dots A_N; B_1\dots B_M\}\,,
\end{equation}
for any state $A_0\in H^c,$ 
\begin{equation}
\langle B_0 ||[ A_1\dots A_N; B_1\dots B_M]^o\rangle= \{ A_1\dots A_N; B_0B_1\dots B_M\}\,,
\end{equation}
for any state $B_0\in H^o,$ and
\begin{equation}
\langle \tilde{A}|c_0^-|[]_{D^2}\rangle= \{\tilde{A}\}_{D^2}\,,
\end{equation}
where $\tilde{A}\in \tilde{H}^c.$ Note that the disk one-point function is defined by inserting a string field in $\tilde{H}^c.$ All the other string vertices are defined with the states in $H^c$ and $H^o.$

\section{Klebanov-Strassler solution and supersymmetry breaking}\label{sec:review KS}
In this section, we shall review the Klebanov-Strassler construction of a holographic pair of warped deformed conifold and a confining four-dimensional gauge theory \cite{Klebanov:2000hb}, and the anti-D3-brane supersymmetry breaking within the same background studied by Kachru-Pearson-Verlinde \cite{Kachru:2002gs}.

As this subject has been extensively studied, we will keep this section to a minimal length. For more complete discussions on the topic, see, for example, \cite{Klebanov:2000hb,Herzog:2001xk,Kachru:2002gs,Strassler:2005qs}.

\subsection{Klebanov-Strassler solution}
The gravity solution of Klebanov-Strassler is obtained by finding the backreacted solution in the presence of the quantized Ramond-Ramond three-form and varying NSNS threeform flux on a deformed conifold. This supergravity solution is expected to be dual to 4d $\mathcal{N}=1$ gauge theory with $SU(N+M)\times SU(M)$ gauge group which confines at the low-energy. As we are mostly concerned with the gravity side of the solution, we will omit the discussion of the confining gauge theory.

Let us start with type IIB string theory compactified on a deformed conifold. The metric of the deformed conifold is well known \cite{Candelas:1989js,Minasian:1999tt,Ohta:1999we,Herzog:2001xk}.  For the deformed conifold embedded in $\Bbb{C}^4$
\begin{equation}
    z_1^2+z_2^2+z_3^2+z_4^2=\epsilon^2\,,
\end{equation}
the metric is given as
\begin{align}
ds_{CY}^2=\frac{1}{2} \epsilon^{4/3} K(\tau)&\biggr[ \frac{1}{3K^3(\tau)} (d\tau^2+(g^5)^2)+\cosh^2\left(\frac{\tau}{2}\right) [(g^3)^2+(g^4)^2]\nonumber\\&+\sinh^2\left(\frac{\tau}{2}\right) [(g^1)^2+(g^2)^2]\biggr]\,,
\end{align}
where $\tau$ describes the radial direction of the deformed conifold, and $g^i$ is a differential form along the $S^2\times S^3.$ For the details on the metric of the deformed conifold, see \S\ref{sec:deformed conifold}. As one approaches the tip of the deformed conifold $\tau=0,$ $S^2$ of $T^{1,1}$ shirinks to zero size, whereas radius of $S^3$ asymptotes to $\epsilon^{2/3}.$ Hence, the tip of the throat is not singular as the singularity is ``deformed."

The Klebanov-Strassler solution is then obtained by finding the warped metric caused by three-form fluxes and five-form Ramond-Ramond flux
\begin{equation}
F_3= \frac{M}{2} \left( g^5 \wedge g^3 \wedge g^4 +d [ F(\tau)(g^1\wedge g^3+g^2\wedge g^4)]\right)\,,
\end{equation}
\begin{equation}
H_3= \frac{g_sM}{2} \left[ d\tau\wedge (f'g^1\wedge g^2+k' g^3\wedge g^4) + \frac{1}{2} (k-f)g^5\wedge(g^1\wedge g^3+g^2\wedge g^4)\right]\,,
\end{equation}
where $F(\tau),$ $f(\tau),$ and $k(\tau)$ are defined as
\begin{align}
&F(\tau)=\frac{\sinh(\tau)-\tau}{2\sinh\tau}=\frac{\tau^2}{12}+\dots\,\\
&f(\tau)=\frac{\tau\coth\tau-1}{2\sinh\tau}(\cosh\tau-1)=\frac{\tau^3}{12}+\dots\,\\
&k(\tau)=\frac{\tau\coth\tau-1}{2\sinh\tau}(\cosh\tau+1)=\frac{\tau}{3}+\frac{\tau^3}{180}+\dots\,.
\end{align}

The metric of the warped deformed conifold is then given as
\begin{equation}
    ds^2=h^{-1/2}(\tau)dx_\mu dx^\mu +h^{1/2}(\tau) ds^2_{CY}\,,
\end{equation}
where
\begin{equation}
    h(\tau)=(g_s M)2^{2/3}\epsilon^{-8/3} I(\tau)\,,
\end{equation}
\begin{equation}
    I(\tau):= \int_\tau^{\infty} dx\frac{x\coth x-1}{\sinh^2x} (\sinh(2x)-2x)^{1/3}\,.
\end{equation}
For large $\tau,$ one can rewrite $\tau$ as
\begin{equation}
    r^2=\frac{3}{2^{5/3}}\epsilon^{4/3}e^{2\tau/3}\,.
\end{equation}
In the new radial coordinate, $h(\tau)$ reads
\begin{equation}
    h(r)=b_0+4\pi \frac{ a(g_sM)^2\log(r/r_0)+a(g_sM)^2/4}{r^4}\,.
\end{equation}
This solution preserves minimal supersymmetry and satisfies the imaginary-self-dual (ISD) condition
\begin{equation}
    g_s\star_6F_3=H_3\,.
\end{equation}
The key insight of \cite{Klebanov:2000hb} was that this logarithmic flow of the warping is dual to a cascade of the Seiberg duality of the dual gauge theory \cite{Strassler:2005qs}. Furthermore, because the warping $h(\tau)$ asymptotes to a finite value of small $\tau,$ the sequence of Seiberg dualities of the dual gauge theory is expected to terminate, leaving the gauge theory confining. This confinement scale is expected to be exponentially small compared to the UV scale, which is reflected in the exponential hierarchy between the UV scale and the scale of the tip in the bulk dual. Therefore, the Klebanov-Strassler solution can be a testing ground for studying ideas related to extreme scale-separation.

\subsection{Anti-D3-brane supersymmetry breaking}
The Klebanov-Strassler solution features an extreme hierarchy between the UV and confining scales. Furthermore, the KS solution can be embedded into a compact Calabi-Yau orientifold \cite{Giddings:2001yu}. Therefore, the KS solution provides a fruitful ground for studying calculable low-energy supersymmetry breaking. 

The idea of the foundational work by Kachru-Pearson-Verlinde \cite{Kachru:2002gs} was to break supersymmetry by placing a stack of p anti-D3-branes at the tip of the KS throat. Because the energy scale of the tip is exponentially red-shifted compared to the $M_{pl},$ the scale of the supersymmetry breaking is exponentially small compared to the Planck scale. Furthermore, when the size of $S^3$ at the tip of the KS throat is large, compared to the string length scale, one can reliably compute the effects of low-energy supersymmetry breaking in the low-energy supergravity approximation. Correspondingly, the analysis of KPV was performed in the large radius limit, $1/g_sM\ll1,$ and the probe limit, $p/g_sM^2\ll 1.$ 

The important observation of \cite{Kachru:2002gs} was that a point-like configuration of the anti-D3-brane stack is unstable, and the anti-D3-brane stack puffs up to a fuzzy NS5-brane due to the Myers effect \cite{Myers:1999ps}. Depending on the value of $p/M, $ there can be a classical instability leading to a complete decay of the anti-D3-brane back to a supersymmetric background. On the other hand, even in the presence of a false vacuum for a small value of $p/M,$ there is a quantum mechanical decay to a supersymmetric vacuum.   

We shall illustrate the metastable non-supersymmetric vacuum from two perspectives: anti-D3-brane picture and NS5-brane picture. 

Let us start with the anti-D3-brane perspective. We shall write the worldvolume action of the S-dual form of the anti-D3-brane
\begin{equation}
    S=-T_{D3} \int d^4x \sqrt{- \det(G_{\|} +2\pi g_s \mathcal{F})\det(Q)}-2\pi\mu_3\int \text{Tr} \mathfrak{i}_\Phi \mathfrak{i}_\Phi B_6 \,,
\end{equation}
where
\begin{equation}
    Q^i_{~j}=\delta^{i}_{~j}+\frac{2\pi i}{g_s}[\Phi^i,\Phi^k](G_{kj}+g_sC_{kj})\,.
\end{equation}
Note that $\Phi^i$ is a $p\times p$ matrix describing the anti-D3-brane position in the deformed conifold. By evaluating the worldvolume action in the probe approximation, one can find the effective potential \cite{Kachru:2002gs}
\begin{equation}
    V_{eff}(\Phi)=g_s^{-1} \sqrt{-\det(G_{\|})}\left(p-i\frac{4\pi^2}{3}F_{ijk}\text{Tr}( [\Phi^i,\Phi^j]\Phi^k)-\frac{\pi^2}{g_s^2} \text{Tr} ([\Phi^i,\Phi^j][\Phi_i,\Phi_j])+\dots \right)\,.
\end{equation}

The effective potential $V_{eff}(\Phi)$ admits a non-trivial minimum away from $\Phi=0$ at
\begin{equation}
    [[\Phi^i,\Phi^j],\Phi_j]-ig_s^2 F_{ijk} [\Phi^j,\Phi^k]=0\,.
\end{equation}
This equation is solved by
\begin{equation}
    \Phi^i=-\frac{g_s^2}{12} F_{abc}\epsilon^{abc} \alpha^i\,, 
\end{equation}
where $\alpha^i$ is a dimension p representation of the generator of the $SU(2)$ algebra
\begin{equation}
    [\alpha^i,\alpha^j]=2i \epsilon^{ijk}\alpha_k\,.
\end{equation}
At this non-trivial minimum, the effective potential is reduced
\begin{equation}
    V_{eff}\simeq \frac{\mu_3}{g_s}\left(p-\frac{\pi^2 (p^2-1)}{3b_0^{12}M^2}\right)\,,
\end{equation}
compared to the original open string vacuum $\Phi^i=0.$ Hence, one can conclude that the fuzzy sphere configuration is energetically favored. One important point is that to find the non-trivial open string vacuum, we ignored the higher order terms in $\Phi.$ This approximation is valid only when $p/M$ is sufficiently small.

We shall now reproduce this non-supersymmetric vacuum from the NS5-brane perspective. The NS5-brane worldvolume action reads \cite{Polchinski:2000uf}
\begin{equation}
    S=-\frac{\mu_5}{g_s^2}\int d^6x \sqrt{-\det(G_{\|})\det (G_\perp+2\pi g_s\mathcal{F})} +\mu_5\int B_6\,,
\end{equation}
where
\begin{equation}
    \mathcal{F}=F_2-\frac{1}{2\pi} C_2\,.
\end{equation}
We shall let the NS5-brane to wrap an $S^2$ inside the $S^3$ at the tip. The angular location of $S^2$ will be denoted by $\psi.$ The p anti-D3-brane is expected to be dual to an NS5-brane with the worldvolume flux
\begin{equation}
    \int_{S^2} F_2=2\pi p\,.
\end{equation}
With the above worldvolume flux, we can evaluate the NS5-brane action in the probe approximation to compute the effective potential
\begin{align}
    V_{eff}(\psi)\propto & M\left(V_0(\psi)-\frac{1}{2\pi}(2\psi-\sin2\psi)\right)\,,\\
    =&\left( p-\frac{4M}{3\pi}\psi^3+\frac{b_0^4M}{2\pi^2p}\psi^4+\dots\right)\,,
\end{align}
where
\begin{equation}
    V_0(\psi):=\frac{1}{\pi}\sqrt{b_0^4\sin^4\psi+\left(\pi\frac{p}{M}-\psi+\frac{1}{2}\sin(2\psi)\right)^2}\,.
\end{equation}
In the small $\psi$ regime, one can find that $V_{eff}(\psi)$ admits a minimum 
\begin{equation}
    V_{eff}(\psi_{min})\simeq \frac{\mu_3p}{g_s}\left( 1-\frac{8\pi^2p^2}{3b_0^{12}M^2}\right)\,,
\end{equation}
which agrees with the anti-D3-brane approach. 

We shall close this section with a few comments. It is important to note that neither of the two approaches, strictly speaking, offers a controlled approximation scheme. Because some of the couplings between anti-D3-branes and the closed string fields are unknown, \cite{Kachru:2002gs} used the S-dual form of the action. Strictly speaking, this S-dual form of the action is well controlled when $1/g_s\ll1,$ whereas one would ideally like to be in a small string coupling regime $g_s\ll1.$ Furthermore, the worldvolume theory of the NS5-brane is expected to be strongly coupled. Therefore, it is of crucial importance to develop a controlled approximation scheme to study the anti-D3-brane supersymmetry breaking. In the remaining sections, we shall consequently develop a systematic, controlled approximation scheme that bypasses the control issue of \cite{Kachru:2002gs}. 

\section{Klebanov-Strassler in SFT}\label{sec:KS in SFT}
In this section, we will study the supergravity solution of Klebanov-Strassler in the large radius limit in the context of string field theory. Before we start, let us first comment on the feasibility of the string field theoretic analysis.

Far away from the tip of the deformed conifold, the spacetime geometry approximates to $AdS_5\times T^{1,1},$  whose radius slowly varies as the distance from the tip changes
\begin{equation}
ds^2= \frac{r^2}{L^2\sqrt{\ln(r/r_s)}}dx_{\|}^2+\frac{L^2\sqrt{\ln(r/r_2)}}{r^2}dr^2+L^2\sqrt{\ln(r/r_s)}ds_{T^{1,1}}^2\,.
\end{equation}
On the other hand, near the tip, the geometry approximates to $\Bbb{R}^{1,3}\times X,$ where $X$ is the deformed conifold \cite{Klebanov:2000hb,Herzog:2001xk}
\begin{equation}
ds^2= \frac{\epsilon^{4/3}}{2^{1/3}a_0^{1/2}g_s M} dx_{\Bbb{R}^{1,3}}^2 +\frac{2^{1/3}a_0^{1/2} g_sM}{ \epsilon^{4/3}}ds^2_{X}\,,
\end{equation}
where
\begin{equation}
ds_X^2=\epsilon^{4/3}\left[2^{-5/3}3^{-1/3}(d\tau^2+(g^5)^2)+((g^3)^2+(g^4)^2)+ 2^{-2}\tau^2 ( (g^1)^2+(g^2)^2)\right]+\mathcal{O}(\alpha'^2)\,,
\end{equation}
and the deformed conifold is defined as 
\begin{equation}
z_1^2+z_2^2+z_3^2+z_4^2=\epsilon^2\,.
\end{equation}
Note that the metric above is string-frame metric. Also, it is important to note that the metric is appropriately scaled compared to the bare metric of $\Bbb{R}^{1,3}\times X.$ 

As strings propagating in a Calabi-Yau background admits a well-defined CFT description, this near-tip limit can furnish a good starting point for string field theoretic analysis, provided that the Ramond-Ramond flux density diminishes in the near-tip limit. However, as the defining data for the worldsheet CFT for the deformed conifold is poorly understood, one may complain that we cannot make much progress. However, as we explained in \S\ref{sec:strategy}, one can take a large radius limit $g_s M \rightarrow \infty$ combined with the double scaling limit, which will make the analysis even more accessible, as we can study the background order by order in the large radius limit where the zeroth order background can be treated as a free field CFT background. The only remaining issue is if the fluxes in the Klebanov-Strassler solution can be understood as a small deformation of a Calabi-Yau compactification amenable to string field theoretic analysis. We will answer this question affirmatively and find the perturbative background solution of string field theory that corresponds to the Klebanov-Strassler solution.

\subsection{Large radius limit of the Klebanov-Strassler solution}\label{sec:large volume limit KS}
The goal of this section is twofold. First, we shall understand if the RR three-form flux and the NSNS threeform flux can be understood as a small perturbation in the large radius limit. Second, we shall identify the properly normalized vertex operators for the threeform fluxes in the large radius coordinates.

The vertex operator for the Ramond-Ramond threeform flux turned on to find the Klebanov-Strassler solution is
\begin{equation}
V_{RR}=\frac{ig_s}{16\pi} \frac{1}{3!} F_{abc} c\bar{c} e^{-\phi/2}\Sigma_\alpha (\Gamma^{abc})^{\alpha\beta} e^{-\bar{\phi}/2}\overline{\Sigma}_\beta\,,
\end{equation}
where
\begin{equation}
F_3= \frac{M}{2} \left( g^5 \wedge g^3 \wedge g^4 +d [ F(\tau)(g^1\wedge g^3+g^2\wedge g^4)]\right)\,,
\end{equation}
such that
\begin{equation}
\frac{1}{(2\pi)^2}\int_{S^3}F_3=M\,.
\end{equation}
The vertex operator for the NSNS threeform flux is
\begin{equation}
V_{NS}= \frac{1}{4\pi}B_{ab} c\bar{c} e^{-\phi}\psi^a e^{-\bar{\phi}}\bar{\psi}^b\,,
\end{equation}
where
\begin{equation}
B_2=\frac{g_sM}{2} [f(\tau)g^1\wedge g^2 +k(\tau)g^3\wedge g^4]\,,
\end{equation}
\begin{equation}
H_3= \frac{g_sM}{2} \left[ d\tau\wedge (f'g^1\wedge g^2+k' g^3\wedge g^4) + \frac{1}{2} (k-f)g^5\wedge(g^1\wedge g^3+g^2\wedge g^4)\right]\,.
\end{equation}
Note that $F(\tau),$ $f(\tau),$ and $k(\tau)$ are defined as
\begin{align}
&F(\tau)=\frac{\sinh(\tau)-\tau}{2\sinh\tau}=\frac{\tau^2}{12}+\dots\,\\
&f(\tau)=\frac{\tau\coth\tau-1}{2\sinh\tau}(\cosh\tau-1)=\frac{\tau^3}{12}+\dots\,\\
&k(\tau)=\frac{\tau\coth\tau-1}{2\sinh\tau}(\cosh\tau+1)=\frac{\tau}{3}+\frac{\tau^3}{180}+\dots\,.
\end{align}

We shall rewrite $F_3$ in the large radius coordinates. We shall first use an identity \cite{Herzog:2001xk}
\begin{equation}
\frac{1}{2}g^5\wedge(g^1\wedge g^2+g^3\wedge g^4)=g^5\wedge g^3\wedge g^4+\frac{1}{2} d(g^1\wedge g^3+g^2\wedge g^4)\,,
\end{equation}
to rewrite $F_3$ 
\begin{align}
F_3=&\frac{M}{2} \left((1-F(\tau))g^5\wedge g^3\wedge g^4+F(\tau)g^5\wedge g^1\wedge g^2+F'(\tau)d\tau\wedge (g^1\wedge g^3+g^2\wedge g^4)  \right)\,,\\
=&\frac{M}{2} \biggr(2^{1/2} \tilde{R}_{S^3}^{-3} d\tilde{X}^5\wedge d\tilde{X}^7\wedge d\tilde{X}^6 +\frac{\sqrt{2}\tau^2}{6}\tilde{R}_{S^3}^{-1}\tilde{R}_{S^2}^{-2} d\tilde{X}^5\wedge d\tilde{X}^9\wedge d\tilde{X}^8\nonumber\\
&\qquad\quad +\frac{\sqrt{2}\tau}{6} \tilde{R}_{S^3}^{-1}\tilde{R}_{S^2}^{-1}d\tau \wedge(d\tilde{X}^9\wedge d\tilde{X}^7+d\tilde{X}^8\wedge d\tilde{X}^6) \biggr)+\mathcal{O}( \tilde{R}_{S^3}^{-3})\,.
\end{align}
We can then use
\begin{equation}
\tau=\sqrt{2}\tilde{R}_{S^3}^{-1} r\,,
\end{equation}
and
\begin{equation}
\frac{\tau}{\tilde{R}_{S^2}}=\frac{\sqrt{2}}{\tilde{R}_{S^3}}\,,
\end{equation}
to express $F_3$ as
\begin{align}\label{eqn:Ramond large coordinates}
g_s F_3=2^{-2/3}3^{1/3}a_0^{-1/2}\tilde{R}_{S^3}^{-1} &\biggr( \sqrt{2}d\tilde{X}^5\wedge d\tilde{X}^7\wedge d\tilde{X}^6+\frac{\sqrt{2}}{3}d\tilde{X}^5\wedge d\tilde{X}^9\wedge d\tilde{X}^8\nonumber\\
&+\frac{\sqrt{2}}{3} d\tilde{X}^4\wedge \left(d\tilde{X}^9\wedge d\tilde{X}^7+d\tilde{X}^8\wedge d\tilde{X}^6\right)\biggr)+\mathcal{O}(\tilde{R}_{S^3}^{-3})\,.
\end{align}
Therefore, we confirmed that in the large radius limit, the Ramond-Ramond threeform flux is suppressed by $\tilde{R}_{S^3}^{-1}.$ Hence, we can treat $F_3$ as a small perturbation. One important remark is in order. As it stands, the flux quantization of \eqref{eqn:Ramond large coordinates} is off by $\tilde{R}_{S^3}^{-3}.$ In string perturbation theory, the worldsheet is agnostic to the quantization of the Ramond flux. Therefore, it is not a fundamental problem that \eqref{eqn:Ramond large coordinates} is not properly quantized. The terms in \eqref{eqn:Ramond large coordinates} that are suppressed by higher orders of $\tilde{R}_{S^3}^{-1}$ will correspond to higher order terms in the perturbative background solution of string field theory.

Now we shall rewrite $H_3$ in the large radius coordinates. We shall write $H_3$ as
\begin{align}
H_3=&\frac{g_sM}{2} \left( d\tau\wedge\left( \frac{\tau^2}{4} g^1\wedge g^2+\frac{1}{3} g^3\wedge g^4\right)+\frac{\tau}{6} g^5 \wedge(g^1\wedge g^3+g^2\wedge g^4)\right)+\dots\,,\\
=&2^{-2/3}3^{1/3}a_0^{-1/2}\tilde{R}_{S^3}^{-1}\biggr( \sqrt{2} d\tilde{X}^4\wedge\left( d\tilde{X}^9\wedge d\tilde{X}^8+\frac{1}{3} d\tilde{X}^7\wedge d\tilde{X}^6\right)\nonumber\\
&\qquad\qquad\qquad\qquad+\frac{\sqrt{2}}{3} d\tilde{X}^5\wedge( d\tilde{X}^9\wedge d\tilde{X}^7+ d\tilde{X}^8\wedge d\tilde{X}^6)\biggr)+\mathcal{O}(\tilde{R}_{S^3}^{-3})\,.\label{eqn:H in large radius}
\end{align}
Same as $F_3,$ we again conclude that $H_3$ is suppressed by $\mathcal{O}(\tilde{R}_{S^3}^{-1}),$ which therefore can be treated as a small perturbation in the large radius limit. The higher-order corrections will be found as solutions to the perturbative background solution of string field theory. Also, note that we have
\begin{equation}
g^2|F_3|^2=|H_3|^2\,,
\end{equation}
which guarantees that the flux choice solves the Killing spinor equations and the equations of motion of the low-energy supergravity.

We shall define the relevant vertex operators for the threeform fluxes in the large radius limit
\begin{equation}
\tilde{V}_{RR}=\frac{ig_s}{16\pi} \frac{1}{3!} \tilde{F}_{abc} c\bar{c} e^{-\phi/2}\Sigma_\alpha (\tilde{\Gamma}^{abc})^{\alpha\beta} e^{-\bar{\phi}/2}\overline{\Sigma}_\beta\,,
\end{equation}
\begin{equation}
\tilde{V}_{NS}= \frac{1}{4\pi}\tilde{B}_{ab} c\bar{c} e^{-\phi}\tilde{\psi}^a e^{-\bar{\phi}}\bar{\tilde{\psi}}^b\,,
\end{equation}
where
\begin{equation}
g_sF_3=\frac{g_s}{3!}\tilde{F}_{abc} d\tilde{X}^a\wedge d\tilde{X}^b\wedge d\tilde{X}^c+\mathcal{O}(\tilde{R}_{S^3}^{-3})\,,
\end{equation}
\begin{equation}
H_3= dB_2=\frac{1}{3!}\tilde{H}_{abc}d\tilde{X}^a\wedge d\tilde{X}^b\wedge d\tilde{X}^c+\mathcal{O}(\tilde{R}_{S^3}^{-3})\,,
\end{equation}
and
\begin{equation}
\tilde{B}_{ab}=\frac{1}{3}\tilde{H}_{abc} \tilde{X}^c\,.
\end{equation}

\subsection{Perturbative background solution}
In this section, we will study the perturbative background solution sourced by the threeform fluxes. 

The closed string field theory equation of motion is given by
\begin{equation}
\frac{4}{g_c^2} Q_B |\Psi\rangle+\sum_n \frac{1}{n!}\mathcal{G}[\Psi^n]=0\,.
\end{equation}
Although we can, in principle, consider $g_s$ corrections, we shall only focus on the tree-level amplitudes. We shall expand the background solution as
\begin{equation}
\Psi=\sum_n \tilde{R}_{S^3}^{-n}\Psi_n\,.
\end{equation}

We shall study the background solution up to the third order in the large radius expansion. The first-order equations read
\begin{equation}
\frac{4}{g_c^2} \tilde{R}_{S^3}^{-1} Q_B\Psi_1=0\,,
\end{equation}
which requires that the first-order deformation is marginal. We shall set
\begin{equation}
\tilde{R}_{S^3}^{-1} \Psi_1^{-1,-1}=\tilde{V}_{NSNS}\,,
\end{equation}
\begin{equation}
\tilde{R}_{S^3}^{-1}\Psi_1^{-\frac{1}{2},-\frac{1}{2}}=\tilde{V}_{RR}\,.
\end{equation}

\subsubsection{Second order equations}\label{sec:KS sec order}
The second-order equations are
\begin{equation}
\frac{4}{g_c^2}\tilde{R}_{S^3}^{-2}Q_B|\Psi_2\rangle=-\frac{1}{2}\tilde{R}_{S^3}^{-2} \mathcal{G} \left[\Psi_1\otimes \Psi_1\right]_{S^2}\,.
\end{equation}
We shall split $\Psi_2$ into the NSNS sector $\Psi_2^{-1,-1}$ and the RR sector $\Psi_2^{-\frac{1}{2},-\frac{1}{2}}.$ We also define a projection operator $\Bbb{P}$ that projects a state into $L_0^+$ nilpotent components.  We then have
\begin{align}
&\frac{4}{g_c^2}\tilde{R}_{S^3}^{-2} Q_B\Bbb{P}|\Psi_2^{-1,-1}\rangle=-\frac{1}{2}\mathcal{G}\Bbb{P}\left[\tilde{V}_{NSNS}\otimes\tilde{V}_{NSNS}+\tilde{V}_{RR}\otimes\tilde{V}_{RR}\right]_{S^2}\,,\\
&\frac{4}{g_c^2}\tilde{R}_{S^3}^{-2}Q_B(1-\Bbb{P})|\Psi_2^{-1,-1}\rangle=-\frac{1}{2}\mathcal{G}(1-\Bbb{P})\left[\tilde{V}_{NSNS}\otimes\tilde{V}_{NSNS}+\tilde{V}_{RR}\otimes\tilde{V}_{RR}\right]_{S^2}\,,\\
&\frac{4}{g_c^2}\tilde{R}_{S^3}^{-2} Q_B\Bbb{P}|\Psi_2^{-\frac{1}{2},-\frac{1}{2}}\rangle=-\mathcal{G}\Bbb{P}\left[\tilde{V}_{NSNS}\otimes\tilde{V}_{RR}\right]_{S^2}\,,\\
&\frac{4}{g_c^2}\tilde{R}_{S^3}^{-2} Q_B(1-\Bbb{P})|\Psi_2^{-\frac{1}{2},-\frac{1}{2}}\rangle=-\mathcal{G}(1-\Bbb{P})\left[\tilde{V}_{NSNS}\otimes\tilde{V}_{RR}\right]_{S^2}\,.
\end{align}

$(1-\Bbb{P})$ components of the equations of motion can be easily solved by using an identity
\begin{equation}
\{Q_B,b_0^+\}=L_0^+\,,
\end{equation}
\begin{align}
&\frac{4}{g_c^2}\tilde{R}_{S^3}^{-2}|\Psi_2^{-1,-1}\rangle=-\frac{b_0^+}{2L_0^+}(1-\Bbb{P})\mathcal{G} \left[\tilde{V}_{NSNS}\otimes\tilde{V}_{NSNS}+\tilde{V}_{RR}\otimes\tilde{V}_{RR}\right]_{S^2}\,,\\
&\frac{4}{g_c^2}\tilde{R}_{S^3}^{-2}|\Psi_2^{-\frac{1}{2},-\frac{1}{2}}\rangle=-\frac{b_0^+}{L_0^+}(1-\Bbb{P})\mathcal{G} \left[\tilde{V}_{NSNS}\otimes \tilde{V}_{RR}\right]_{S^2}\,.
\end{align}

On the other hand, we shall explicitly find the form of $\Bbb{P}\Psi_2$ to solve the $L_0^+$ nilpotent components of the equations of motion. By using the results of \cite{Cho:2023mhw}, we find
\begin{align}
\frac{4}{g_c^2}\tilde{R}_{S^3}^{-2}Q_B\Bbb{P}|\Psi_2^{-1,-1}\rangle=&\frac{\pi}{2\kappa_{10}^2g_s^2} (\partial c+\bar{\partial}\bar{c})c\bar{c} H_{acd}H_{bef} \eta^{ce}\eta^{df} e^{-\phi}\tilde{\psi}^a e^{-\bar{\phi}}\bar{\tilde{\psi}}^b\nonumber\\
&+\frac{\pi}{2\kappa_{10}^2} (\partial c+\bar{\partial}\bar{c}) c\bar{c}\left(F_{Acd}F_{Bef}\eta^{ce}\eta^{df}-\frac{\eta_{AB}}{3!}|F|^2\right) e^{-\phi}\tilde{\psi}^Ae^{-\bar{\phi}}\bar{\tilde{\psi}}^B\,,\label{eqn:sec ord eqn NS}
\end{align}
\begin{align}
\frac{4}{g_c^2}\tilde{R}_{S^3}^{-2}Q_B\Bbb{P}|\Psi_2^{-\frac{1}{2},-\frac{1}{2}}\rangle=\frac{\pi}{4g_s\kappa_{10}^2} \frac{1}{(3!)^2} H_{abc}F_{def} c\bar{c} &\biggr(\eta e^{\phi/2}\Sigma^\alpha P_-^{10} (\tilde{\Gamma}^{abc}\tilde{\Gamma}^{def})_{\alpha}^{~\beta} e^{-\bar{\phi}/2}\overline{\Sigma}_\beta\nonumber\\
&-\bar{\eta}e^{-\phi/2}\Sigma_\alpha P_+^{10}(\tilde{\Gamma}^{abc}\tilde{\Gamma}^{def})^\alpha_{~\beta} e^{\bar{\phi}/2}\overline{\Sigma}^\beta\biggr)\,.
\end{align}

We shall first solve $\Bbb{P}\Psi_2^{-1,-1}.$ It is first instructive to note that the right-hand side of \eqref{eqn:sec ord eqn NS} is identical to that of
\begin{equation}
    4\pi \biggr[\frac{\partial}{\partial g^{AB}} \left(\sqrt{-G}\mathcal{L}\right) +\frac{1}{4}g_{AB} \frac{\partial}{\partial\Phi} \left(\sqrt{-G}\mathcal{L}\right)\biggr] (\partial c+\bar{\partial}\bar{c}) c\bar{c} e^{-\phi}\tilde{\psi}^Ae^{-\bar{\phi}}\bar{\tilde{\psi}}^B\,,
\end{equation}
in the weak field approximation, where $\mathcal{L}$ is the low-energy supergravity action. We shall write the second order solution $\Psi_2^{-1,-1}$ as
\begin{align}
\tilde{R}_{S^3}^{-2}\Bbb{P}\Psi_2^{-1,-1}=&\mathcal{G}_{AB} c\bar{c} e^{-\phi}\tilde{\psi}^Ae^{-\bar{\phi}}\bar{\tilde{\psi}}^B+\mathcal{D} c\bar{c}(\eta \bar{\partial}\bar{\xi}e^{-2\bar{\phi}}-\partial\xi e^{-2\phi}\bar{\eta})\nonumber\\
&+\frac{i}{2\sqrt{2}}\mathcal{F}_A(\partial c+\bar{\partial}\bar{c}) c\bar{c} (e^{-\phi}\tilde{\psi}^Ae^{-2\bar{\phi}}\bar{\partial}\bar{\xi}+e^{-2\phi}\partial\xi e^{-\bar{\phi}}\bar{\tilde{\psi}}^A)\,. 
\end{align}
We shall assume that the functions $\mathcal{G}_{AB},$ $\mathcal{D},$ and $\mathcal{F}_A$ only depend on the internal coordinates. We find
\begin{align}
Q_B\left( \tilde{R}_{S^3}^{-2}\Bbb{P}\Psi_2^{-1,-1}\right)=&\mathcal{A}_{AB} (\partial c+\bar{\partial}\bar{c})c\bar{c} e^{-\phi}\tilde{\psi}^A e^{-\bar{\phi}}\bar{\tilde{\psi}}^B+\mathcal{B}_Ac\bar{c}( \eta e^{-\bar{\phi}}\bar{\tilde{\psi}}^A+c.c.)\nonumber\\
&+\mathcal{C} (\partial c+\bar{\partial}\bar{c}) c\bar{c}\left(\eta e^{-2\phi}\bar{\partial}\bar{\xi}-e^{-2\phi}\partial\xi\bar{\eta}\right)\,,
\end{align}
where
\begin{equation}
\mathcal{A}_{AB}=-\frac{1}{4} \partial^2 \mathcal{G}_{AB}-\frac{1}{4} (\partial_B\mathcal{F}_A+\partial_A\mathcal{F}_B)\,, 
\end{equation}
\begin{equation}
\mathcal{B}_A=i\frac{1}{\sqrt{2}} \partial^B\mathcal{G}_{BA}+\frac{i}{\sqrt{2}} \mathcal{F}_A+i\frac{1}{\sqrt{2}} \partial_A\mathcal{D}\,,
\end{equation}
\begin{equation}
\mathcal{C}=\frac{1}{4}\partial^A\mathcal{F}_A-\frac{1}{4}\partial^2\mathcal{D}\,.
\end{equation}

We shall choose the following ansatz
\begin{equation}
\mathcal{F}_A=-\partial_A \mathcal{D}-\partial^B \mathcal{G}_{BA}\,,\quad \partial^2 \mathcal{D}=\partial^A \mathcal{F}_A\,,\quad \mathcal{D}=-\frac{1}{4} \mathcal{G}_{AB}\eta^{AB}+\delta\mathcal{D}\,.
\end{equation}
Then, the equation of motion for $\mathcal{A}_{AB}$ can be rewritten as
\begin{align}\label{eqn:eom aab}
    -\frac{1}{4}\partial^2\mathcal{G}_{AB}+\frac{1}{4} (\partial_B\partial^C\mathcal{G}_{CA}+\partial_A\partial^C\mathcal{G}_{CB}) +\frac{1}{2} \partial_A\partial_B \mathcal{D} =&\pi g_c^2  \left(T_{AB}-\frac{1}{8}\eta_{AB}T\right)\,. 
\end{align}
As we shall show in the next section \S\ref{sec:spacetime susy}, the flux choice we made preserves minimal spacetime supersymmetry, namely the imaginary-self-dual (ISD) condition. The ISD condition implies
\begin{equation}
    |H|^2=g_s^2|F|^2\,,
\end{equation}
which is also visible from the form of the fluxes we found in the previous section. Also, ISD condition implies that $T_{ab}=0.$ \footnote{This identity can be shown by rewriting $G= G_++G_-,$ and noting that the term in the action of the form $\int \sqrt{-g_4} dx^0\wedge dx^1\wedge dx^2 \wedge dx^3 \wedge A\wedge B,$ where $A$ and $B$ are threeforms, is trivial under the variation of the internal metric.} 

As one can check, the equation \eqref{eqn:eom aab} generalizes the beta function of the linear sigma model. The difference here is that the Ramond-Ramond fluxes are also included in the equation, which goes beyond the non-linear sigma model. This also indicates that $\mathcal{G}_{AB}$ denotes the string-frame metric.

We find the solutions to be
\begin{equation}
    \mathcal{G}_{\mu\nu}=-\frac{\pi g_c^2}{4}\eta_{\mu\nu}(\tilde{X}^4)^2 T\,,\quad \mathcal{G}_{ab} =\frac{3\pi g_c^2}{10}\eta_{ab} (\tilde{X}^4)^2 T\,,
\end{equation}
\begin{equation}
\mathcal{F}_\mu=0\,,\quad \mathcal{F}_a=-\frac{\pi g_c^2}{20} T\tilde{X}_a\,,\quad \mathcal{D}= -\frac{3\pi g_c^2}{10} T(\tilde{X}^4)^2 +\frac{\pi g_c^2}{40} T\tilde{X}_a\tilde{X}^a\,.
\end{equation}

For the detailed form and the derivation of the background solution, see \S\ref{sec:sec back app}. Note that we chose the profile of the metric to only depend on $\tilde{X}^4$ non-trivially, as $\tilde{X}^4$ corresponds to the radial direction in the low-energy supergravity background. We could've chosen a different profile. However, that will correspond to a background different from the Klebanov-Strassler solution. Note also that we added the Riemann tensor term to correctly capture the fact that the internal space is the deformed conifold. This, again, is in agreement with the non-linear sigma model approach to studying curved backgrounds, where one can treat the metric as a deformation of the flat metric by the second-order term of the normal coordinate expansion of the metric. As one can check, the curvature term we turned on is not exactly marginal. This, in turn, also agrees with that if we treat $h_{AB}$ in the linearized gravity $g_{AB}=\eta_{AB}+h_{AB}$ as the first order term in the normal coordinate expansion, the Ricci curvature evaluated at the higher order in the expansion does not vanish. To make the Ricci curvature vanish, one needs to turn on a new term at the second order in the expansion. This new term will start to show up in the fourth order in the large radius expansion, which is beyond the scope of this work.

We shall now study the RR sector. We write
\begin{equation}
\tilde{R}_{S^3}^{-2} \Bbb{P}|\Psi_2^{-\frac{1}{2},-\frac{1}{2}}\rangle= \mathfrak{F}^{\alpha\beta}c\bar{c} e^{-\phi/2}\Sigma_\alpha e^{-\bar{\phi}}\overline{\Sigma}_\beta\,.
\end{equation}
Then, we compute
\begin{align}
\tilde{R}_{S^3}^{-2}Q_B\Bbb{P}|\Psi_2^{-\frac{1}{2},-\frac{1}{2}}\rangle=& -\frac{1}{4}\partial^2\mathfrak{F}^{\alpha\beta}(\partial c+\bar{\partial}\bar{c})c\bar{c} e^{-\phi/2}\Sigma_\alpha e^{-\bar{\phi}/2}\overline{\Sigma}_\beta\nonumber\\
&+\frac{i}{2}c\bar{c} \left[ (\slashed{\partial} \mathfrak{F})_{\alpha}^{~\beta}\eta e^{\phi/2}\Sigma^\alpha e^{-\bar{\phi}/2}\overline{\Sigma}_\beta- (\partial^a\mathfrak{F}\Gamma_a)^\alpha_{~\beta}e^{-\phi/2}\Sigma_\alpha \bar{\eta}e^{\bar{\phi}/2}\overline{\Sigma}^\beta\right]\,.
\end{align}
We find the solution to be
\begin{equation}
\mathfrak{F}^{\alpha\beta}= P_+^{10} f_a(\tilde{X})\Gamma^{0123}\Gamma^a\,,
\end{equation}
where $f_a(\tilde{X})$ is a linear function in $\tilde{X}$ such that
\begin{equation}
\frac{4}{g_c^2}\partial^a f_a(\tilde{X})= i\frac{\pi}{2g_s\kappa_{10}^2}[H_3\wedge F_3]\,.
\end{equation}
In accordance with the low-energy supergravity, we shall choose 
\begin{equation}
f_4(\tilde{X})=i\frac{\pi g_c^2}{8g_s\kappa_{10}^2}[H_3\wedge F_3] \tilde{X}_4\,,
\end{equation}
and
\begin{equation}
f_i(\tilde{X})=0\,,
\end{equation}
for $i\neq4.$

\subsubsection{Third order equations}\label{sec:third order KS}
In this section, we shall study the third-order equations of motion. For our purposes, we do not need the complete detail of the third-order solutions other than their existence and some of their properties. We shall assume that the third-order solution exists. We find this assumption reasonable as the supergravity solution already exists to this order. We shall show that only the three form components of the RR tadpole survive. We shall also comment a particular form of the modulus at the third order, whose existence will play an important role later.

We write the third-order equations of motion. We shall only write the $\Bbb{P}$ projected part
\begin{align}
\frac{4}{g_c^2}\tilde{R}_{S^3}^{-3}Q_B\Bbb{P}|\Psi_3^{-1,-1}\rangle=&-\mathcal{G}\Bbb{P} \left[\frac{1}{3!}  \tilde{V}_{NSNS}^3+\frac{1}{2}\tilde{V}_{NS}\otimes\tilde{V}_{RR}^2\right]_{S^2}-\mathcal{G}\Bbb{P}\left[ \tilde{V}_{NSNS}\otimes \tilde{R}_{S^2}^{-2} \Psi_{2}^{-1,-1}\right]\nonumber\\
&-\mathcal{G}\Bbb{P} \left[ \tilde{V}_{RR}\otimes\tilde{R}_{S^3}^{-2} \Psi_{2}^{-\frac{1}{2},-\frac{1}{2}}\right]\,,
\end{align}
\begin{align}
    \frac{4}{g_c^2}\tilde{R}_{S^3}^{-3}Q_B\Bbb{P}|\Psi_3^{-\frac{1}{2},-\frac{1}{2}}\rangle=&-\mathcal{G}\Bbb{P}\left[ \frac{1}{3!}\tilde{V}_{RR}^3+\frac{1}{2}\tilde{V}_{NSNS}^2\otimes\tilde{V}_{RR} \right]-\mathcal{G}\Bbb{P}\left[ \tilde{V}_{NSNS}\otimes \tilde{R}_{S^3}^{-2}\Psi_2^{-\frac{1}{2},-\frac{1}{2}}\right]\nonumber\\
    &-\mathcal{G}\Bbb{P}\left[\tilde{V}_{RR}\otimes\tilde{R}_{S^3}^{-2}\Psi_2^{-1,-1}\right]\,.\label{eqn:third ord RR closed}
\end{align}

We shall first show that $\Psi_3^{-1,-1}$ enjoys moduli. We write a candidate modulus
\begin{align}
    \Psi_{3,m}^{-1,-1}= &c\bar{c} \biggr[\mathcal{G}^{3,m}_{AB} e^{-\phi}\tilde{\psi}^A e^{-\bar{\phi}}\bar{\tilde{\psi}}^B +\mathcal{D}^{3,m} (\eta e^{-2\bar{\phi}} \bar{\partial}\xi-e^{-2\phi}\partial\xi \bar{\eta}) \nonumber\\
    &\quad+F_A^{3,m} (\partial c+\bar{\partial}\bar{c}) (e^{-\phi}\tilde{\psi}^A e^{-2\bar{\phi}}\bar{\partial}\bar{\xi}+ (\partial c+\bar{\partial}\bar{c}) e^{-2\phi}\partial\xi e^{-\bar{\phi}}\bar{\tilde{\psi}}^A)\biggr]\,.
\end{align}
We find that if the following equations are met
\begin{equation}
    \partial^B\mathcal{G}_{AB}^{3,m}+\mathcal{F}_A^{3,m}+\partial_A\mathcal{D}^{3,m}=0\,,
\end{equation}
for linear functions $\mathcal{G}_{AB},$ $\mathcal{D}$ and a constant $\mathcal{F}_A,$ $\Psi_{3,m}^{-1,-1}$ is BRST-closed. Note that we assumed $\mathcal{G}_{AB}$ is symmetric.

We shall now show that the only non-trivial tadpole in the equation \eqref{eqn:third ord RR closed} is of the three-form flux. This implies that imposing the flux quantization by hand, $\Bbb{P}\Psi_3^{-\frac{1}{2},-\frac{1}{2}}$ only contains the three form flux components. We shall use the test field
\begin{equation}
    V_{t,R}= \mathcal{F}^{\alpha\beta} c\bar{c}e^{-\phi/2}\Sigma_\alpha e^{-\bar{\phi}/2}\overline{\Sigma}_\beta\,,
\end{equation}
where
\begin{equation}
    \mathcal{F}^{\alpha\beta}=(\mathcal{F}_1)_A (\Gamma^A)^{\alpha\beta}+\frac{1}{5!} (\mathcal{F}_5)_{ABCDE} (\Gamma^{ABCDE})^{\alpha\beta}\,.
\end{equation}

Let us first study
\begin{equation}
    \mathcal{A}_{3,R,1}=-\frac{1}{3!}\left\{ \tilde{V}_{RR}^3\otimes V_{t,R} \right\}\,.
\end{equation}
The matter CFT correlator takes the following form
\begin{align}
    &F_{abc}F_{def}F_{ghi}  (\Gamma^{abc})^{\alpha_1\beta_1} (\Gamma^{def})^{\alpha_2\beta_2}(\Gamma^{ghi})^{\alpha_3\beta_3} \left(\mathcal{F}_{1,A} \Gamma^A+\frac{1}{5!}\mathcal{F}_{5,ABCGH}\Gamma^{ABCGH} \right)^{\alpha_4\beta_4} \nonumber\\
    &\times\left( A_1(z) (\Gamma_D)_{\alpha_1\alpha_2} (\Gamma^D)_{\alpha_3\alpha_4} +A_2(z) (\Gamma_D)_{\alpha_1\alpha_3} (\Gamma^D)_{\alpha_2\alpha_4}+A_3(z) (\Gamma_D)_{\alpha_1\alpha_4}(\Gamma^D)_{\alpha_2\alpha_3} \right)\nonumber\\
    &\times \left( A_1(\bar{z}) (\Gamma_E)_{\beta_1\beta_2} (\Gamma^E)_{\beta_3\beta_4} +A_2(\bar{z}) (\Gamma_E)_{\beta_1\beta_3} (\Gamma^E)_{\beta_2\beta_4}+A_3(\bar{z}) (\Gamma_E)_{\beta_1\beta_4}(\Gamma^E)_{\beta_2\beta_3} \right)\,,
\end{align}
where $z$ is the cross-ratio. There are essentially two types of spinor index contractions. First,
\begin{equation}
    F_{abc}F_{def}F_{ghi}\text{Tr}\left( \Gamma^D \Gamma^{abc} \Gamma_D \Gamma^{def}\right) \text{Tr} \left(\Gamma^E \gamma^{ghi}\Gamma_E \left(\mathcal{F}_{1,A}\Gamma^A+\frac{1}{5!}\mathcal{F}_{5,ABCGH} \Gamma^{ABCGH}\right)\right)\,.
\end{equation}
Second,
\begin{equation}
    F_{abc}F_{def}F_{ghi}   \text{Tr}\left( \Gamma^E \Gamma^{def}\Gamma_D\Gamma^{abc}\Gamma_E \Gamma^{ghi}\Gamma^D \left(\mathcal{F}_{1,A}\Gamma^A+\frac{1}{5!}\mathcal{F}_{5,ABCGH} \Gamma^{ABCGH}\right)\right)\,.
\end{equation}
Both of the traces vanish. This result was already expected from the spacetime covariance. If the above contractions weren't trivial, there must be a way to write a term involving three $F_3$ and $F_i,$ for $i=1$ or $5,$ without further introducing derivatives. However, because Ramond-Ramond fluxes are completely anti-symmetric tensors, there is simply no way to write such a term in the action. This is in agreement with the low-energy supergravity \cite{Policastro:2008hg}. 

Now, we shall, in turn, comment on the rest of the terms 
\begin{equation}
    \mathcal{A}_{3,R,2}:=-\frac{1}{2!} \left\{ \tilde{V}_{NSNS}^2\otimes V_{t,R}\otimes \tilde{V}_{RR}\right\}\,,\quad \mathcal{A}_{3,R,3}:=-\left\{ \tilde{V}_{NSNS}\otimes \tilde{R}_{S^3}^{-2}\Psi_2^{-\frac{1}{2},-\frac{1}{2}}\otimes V_{t,R}\right\}\,,
\end{equation}
\begin{equation}
    \mathcal{A}_{3,R,4}=-\left\{ \tilde{V}_{RR} \otimes\tilde{R}_{S^3}^{-2}\Psi_2^{-1,-1}\otimes V_{t,R} \right\}\,.
\end{equation}
Note that the $\tilde{V}_{NSNS}$ stands for anti-symmetric two-form, $\Psi_2^{-\frac{1}{2},-\frac{1}{2}}$ is a fully anti-symmetric five-form flux. Therefore, $\mathcal{A}_{3,R,2}$ vanishes as well because there is no way to write down a term in the action including three three-form fluxes and either one-form or five-form without introducing derivatives, as we showed already. For $i=3,4$ the structure of the off-shell amplitude is rather simple, as they are three-point amplitudes. As it is well known, there is no three-point coupling involving  $B_2,$ $F_5,$ and $F_i$ for $i=1,~5.$ This is again due to the spacetime covariance. Therefore, we conclude $\mathcal{A}_{3,R,3}=0$ as well. Lastly, let us discuss $\mathcal{A}_{3,R,4}.$ The only non-trivial contribution to $\mathcal{A}_{3,R,4}$ can be potentially generated by
\begin{equation}
    \mathcal{A}_{3,R,4}=-\left\{ \tilde{V}_{RR}\otimes V_{t,R}\otimes \mathcal{G}_{AB} e^{-\phi}\tilde{\psi}^Ae^{-\bar{\phi}}\bar{\tilde{\psi}}^B\right\}\,,
\end{equation}
where the matter CFT correlator is proportional to the following factor
\begin{equation}
    \frac{1}{3!}\mathcal{G}_{AB}F_{abc}\text{Tr}\left( \Gamma^A \Gamma^{abc}\Gamma^B  \left(\mathcal{F}_{1,A}\Gamma^A+\frac{1}{5!}\mathcal{F}_{5,CDEFG} \Gamma^{CDEFG}\right)\right)\,.
\end{equation}
The above expression again vanishes because $\mathcal{G}_{AB}$ is a symmetric rank 2 tensor. Therefore, we conclude that only the Ramond-Ramond threeform in the eom receives non-trivial source terms.

\subsection{Spacetime supersymmetry}\label{sec:spacetime susy}
In this section, we shall study the spacetime supersymmetry of the background up to the first order in the large radius expansion. For a more detailed study of spacetime supersymmetry of flux vacua, see, for example, \cite{Cho-Kim2}.  We shall also comment on the distinction between D3-branes and anti-D3-branes.

In string field theory, spacetime supersymmetry is a fermionic gauge symmetry that does not alter the background solution \cite{Sen:2015uoa}. We write the spacetime supercharge as
\begin{equation}
    \Lambda=\epsilon^\alpha c\bar{c} e^{-\phi/2}\Sigma_\alpha e^{-2\bar{\phi}}\bar{\partial}\bar{\xi}+\bar{\epsilon}^\alpha c\bar{c} e^{-2\phi}\partial\xi e^{-\bar{\phi}/2}\overline{\Sigma}_\alpha\,.
\end{equation}
We can decompose the ten-dimensional spinor $\epsilon^\alpha$ into four-dimensional and six-dimensional spinors $\epsilon_4^{\alpha_{(4)}}$ and $\epsilon_6^{\alpha_{(6)}}$
\begin{equation}
    \epsilon^\alpha=\epsilon_4^{\alpha_{(4)}}\otimes \epsilon_6^{\alpha_{(6)}}+\epsilon_4^{\dot{\alpha}_{(4)}}\otimes \epsilon_6^{\dot{\alpha}_{(6)}}\,,
\end{equation}
\begin{equation}
    \bar{\epsilon}^\alpha=\bar{\epsilon}_4^{\alpha_{(4)}}\otimes \bar{\epsilon}_6^{\alpha_{(6)}}+\bar{\epsilon}_4^{\dot{\alpha}_{(4)}}\otimes \bar{\epsilon}_6^{\dot{\alpha}_{(6)}}\,.
\end{equation}
As we have defined $\Sigma_\alpha$ as a chiral spin field
\begin{equation}
    (\Gamma^{0\dots9})_\alpha^{~\beta}\Sigma_\beta=\Sigma_\alpha\,,
\end{equation}
$\epsilon^\alpha$ must satisfy the following chirality condition
\begin{equation}
    \epsilon^\alpha (\Gamma^{0\dots 9})_\alpha^{~\beta} =\epsilon^\beta\,,\quad (\Gamma^{0\dots9})^\alpha_{~\beta} \epsilon^\beta=-\epsilon^\alpha\,.
\end{equation}
We choose the chiralities of $\alpha_{(4)}$ and $\alpha_{(6)}$ such that
\begin{equation}
    i(\Gamma^{0123})^{\beta_{(4)}}_{~\alpha_{(4)}}\epsilon_4^{\alpha_{(4)}}= \epsilon_4^{\beta_{(4)}}\,, -i(\Gamma^{4\dots9})^{\alpha_{(6)}}_{~\beta_{(6)}} \epsilon_6^{\beta_{(6)}}=\epsilon_6^{\alpha_{(6)}}\,,
\end{equation}
and
\begin{equation}
    i(\Gamma^{0123})^{\dot{\alpha}_{(4)}}_{~\dot{\beta}_{(4)}}\epsilon_4^{\dot{\beta}_{(4)}}= -\epsilon_4^{\dot{\alpha}_{(4)}}\,, -i(\Gamma^{4\dots9})^{\dot{\alpha}_{(6)}}_{~\dot{\beta}_{(6)}} \epsilon_6^{\dot{\beta}_{(6)}}=-\epsilon_6^{\dot{\alpha}_{(6)}}\,.
\end{equation}

Given a background solution $\Psi,$ the requirement that the supersymmetry transformation does not alter the background is phrased as
\begin{equation}
    \frac{4}{g_c^2}Q_B|\Lambda\rangle+\sum_n \frac{1}{n!} \mathcal{G}\left[ \Lambda\otimes (\Psi)^{\otimes n}\right]^c=0\,.
\end{equation}
The above equation can be solved perturbatively by expanding
\begin{equation}
    \Lambda=\sum_i \tilde{R}_{S^3}^{i} \Lambda_i\,.
\end{equation}

The zeroth order equation is
\begin{equation}
    Q_B|\Lambda_0\rangle= -\frac{1}{4} (\partial c+\bar{\partial}\bar{c}) \partial^2 \Lambda_0 +\frac{i}{2} (\slashed{\partial}\epsilon)_\alpha\eta c\bar{c}e^{\phi/2}\Sigma^\alpha e^{-2\bar{\phi}}\bar{\partial}\bar{\xi}-\frac{i}{2}(\slashed{\partial}\bar{\epsilon})_\alpha\bar{\eta}c\bar{c}e^{-2\phi}\partial\xi e^{\bar{\phi}/2}\overline{\Sigma}^\alpha\,.
\end{equation}
We can, therefore, conclude that constant spinors $\epsilon$ and $\bar{\epsilon}$ are valid supercharges to the leading order in the large radius expansion.

A few comments are in order. As we are working in a small local patch of the Klebanov-Strassler throat in the large radius expansion, to the leading order in the expansion, all constant components of $\epsilon_6^{\alpha_{(6)}}$ are valid supercharges. However, the background solution to higher orders in $1/\tilde{R}_{S^3}$ creates a non-trivial tadpole for some of the internal spinors. In particular, in the fluxless background, the curvature corrections will force the spinor to be covariantly constant. And, the fluxes, as we shall see, pick out, again, the covariantly constant spinor. 

The first-order equation for the spacetime supersymmetry reads \cite{Cho-Kim2}
\begin{equation}
    \frac{4}{g_c^2}\tilde{R}_{S^3}^{-1}Q_B\Bbb{P}|\Lambda_1\rangle=-i\frac{C_{S^2}}{\sqrt{2}} \left[\frac{1}{16\pi} H_{ijA}(\Gamma^{ij})^\alpha_{~\beta}\epsilon^\beta -\frac{1}{3!}\frac{g_s}{16\pi}F_{ijk} (\Gamma^{ijk} \Gamma_A)^{\alpha}_{~\gamma}\bar{\epsilon^\gamma} \right] c\bar{c}e^{-\phi/2}\Sigma_\alpha e^{-\bar{\phi}}\bar{\tilde{\psi}}^A+c.c\,.
\end{equation}
We have chosen the fluxes such that
\begin{equation}
    \frac{1}{3!}g_s F_{ijk}\Gamma^{ijk}=\frac{1}{3!} H_{ijk}\Gamma^{ijk}\Gamma_6 \,,
\end{equation}
which is equivalent to the ISD condition. We are thus led to conclude that the spinors that satisfy
\begin{equation}
    \epsilon^\beta = -(\Gamma_6)^\beta_{~\gamma} \bar{\epsilon}^\gamma\,,
\end{equation}
solves the spacetime supersymmetry eom, and therefore, they correspond to the spacetime supercharges. The preserved supercharges can also be written as
\begin{equation}
    \epsilon^\alpha= -(\Gamma^{0123})^\alpha_{~\beta}\bar{\epsilon}^\beta= \bar{\epsilon}^\beta (\Gamma^{0123})_\beta^{~\alpha}\,.
\end{equation}
We shall henceforth declare that D3-branes that respect the same supercharges supersymmetric D3-branes and the ones that preserve the opposite supercharges anti-D3-branes. Therefore, for the anti-D3-branes, the boundary condition for the spin fields is determined to be
\begin{equation}
    \Sigma_\alpha(z)=-(\Gamma^{0123})_\alpha^{~\beta}\overline{\Sigma}_\beta(\bar{z})\,.
\end{equation}

\section{Anti-D3-branes in KS throat}\label{sec:anti SFT}
In the previous section, we studied the perturbative background solution in string field theory that corresponds to the Klebanov-Strassler solution in the large radius expansion. We will add a stack of p anti-D3-branes into the spectrum and compute the open string background solution up to the third order in $\tilde{R}_{S^3}^{-1}$ expansion. Then, we shall compute the spectrum of the radial mode of the puffed anti-D3-brane to study the stability of the open string background.

\subsection{Perturbative background solution with anti-D3-branes}
To understand the perturbation by $p$ anti-D3-branes, we shall first study the overall normalization of the disk diagram. The overall normalization of the disk amplitude is suppressed by $\mathcal{O}(g_s \tilde{R}_{S^3}^{-6}).$ Note that an additional factor of $p$ shall be included for the computation of the closed string disk-one-point and the corresponding string bracket. Therefore, the backreaction to the geometry due to the p anti-D3-brane is suppressed by a factor of
\begin{equation}
\frac{pg_s}{R^6}\,,
\end{equation}
compared to $C_{S^2}.$ We shall define a new parameter
\begin{equation}
\Lambda:=\frac{p g_s}{R^4}=\frac{p}{g_sM^2}\,,
\end{equation}
which measures the size of the backreaction caused by the $p$ anti-D3-branes at order $R^{-2}$ compared to the bare sphere diagram. A cautionary remark is in order. When computing the tension of the anti-D3-brane stack in the low-energy supergravity, one obtains the following scaling
\begin{equation}
\frac{p g_s \epsilon^{8/3}}{ (g_sM)^2}\,,
\end{equation}
compared to the bare sphere coupling. It should be noted that the naive scaling from our counting seems to differ because we rescaled the spacetime coordinates such that the sphere diagram comes with the conventional overall normalization, which is a convenient choice to ensure that the factorization works properly. Had we not rescaled the spacetime coordinates, we would have obtained the same scaling as the power counting done in the low-energy supergravity. However, we shall stick to the convention we are using.

With this introduction of new perturbative expansion parameter $\Lambda,$ we can employ a double expansion
\begin{equation}
\Psi^c=\sum_{n,m}\tilde{R}_{S^3}^{-n}\Lambda^m \Psi_{n,m}^c\,,
\end{equation} 
\begin{equation}
\Psi^o=\sum_{n,m} \tilde{R}_{S^3}^{-n}\Lambda^m \Psi_{n,m}^o\,.
\end{equation}
We shall illustrate how to study the open string background. We shall choose to work up to $n=3$ and $m=0$, which corresponds to $\alpha'^{3/2}$ corrected open string background. The spacetime action computed with this background solution is order $\alpha'^2.$ The determination of the stability up to higher orders in $\alpha'$ is an important problem left for future work. Also, we shall assume that $\Psi_{3,0}^c$ exists. As the low-energy supergravity solution already exists up to $\alpha'^{2},$ this assumption is reasonable. For our analysis, the precise form of $\Psi_{3,0}^c$ won't be needed. 

The equations of motion we shall solve are
\begin{align}
&\frac{4}{g_c^2} Q_B|\Psi^c\rangle+ \sum_{n,m} \frac{1}{n!m!}\mathcal{G}\left[ (\Psi^c)^n;(\Psi^o)^m\right]^c+\mathcal{G}[]_{D^2}^c=0\,,\\
&\frac{1}{g_o^2}Q_B|\Psi^o\rangle + \sum_{n,m}\frac{1}{n!m!}\mathcal{G}\left[ (\Psi^c)^n;(\Psi^o)^m\right]^o=0\,.
\end{align}

The equations at order $(n,m)=(1,0)$ are
\begin{align}
&\frac{4}{g_c^2} \tilde{R}_{S^3}^{-1}Q_B|\Psi_{1,0}^c\rangle=0\,,\\
&\frac{1}{g_o^2} \tilde{R}_{S^3}^{-1} Q_B|\Psi_{1,0}^o\rangle + \tilde{R}_{S^3}^{-1}\mathcal{G}[\Psi_{1,0}^c]^o_{D^2}=0\,.
\end{align}
The equations at order $(n,m)=(2,0)$ are
\begin{align}
&\frac{4}{g_c^2} \tilde{R}_{S^3}^{-2} Q_B|\Psi_{2,0}^c\rangle+\frac{\tilde{R}_{S^3}^{-2}}{2} \mathcal{G}\left[ \Psi_{1,0}^c\otimes \Psi_{1,0}^c\right]_{S^2}^c =0\,,\\
&\frac{4}{g_o^2} \tilde{R}_{S^3}^{-2} Q_B|\Psi_{2,0}^o\rangle+ \tilde{R}_{S^3}^{-2}\mathcal{G}\left[\frac{1}{2}\Psi_{1,0}^c\otimes\Psi_{1,0}^c+\frac{1}{2}\Psi_{1,0}^o\otimes\Psi_{1,0}^o+\Psi_{1,0}^c\otimes \Psi_{1,0}^o+\Psi_{2,0}^c\right]_{D^2}^o=0\,.
\end{align}
The equations at order $(n,m)=(3,0)$ are
\begin{align}
&\frac{4}{g_c^2}\tilde{R}_{S^3}^{-3}  Q_B|\Psi_{3,0}^c\rangle+\tilde{R}_{S^3}^{-3} \mathcal{G}\left[\frac{1}{3!}(\Psi_{1,0}^c)^3+  \Psi_{1,0}^c\otimes\Psi_{2,0}^c\right]_{S^2}^c =0\,,\\
&\frac{4}{g_o^2} \tilde{R}_{S^3}^{-3} Q_B|\Psi_{3,0}^o\rangle+\tilde{R}_{S^3}^{-3}\mathcal{G}\left[\frac{1}{3!}(\Psi_{1,0}^c+\Psi_{1,0}^o)^3+(\Psi_{1,0}^c+\Psi_{1,0}^o)\otimes(\Psi_{2,0}^c+\Psi_{2,0}^o)+\Psi_{3,0}^c\right]^o_{D^2}=0\,.
\end{align}

As we already solved the closed string field equations of motion up to the second order, and we won't be needing the detailed form of $\Psi_{3,0}^c,$ we shall proceed to study the open string background with one comment on $\Psi_{3,0}^c.$ $\Psi_{3,0}^c$ may have constant terms that are closed under $Q_B.$ This represents moduli at order $\tilde{R}_{S^3}^{-3}.$ We shall choose the moduli such that the anti-D3-brane located at the origin of the throat solves the equation of motion for diagonal D3-brane ``moduli", as expected from the low-energy supergravity. 

\subsection{First order equation}\label{sec:first order KPV}
We shall make an ansatz for the first-order open string background solution
\begin{equation}
\tilde{R}_{S^3}^{-1}\Bbb{P}\Psi_{1,0}^o= f_i ce^{-\phi}\psi^i\,,
\end{equation}
where $f_i\propto \alpha_i$ is a constant that will be determined later, and $\alpha^i$ is a generator of $p$ dimensional representation of the $SU(2)$ Lie algebra. $\alpha^i$ satisfies
\begin{equation}
[\alpha^i,\alpha^j]=2i \epsilon^{ijk}\alpha_k\,,
\end{equation}
where $i,j,k\in [5,6,7].$ 

Let us solve the first-order equation for the open string field
\begin{equation}
\frac{4}{g_o^2}\tilde{R}_{S^3}^{-1}Q_B |\Psi_{1,0}^o\rangle =-\tilde{R}_{S^3}^{-1}\mathcal{G} [\Psi_{1,0}^c]_{D^2}^o\,.
\end{equation}
$(1-\Bbb{P})$ projected component of the equation can be easily solved
\begin{equation}
\frac{4}{g_o^2}\tilde{R}_{S^3}^{-1} (1-\Bbb{P}) |\Psi_{1,0}^o\rangle=-\frac{b_0}{L_0} \tilde{R}_{S^3}^{-1}(1-\Bbb{P})\mathcal{G}[\Psi_{1,0}^c]_{D^2}^o\,.
\end{equation}
For $L_0$ nilpotent component of the background solution to exist, because $Q_B f_i \alpha^i=0,$ the open string tadpole must be absent. To check that the tadpole is absent, we can compute 
\begin{equation}
\{ \Psi_{1,0}^c; V_t^o\}_{D^2}\,,
\end{equation}
where $V_t^o$ is a test field that can take the following forms
\begin{equation}
V_t^o=c e^{-\phi}\tilde{\psi}^a\,,\quad c\partial c e^{-2\phi}\partial\xi\,.
\end{equation}
Due to the ghost structure, $c\partial c e^{-2\phi}\partial\xi$ cannot have a non-trivial overlap. Therefore, we shall study $V_t^o=ce^{-\phi}\tilde{\psi}^a.$ Let us first study
\begin{equation}
\{\tilde{V}_{RR};V_t^o\}_{D^2}\,.
\end{equation}
As the total picture number is $-2,$ we shall not insert a PCO. As there are no remaining moduli, we shall place the closed string vertex at the origin of the disk and fix the location of $V_t^o.$ The tensor structure of the correlator contains a factor of
\begin{equation}
F_{abc}\text{Tr}\left( \tilde{\Gamma}^{0123} \tilde{\Gamma}^{abc}\tilde{\Gamma}^d\right) =0\,,
\end{equation}
as $F$ is an anti-symmetric tensor. Therefore, the Ramond-Ramond flux does not yield a non-trivial result. For the similar reason, $\{\tilde{V}_{NSNS};V_t^o\}_{D^2}$ vanishes as well. The only non-trivial contraction involves the following term of the PCO
\begin{equation}
\mathcal{X}\supset e^\phi T_F\,.
\end{equation}
Because the amplitude is on-shell and all the vertex operators are primary, we can place the PCO anywhere without loss of generality. We shall, therefore, place the PCO at $i.$ Then the correlator has the following tensor structure
\begin{equation}
H_{abc} \delta^{a}_{e} (\eta^{bc}\eta^{de} \oplus \eta^{bd}\eta^{ce}\oplus\eta^{be}\eta^{cd})\,.
\end{equation}
As every contraction involves contracting anti-symmetric indices, we conclude
\begin{equation}
\{\Psi_{1,0}^c;V_t^o\}_{D^2}=0\,,
\end{equation}
which solves the first-order equation.

\subsection{Second order equation}
Let us now study the second-order equation. Same as before $(1-\Bbb{P})$ projected component is solved as
\begin{equation}
\frac{4}{g_o^2}\tilde{R}_{S^3}^{-2} (1-\Bbb{P})|\Psi_{2,0}^o\rangle= -\frac{b_0}{L_0} (1-\Bbb{P})\tilde{R}_{S^3}^{-2}\mathcal{G}\left[\frac{1}{2}\Psi_{1,0}^c\otimes\Psi_{1,0}^c+\frac{1}{2}\Psi_{1,0}^o\otimes\Psi_{1,0}^o+\Psi_{1,0}^c\otimes \Psi_{1,0}^o+\Psi_{2,0}^c\right]_{D^2}^o\,.
\end{equation}

We shall now solve the $\Bbb{P}$ projected component of the equation 
\begin{equation}
\frac{4}{g_o^2} \tilde{R}_{S^3}^{-2} Q_B\Bbb{P}|\Psi_{2,0}^o\rangle+ \tilde{R}_{S^3}^{-2}\mathcal{G}\Bbb{P}\left[\frac{1}{2}\Psi_{1,0}^c\otimes\Psi_{1,0}^c+\frac{1}{2}\Psi_{1,0}^o\otimes\Psi_{1,0}^o+\Psi_{1,0}^c\otimes \Psi_{1,0}^o+\Psi_{2,0}^c\right]_{D^2}^o=0\,.
\end{equation}
To make various degeneration limits more manifest, we shall rewrite the above equation as
\begin{align}\label{eqn:second order open eom}
\frac{4}{g_o^2}\tilde{R}_{S^3}^{-2}Q_B\Bbb{P}|\Psi_{2,0}^o\rangle=&-\tilde{R}_{S^3}^{-2}\mathcal{G}\Bbb{P}\left[\frac{1}{2} (\Psi_{1,0}^c)^2\right]_{D^2}^o-\tilde{R}_{S^3}^{-2}\mathcal{G}\Bbb{P}\left[ \Psi_{1,0}^c\otimes \left( -g_o^2\frac{b_0}{L_0} (1-\Bbb{P})\mathcal{G} [\Psi_{1,0}^c]_{D^2}^o\right)\right]_{D^2}^o\nonumber\\
&-\frac{1}{2} \tilde{R}_{S^3}^{-2}\mathcal{G}\Bbb{P} \left[ \left( -g_o^2\frac{b_0}{L_0} (1-\Bbb{P})\mathcal{G} [\Psi_{1,0}^c]_{D^2}^o\right)\otimes \left( -g_o^2\frac{b_0}{L_0} (1-\Bbb{P})\mathcal{G} [\Psi_{1,0}^c]_{D^2}^o\right)\right]_{D^2}^o\nonumber\\
&-\tilde{R}_{S^3}^{-2} \mathcal{G}\Bbb{P}\left[ \Bbb{P} \Psi_{2,0}^c\right]_{D^2}^o-\tilde{R}_{S^3}^{-2}\mathcal{G}\Bbb{P}\left[ -\frac{g_c^2}{4}\frac{b_0^+}{L_0^+}(1-\Bbb{P})\mathcal{G} \left[ \frac{1}{2}(\Psi_{1,0}^c)^2 \right]_{S^2}^c\right]_{D^2}^o\nonumber\\
&-\tilde{R}_{S^3}^{-2}\mathcal{G}\Bbb{P}\left[ \frac{1}{2} \Bbb{P}\Psi_{1,0}^o\otimes\Bbb{P}\Psi_{1,0}^o+\Psi_{1,0}^c\otimes\Bbb{P}\Psi_{1,0}^o\right]_{D^2}^o\,,
\end{align}
where the first three lines affect the center of mass of the D-branes, and the last line affects the relative positions of the D-branes.

To compute the source terms, we shall compute an overlap between the source terms and the test field $V_t$ that can take the following forms
\begin{equation}
V_t^o=ce^{-\phi}\tilde{\psi}^A\,,\quad c\partial c e^{-2\phi}\partial\xi\,.
\end{equation}

We shall first study the first three lines in the large stub limit.  Because $\Psi_{1,0}^c$ in the string bracket are zero momentum states, states with $L_0<0$ cannot propagate in the Feynman diagrams. Therefore, only the following two terms actually produce non-trivial results in the large stub limit
\begin{equation}
-\tilde{R}_{S^3}^{-2}\mathcal{G}\Bbb{P}\left[\frac{1}{2}(\Psi_{1,0}^c)^2\right]_{D^2}^o\,,\quad -\tilde{R}_{S^3}^{-2}\mathcal{G}\Bbb{P}\left[\Bbb{P}\Psi_{2,0}^c\right]_{D^2}^o\,.
\end{equation}

Let us start by evaluating
\begin{equation}
-\tilde{R}_{S^3}^{-2}\mathcal{G}\Bbb{P}\left[\frac{1}{2}(\Psi_{1,0}^c)^2\right]_{D^2}^o=-\mathcal{G}\Bbb{P}\left[ \frac{1}{2}\tilde{V}_{NSNS}^2+\frac{1}{2}\tilde{V}_{RR}^2+\tilde{V}_{NSNS}\otimes\tilde{V}_{RR}\right]_{D^2}^o\,.
\end{equation}
We compute
\begin{equation}
\left\{ V_{t,1}^o \otimes\frac{1}{2} \tilde{V}_{NSNS}^2\right\}_{D^2}\,,
\end{equation}
where $V_{t,1}^o=ce^{-\phi}\psi^A.$ As the total picture number is $-5,$ we need to insert in total three PCOs. Furthermore, there are a total of 5 worldsheet fermion fields to contract. An odd number of $\mathcal{X}\supset e^\phi T_F$ terms should be contracted to have a non-vanishing contraction. We shall argue that all of these contractions vanish. When all of the $e^\phi T_F$ terms in the three PCOs are contracted, at least two of the bosons in $e^\phi T_F$ shall be contracted against each other. Otherwise, we will obtain at least a second derivative of $B$ that vanishes. As a result, only one of the spacetime bosons in $\mathcal{X}\supset e^\phi T_F$ can be contracted against $\tilde{V}_{NSNS},$ and the other $\tilde{V}_{NSNS}$ shall be evaluated to zero as we are placing the anti-D3-brane stack at the origin where the $B$ fields vanish. Hence, we conclude that in the interior of the moduli space, $\{V_{t,1}^o \otimes \tilde{V}_{NSNS}^2\}$ vanishes. Similarly, we can argue that the boundary contribution also vanishes. At the boundary of the moduli space, to respect the boundary conditions on the PCOs, we must perform vertical integrations \cite{Sen:2015hia}. If the initial position of the PCOs were $p_1,~p_2,$ and $p_3,$ we can sequentially move them to $W_1,~W_2,$ and $W_3.$ The effect of the vertical integration can be attained by replacing the integral measure in the interior of the moduli space
\begin{equation}
dt_1 dt_2 \oint dwb(w) \oint dw' b(w') \mathcal{X}(p_1)\mathcal{X}(p_2)\mathcal{X}(p_3)\,,
\end{equation}
with
\begin{align}
&dt_\partial \oint dw b(w) \biggr[ (\xi(p_1)-\xi(W_1))\mathcal{X}(p_2)\mathcal{X}(p_3)+\mathcal{X}(W_1) (\xi(p_2)-\xi(W_2))\mathcal{X}(p_3)\nonumber\\&+\mathcal{X}(W_1)\mathcal{X}(W_2)(\xi(p_3)-\xi(W_3))\biggr]\nonumber\\
&dt_\partial  \mathcal{X}(W_1) \frac{W_2}{\partial t_\partial} \partial\xi(W_1)(\xi(p_3)-\xi(W_3))\,.
\end{align}
To yield a non-trivial contribution, we still need to contract $e^\phi T_F$ from one of the PCOs. But, there are not enough spacetime bosons to contract against $B.$ Therefore, we obtain zero from the vertical integration as well. 

We shall now compute
\begin{equation}
\left\{ V_{t,2}^o\otimes\frac{1}{2}\tilde{V}_{NSNS}^2\right\}_{D^2}\,,
\end{equation}
where $V_{t,2}^o= c\partial c e^{-2\phi}\partial\xi.$ To have a non-trivial correlation, $-\partial \eta be^{2\phi}-\partial(\eta be^{2\phi})$ from one of the PCOs and $e^\phi T_F$ from two of the PCOs must be contracted. We shall fix the location of one $\tilde{V}_{NSNS}$ at $i,$ the location of $V_{t,2}^o$ at $0,$ and we shall choose the position of the remaining vertex as the modulus. Let us place one of the PCOs at $V_{t,2}^o,$ and two other PCOs at the movable vertex operator. We, therefore, find the following correlator contribution from the ghost sector
\begin{equation}
\langle c(i)c(-i)\partial c(0)\rangle\,,
\end{equation}
which vanishes. Note that the vertical integration also vanishes. To saturate the background $\phi$ charge, the vertical integration will involve contraction of $\xi$ due to one jumping PCO, and two $-\partial \eta be^{2\phi}-\partial(\eta be^{2\phi})$ from two other PCOs. However, this contribution vanishes because we don't have a worldsheet bosons to contract against $\tilde{V}_{NSNS}.$ Therefore, we find
\begin{equation}\label{eqn:sec open 1 van1}
\mathcal{G}\Bbb{P}\left[ \frac{1}{2}\tilde{V}_{NSNS}^2\right]_{D^2}^o=0\,.
\end{equation}

We compute
\begin{equation}
\left\{V_{t,1}^o\otimes\frac{1}{2}\tilde{V}_{RR}^2\right\}_{D^2}\,.
\end{equation}
Because the total picture number provided by the vertex operators is $-3,$ we shall insert one PCO. To saturate the background $\phi$ charge, the only admissible term in the PCO is $e^\phi T_F.$ As the RR threeform in the leading large radius approximation is constant, contraction against the spacetime boson yields zero. The vertical integration is also zero as well since there is no way to saturate the background $\phi$ charge.

We compute
\begin{equation}
\left\{V_{t,2}^o\otimes \frac{1}{2}\tilde{V}_{RR}^2\right\}_{D^2}\,.
\end{equation}
To have a non-vanishing contribution, $\mathcal{X}\supset-\partial \eta be^{2\phi}-\partial(\eta be^{2\phi})$ shall be contracted. We can freely move the PCO as the correlator is effectively on-shell. We shall place the PCO at $V_{t,2}^o,$ and as a result, the ghost sector correlator is written as
\begin{equation}
\langle c(i)c(-i)\partial c(0)\rangle=0\,.
\end{equation}
Note that we placed $\tilde{V}_{RR}$ at $i,$ and $V_{t,2}^o$ at $0.$ The vertical integration vanishes again. Therefore, we find
\begin{equation}\label{eqn:sec open 1 van2}
\mathcal{G}\Bbb{P}\left[\frac{1}{2}\tilde{V}_{RR}^2\right]_{D^2}^o=0\,.
\end{equation}

We compute
\begin{equation}
\left\{V_{t,1}^o\otimes \tilde{V}_{NSNS}\otimes\tilde{V}_{RR}\right\}_{D^2}\,.
\end{equation}
The total picture number before the insertion of PCOs is $-4.$ Therefore, we shall insert two PCOs. There are two choices of the PCO contractions that can saturate the background $\phi$ charge. Either we contract $e^\phi T_F$ from both of the PCOs, or we contract $c\partial\xi$ from one PCO and $-\partial\eta be^{2\phi}-\partial(\eta b e^{2\phi})$ from the other. Both of them vanish for a slightly different reason. Let us start with the first case. Because the B-field was chosen to vanish at the anti-D3-brane, at least one of the spacetime boson $\partial X$ should be contracted against the B-field. Then, the remaining $\partial X$ from the other PCO cannot be contracted without yielding zero as both $H_3$ and $F_3$ are constants in the leading large radius limit. Therefore, the first case vanishes. In the second case, because there is no spacetime boson $\partial X$ that can be contracted against the B-field, the contraction is automatically zero. For the same reason, the vertical integration also vanishes.

Let us compute
\begin{equation}
\left\{V_{t,2}^o\otimes \tilde{V}_{NSNS}\otimes\tilde{V}_{RR}\right\}_{D^2}\,.
\end{equation}
To saturate the background $\phi$ charge and contract $\partial X$ against the B-field, one $e^\phi T_F$ should be used from one of the PCOs, and $-\partial \eta be^{2\phi}-\partial(\eta be^{2\phi})$ from the other PCO shall be contracted. We shall fix the location of one $\tilde{V}_{NSNS}$ at $i,$ the location of $V_{t,2}^o$ at $0,$ and we shall choose the position of the remaining vertex as the modulus. Same as before, we shall bring one of the PCOs to $V_{t,2}^o,$ and the other PCO to the movable vertex. We therefore find the following correlator contribution from the ghost sector
\begin{equation}
\langle c(i)c(-i)\partial c(0)\rangle\,,
\end{equation}
which vanishes. Also note that there can be, in principle, an additional contribution from the vertical integration. However, such a contribution actually vanishes, as to saturate the background $\phi$ charge the vertical integration cannot involve contractions of $e^\phi T_F$ in one of the PCOs that are not jumping. Therefore, we conclude
\begin{equation}\label{eqn:sec open 1 van3}
\mathcal{G}\Bbb{P}\left[\tilde{V}_{NSNS}\otimes\tilde{V}_{RR}\right]_{D^2}^o=0\,.
\end{equation}

By collecting \eqref{eqn:sec open 1 van1}, \eqref{eqn:sec open 1 van2}, and \eqref{eqn:sec open 1 van3}, we conclude
\begin{equation}
\tilde{R}_{S^3}^{-2}\mathcal{G}\Bbb{P}\left[\frac{1}{2}(\Psi_{1,0}^c)^2\right]_{D^2}^o=0\,.
\end{equation}

Let us now study
\begin{equation}
\tilde{R}_{S^3}^{-2}\mathcal{G}\Bbb{P}\left[\Bbb{P}\Psi_{2,0}^c\right]_{D^2}^o\,.
\end{equation}
Because the spacetime has the supersymmetry that is preserved by a stack of D3-branes, we can simply evaluate the RR contribution and double it to evaluate the above string bracket. To compute
\begin{equation}
\{\Bbb{P}(\Psi_{2,0}^c)^{-\frac{1}{2},-\frac{1}{2}};V_t^o\}_{D^2}\,,
\end{equation}
we do not need to insert a PCO. As $\Psi_{2,0}^c$ was chosen such that it vanishes at the origin, and because there is no spacetime boson to contract against $\Psi_{2,0}^c,$ we conclude that the above string vertex vanishes. Similarly, the insertion of $\Bbb{P}(\Psi_{2,0}^c)^{-1,-1}$ vanishes as well. Therefore, we conclude
\begin{equation}
\tilde{R}_{S^3}^{-2}\mathcal{G}\Bbb{P}\left[\frac{1}{2}(\Psi_{1,0}^c)^2\right]_{D^2}^o+\tilde{R}_{S^3}^{-2}\mathcal{G}\Bbb{P}\left[\Bbb{P}\Psi_{2,0}^c\right]_{D^2}^o=0\,.
\end{equation}
Note that this result is expected, as we are deliberately placing the stack of anti-D3-brane at the origin of the throat where its position modulus is stabilized. 

Now we are ready to study the last line of \eqref{eqn:second order open eom}
\begin{equation}
-\tilde{R}_{S^3}^{-2}\mathcal{G}\Bbb{P}\left[ \frac{1}{2} \Bbb{P}\Psi_{1,0}^o\otimes\Bbb{P}\Psi_{1,0}^o+\Psi_{1,0}^c\otimes\Bbb{P}\Psi_{1,0}^o\right]_{D^2}^o\,.
\end{equation}
We shall argue that even the above terms do vanish. We can see how the first term vanishes as follows. First, we shall insert one PCO to evaluate 
\begin{equation}
\left\{ V_{t,1}^o \otimes \Bbb{P}\Psi_{1,0}^o\otimes\Bbb{P}\Psi_{1,0}^o\right\}\,.
\end{equation}
To saturate the background $\phi$ charge, $\mathcal{X}\supset e^\phi T_F$ should be contracted. However, as the string field in $\Bbb{P}\Psi_{1,0}^o$ is constant, the contraction against $\partial X$ will necessarily vanish. Second, the following term 
\begin{equation}
\left\{V_{t,2}^o\otimes \Bbb{P}\Psi_{1,0}^o \otimes \Bbb{P}\Psi_{1,0}^o\right\}\,,
\end{equation}
to be non-trivial, $\mathcal{X}\supset -\partial \eta be^{2\phi}-\partial(\eta be^{2\phi})$ must be contracted. This contribution, however, vanishes because 
\begin{equation}
[f_i,f^i]=0\,.
\end{equation}
The second term can be shown to vanish as follows. To evaluate
\begin{equation}
\{\Psi_{1,0}^c;  V_t^o\otimes\Bbb{P}\Psi_{1,0}^o\}_{D^2}\,,
\end{equation}
we shall insert two PCOs. The only would be non-trivial contraction involves one $e^\phi T_F,$ and the other term from the other PCO. However, there is no way to saturate the c-ghost number to do so. Therefore, we again conclude that the second term vanishes. The Ramond-Ramond flux contribution vanishes as well for the same reason.

By collecting the previous results, we find
\begin{equation}
\frac{4}{g_o^2}\tilde{R}_{S^3}^{-2}Q_B\Bbb{P}|\Psi_{2,0}^o\rangle=0\,.
\end{equation}
We can therefore set $\Bbb{P}\Psi_{2,0}^o=0.$

\subsection{Third order equation}
We are finally ready to study the third-order equation, from which we will obtain a constraint that sets the vev of $\Psi_{1,0}^o.$ 
\begin{equation}
\frac{4}{g_o^2} \tilde{R}_{S^3}^{-3} Q_B|\Psi_{3,0}^o\rangle+\tilde{R}_{S^3}^{-3}\mathcal{G}\left[\frac{1}{3!}(\Psi_{1,0}^c+\Psi_{1,0}^o)^3+(\Psi_{1,0}^c+\Psi_{1,0}^o)\otimes(\Psi_{2,0}^c+\Psi_{2,0}^o)+\Psi_{3,0}^c\right]^o_{D^2}=0\,.
\end{equation}
Same as before, $1-\Bbb{P}$ projected equation of motion is trivially solved. As such, we shall focus on the $L_0^+$ nilpotent component. To make various degeneration limits more manifest, we shall expand the above equation as
\begin{align}
\frac{4}{g_o^2}\tilde{R}_{S^3}^{-3}&Q_B\Bbb{P}|\Psi_{3,0}^o\rangle=\mathcal{S}_1+\mathcal{S}_2+\mathcal{S}_3+\mathcal{S}_4+\mathcal{S}_5+\mathcal{S}_6\,,
\end{align}
where
\begin{align}
\mathcal{S}_1:=&-\tilde{R}_{S^3}^{-3} \mathcal{G}\Bbb{P} \left[ \frac{1}{3!} (\Bbb{P}\Psi_{1,0}^o)^3+\frac{1}{2} \Psi_{1,0}^c\otimes (\Bbb{P}\Psi_{1,0}^o)^2 +\Bbb{P}\Psi_{1,0}^o \otimes \left( -g_o^2\frac{b_0}{L_0}\left[ \frac{1}{2} (\Bbb{P}\Psi_{1,0}^o)^2\right]_{D^2}^o\right)\right]_{D^2}^o\nonumber\\
&-\tilde{R}_{S^3}^{-3}\mathcal{G}\Bbb{P}\left[ \left( \Psi_{1,0}^c -g_o^2\frac{b_0}{L_0} (1-\Bbb{P})\mathcal{G}\left[ \Psi_{1,0}^c\right]_{D^2}^o\right) \otimes\left( -g_o^2\frac{b_0}{L_0} (1-\Bbb{P})\left[\frac{1}{2} (\Bbb{P}\Psi_{1,0}^o)^2 \right]_{D^2}^o\right)\right]_{D^2}^o\nonumber\\
&-\tilde{R}_{S^3}^{-3} \mathcal{G}\Bbb{P} \left[ -\frac{g_o^2}{2}\frac{b_0}{L_0}(1-\Bbb{P}) \left[ \Psi_{1,0}^c\right]_{D^2}^o\otimes (\Bbb{P}\Psi_{1,0}^o)^2\right]_{D^2}^o\,,
\end{align}
\begin{align}
\mathcal{S}_2:=-\tilde{R}_{S^3}^{-3}\mathcal{G}\Bbb{P} \left[ \frac{1}{2} (\Psi_{1,0}^c)^2\otimes \Bbb{P}\Psi_{1,0}^o\right]_{D^2}^o +\text{degeneration diagrams}\,,
\end{align}
\begin{equation}
\mathcal{S}_3=-\tilde{R}_{S^3}^{-3}\mathcal{G}\Bbb{P} \left[\frac{1}{3!}(\Psi_{1,0}^c)^3\right]_{D^2}^o+\text{degeneration diagrams}
\end{equation}
\begin{align}
\mathcal{S}_4=&-\tilde{R}_{S^3}^{-3}\mathcal{G}\Bbb{P} \left[ \Psi_1^c \otimes \Bbb{P}\Psi_{2,0}^c \right]_{D^2}^o+\text{degeneration diagrams}\,,
\end{align}
\begin{equation}
\mathcal{S}_5= -\tilde{R}_{S^3}^{-3}\mathcal{G}\Bbb{P} \left[ \Bbb{P}\Psi_{1,0}^o\otimes \Bbb{P}\Psi_{2,0}^c \right]_{D^2}^o+\text{degeneration diagrams}\,,
\end{equation}
\begin{equation}
\mathcal{S}_6=-\tilde{R}_{S^3}^{-3} \mathcal{G}\Bbb{P} \left[ \Bbb{P}\Psi_{3,0}^c\right]_{D^2}^o\,.
\end{equation}
Note that degeneration diagrams contain terms that cover the rest of the moduli space of the corresponding diagrams that are not covered by the explicit string vertices we wrote. 

As the computation of each source term is rather long and technical, we refer to the appendices \S\ref{app:off diag} and \S\ref{app:diag} for the computation of the source terms. Once the dust settles, we find that 
\begin{equation}\label{eqn:sources vanish third order}
\sum_{i=2}^6\mathcal{S}_i=0\,,
\end{equation}
and
\begin{equation}\label{eqn:third order eom anti d3}
\frac{4}{g_o^2} \tilde{R}_{S^3}^{-3}Q_B\Bbb{P}|\Psi_{3,0}^o\rangle= -\frac{1}{2}C_{D^2}[[f^i,f_j]f^j]c\partial ce^{-\phi}\tilde{\psi}^j-i\frac{\sqrt{2}}{2}g_s C_{D^2}\frac{1}{3!} [f_i,f_j]\tilde{F}_{abc}\epsilon^{abcijk}c\partial ce^{-\phi}\tilde{\psi}_k\,.
\end{equation}
Note that \eqref{eqn:sources vanish third order} shouldn't come as a surprise. The source terms in \eqref{eqn:sources vanish third order} correspond to the attractive forces that the anti-D3-branes feel as the source terms in \eqref{eqn:sources vanish third order} source the tadpole for the diagonal open string modes. As we placed the anti-D3-branes where the warping is maximal, the anti-D3-branes are already at the minimum of the potential. Also, in this work, we did not use the S-dual anti-D3-brane; rather, we worked with the anti-D3-brane whose validity lies in $g_s\ll1.$ Therefore, instead of $\tilde{F}_{abc},$ we have found that the second source term depends on the dual of the threeform flux $g_s\tilde{F}_{abc}\epsilon^{abcdef}/3!= H^{def}.$ 

We can then find that the equation \eqref{eqn:third order eom anti d3} admits a solution when
\begin{equation}
    [[f^i,f_j],f^j]+i\sqrt{2} H^{ijk}[f_j,f_k]=0\,,
\end{equation}
which can be solved for
\begin{equation}
    f^i=-\frac{\sqrt{2}}{12}H_{abc}\epsilon_{S^3}^{abc} \alpha^i\,.
\end{equation}
This background solution of the open string field theory agrees with the analysis of \cite{Kachru:2002gs}, modulo the difference that originates from the S-dual treatment of \cite{Kachru:2002gs}.

\section{Conclusions}\label{sec:conclusions}
In this work, we developed a systematic tool to study the string perturbation theory of the supergravity solution of Klebanov-Strassler \cite{Klebanov:2000hb} and the anti-D3-brane supersymmetry breaking therein of \cite{Kachru:2002gs} with the help of open-closed superstring field theory. We took the large radius limit and the double scaling limit explained in \S\ref{sec:strategy} to perform controlled approximations.

Many important problems are left for follow-ups. In particular, studying the backreaction in the closed string sector and extending the open string solution to a higher order in the large radius limit to better understand the $g_s$ and $\alpha'$ corrections appear is a very urgent problem.\footnote{For recent progress on computation of $g_s$ and $\alpha'$ corrections in orientifold compactifications, see, for example, \cite{Kim:2023eut,Kim:2023sfs}.} Relatedly, understanding the asymptotic growth of the coefficients in the large radius expansion appears to be an important problem. Also, it would be important to carefully determine the string spectrum in the string field theory to understand at which value of the $g_sM$ possible instabilities start to appear. It would also be important to attempt to find a non-perturbative open-string vacuum, perhaps assisted by more non-perturbative approaches to open-closed string field theory. Lastly, developing theoretical tools to access the small $g_sM$ regime is a very important open problem. 

\section*{Acknowledgements}
We thank Minjae Cho, Liam McAllister, Andreas Schachner, Juan Maldacena for interesting discussions. We thank Minjae Cho and Arthur Hebecker for comments on the draft. We thank Renata Kallosh for encouragements. We thank the warm hospitality of the organizers of the Workshop on Matrix Models and String Field Theory held in Benasque. We also thank the KITP program “What is String Theory? Weaving Perspectives Together”, supported in part by grant NSF PHY-2309135 to the Kavli Institute for Theoretical Physics (KITP).

\newpage
\appendix
\section{Metric of the deformed conifold}\label{sec:deformed conifold}
In this section, we summarize metric of the deformed conifold and collect some useful results on the near tip limit and the double scaling limit defined in \S\ref{sec:strategy}. For the deformed conifold embedded in $\Bbb{C}^4$
\begin{equation}
 z_1^2+z_2^2+z_3^2+z_4^2=\epsilon^2\,,
\end{equation}
we write the metric as \cite{Candelas:1989js,Minasian:1999tt,Ohta:1999we,Herzog:2001xk}
\begin{align}
ds_{CY}^2=\frac{1}{2} \epsilon^{4/3} K(\tau)&\biggr[ \frac{1}{3K^3(\tau)} (d\tau^2+(g^5)^2)+\cosh^2\left(\frac{\tau}{2}\right) [(g^3)^2+(g^4)^2]\nonumber\\&+\sinh^2\left(\frac{\tau}{2}\right) [(g^1)^2+(g^2)^2]\biggr]\,.
\end{align}
Note we defined
\begin{equation}
g^1=\frac{e^1-e^3}{\sqrt{2}}\,,\quad g^2=\frac{e^2-e^4}{\sqrt{2}}\,,\quad g^3=\frac{e^1+e^3}{\sqrt{2}}\,,\quad g^4=\frac{e^2+e^4}{\sqrt{2}} \,,\quad g^5=e^5\,,
\end{equation}
\begin{equation}
e^1=-\sin\theta_1d\phi_1\,,\quad e^2=d\theta_1\,,\quad e^3=\cos\psi\sin\theta_2d\phi_2-\sin\psi d\theta_2\,,
\end{equation}
\begin{equation}
e^4=\sin\psi\sin\theta_2d\phi_2+\cos\psi d\theta_2\,,\quad e^5=d\psi+\cos\theta_1d\phi_1+\cos\theta_2d\phi_2\,,
\end{equation}
\begin{equation}
K(\tau)=\frac{(\sinh(2\tau)-2\tau)^{1/3}}{2^{1/3}\sinh\tau}\,,
\end{equation}
where $0\leq \theta_i\leq \pi$ and $0\leq\phi_i\leq2\pi$ for $ i=1,2$ parametrize $S^2,$ and $0\leq \psi\leq 4\pi$ denotes the $U(1)$ fiber direction of $T^{1,1}.$ 

In the near-tip limit, the metric approaches
\begin{equation}
ds^2_{tip}=\frac{1}{2}\epsilon^{4/3}\left(\frac{2}{3}\right)^{1/3} \left[ \frac{1}{2}d\tau^2 + \left(\frac{1}{2}(g^5)^2+(g^3)^2+(g^4)^2\right) + \frac{1}{4} \tau^2 \left( (g^1)^2+(g^2)^2\right)\right]\,.
\end{equation}
We can rewrite the metric as
\begin{equation}
ds_{tip}^2= \epsilon^{4/3}  \left( \frac{1}{96} \right)^{1/3} d\tau^2+ R_{S^3}^2 d\Omega_3^2+ R_{S^2}^2d\Omega_2^2\,,
\end{equation}
where
\begin{equation}
R_{S^3}^2= \epsilon^{4/3}\left(\frac{1}{12} \right)^{1/3}\,,\quad R_{S^2}^2= \epsilon^{4/3}\tau^2\left(\frac{1}{96}\right)^{1/3}\,.
\end{equation}
Note that $S^3$ should be viewed as a fiber of the fiber bundle $T^{1,1}:= S^3\rightarrow S^2.$ Note also that the metric is normalized such that the volume of $S^3$ and $S^2$ are $4\sqrt{2}\pi^2$ and $4\pi,$ respectively.

As was noted in \cite{Minasian:1999tt}, metric of $S^3$ can also be written as follows. We shall first define two matrices
\begin{equation}
L_i:=\left(\begin{array}{cc} \cos\frac{\theta_i}{2} e^{i(\psi/2+\phi_i)/2} & -\sin\frac{\theta_i}{2}e^{-i(\psi/2-\phi_i)/2}\\
\sin\frac{\theta_i}{2}e^{i(\psi/2-\phi_i)/2}&\cos\frac{\theta_i}{2} e^{-i(\psi/2+\phi_i)/2}
\end{array}\right)\,.
\end{equation}
Then, using 
\begin{equation}
T:= L_1\sigma_1 L_2^\dagger \sigma_1\,,
\end{equation}
\cite{Minasian:1999tt} showed that the following identity holds
\begin{equation}
\frac{1}{2} \text{Tr}(dT^\dagger dT) =\frac{1}{2} (g^5)^2+(g^3)^2+(g^4)^2\,,
\end{equation}
which proves that the three-sphere metric is given by 
\begin{equation}
\frac{1}{2} (g^5)^2+(g^3)^2+(g^4)^2\,.
\end{equation}
And, we defined $d\Omega_2^2$ as
\begin{equation}
d\Omega_2^2:=\frac{1}{2} (g^1)^2+\frac{1}{2}(g^2)^2\,.
\end{equation}

We shall define four coordinates $x_i,$ for $i=1,\dots,4,$ to rewrite
\begin{equation}
T= \left(\begin{array}{cc}x_1+ix_2&x_3+ix_4\\-x_3+ix_4&x_1-ix_2 \end{array}\right)\,,
\end{equation}
where $S^3$ is embedded as
\begin{equation}
x^2_1+x_2^2+x^2_3+x^2_4=1\,.
\end{equation}
The coordinates $x_i$ are related to the hyperspherical coordinates $\varphi_1,~\varphi_2,~\varphi_3$ as
\begin{align}
&x_1=\cos\varphi_1\,,\\
&x_2=\sin\varphi_1\cos\varphi_2\,\\
&x_3=\sin\varphi_1\sin\varphi_2\cos\varphi_3\,,\\
&x_4=\sin\varphi_1\sin\varphi_2\sin\varphi_3\,,
\end{align}
where $\varphi_1$ and $\varphi_2$ run over the range of $[0,\pi],$ and $\varphi_3$ runs over $[0,2\pi].$ Coventionally, a stack of anti-D3-branes is placed at $\varphi_1=0.$ Therefore, we shall relate $\varphi_1$ to the conifold coordinates momentarily.

Now, we shall study the near-tip limit of the Klebanov-Strassler solution. Near the tip, the metric of the Klebanov-Strassler solution is a rescaled metric of the deformed conifold
\begin{align}
ds_{KS}^2=&\left(\frac{2^{1/3}a_0^{1/2}g_s M}{\epsilon^{4/3}}\right) ds_{tip}^2\,,\\
=&a_0^{1/2}g_sM \left(\frac{1}{48}\right)^{1/3} d\tau^2+\tilde{R}_{S^3}^2d\Omega_3^2+\tilde{R}_{S^2}^2d\Omega_2^2\,,
\end{align}
where 
\begin{equation}
\tilde{R}^2_{S^3}= a_0^{1/2}6^{-1/3} g_sM\,,\quad \tilde{R}^2_{S^2}=a_0^{1/2}2^{-1}6^{-1/3}\tau^2 g_sM\,.
\end{equation}

One can then take a double scaling limit combined with the large radius limit $g_sM\rightarrow \infty$ such that
\begin{equation}
\tau^2 g_s M \,,
\end{equation}
is treated as a fixed number. We shall define $r$
\begin{equation}
r:=a_0^{1/4} \sqrt{g_sM} 48^{-1/6} \tau\,,
\end{equation}
to rewrite the metric as
\begin{equation}
ds_{KS}^2=dr^2+\tilde{R}^2_{S^3}d\Omega_3^2+\tilde{R}^2_{S^2}d\Omega_2^2\,.
\end{equation}
We shall redefine the coordinates
\begin{equation}
\theta_1=\frac{\pi}{2}+\frac{\tilde{R}_{S^3}^{-1}}{\sqrt{2}}\theta_\perp+\tilde{R}_{S^2}^{-1} \theta_{\|}\,,\quad \theta_2=\frac{\pi}{2}+\frac{\tilde{R}_{S^3}^{-1}}{\sqrt{2}}\theta_\perp-\tilde{R}_{S^2}^{-1} \theta_{\|}\,,
\end{equation}
\begin{equation}
\phi_1=-\frac{\tilde{R}_{S^3}^{-1}}{\sqrt{2}}\phi_\perp -\tilde{R}_{S^2}^{-1} \phi_{\|}\,,\quad \phi_2=\frac{\tilde{R}_{S^3}^{-1}}{\sqrt{2}}\phi_\perp-\tilde{R}_{S^2}^{-1} \phi_{\|}\,,
\end{equation}
\begin{equation}
\psi=\tilde{R}_{S^3}^{-1}\sqrt{2} \psi_\perp\,.
\end{equation}
Then, the metric locally looks like $\Bbb{R}^6,$ provided that $\tilde{R}^2_{S^2}$ is large numerically
\begin{equation}
ds_{KS}^2= dr^2+d\psi_\perp^2+d\theta_\perp^2+d\phi_\perp^2+d\theta_{\|}^2+d\phi_{\|}^2+\mathcal{O}(\tilde{R}_{S^3}^{-2},\tilde{R}_{S^2}^{-2})\,.
\end{equation}
Precisely in this regime, we can properly treat the string perturbation theory in the large volume expansion. We also note that around $\theta_1=\theta_2=\pi/2,$ $\psi=0,$ and $\phi_1=\phi_2=0,$ $\varphi_1$ is related to $\psi$ as
\begin{equation}
\varphi_1=\frac{\psi}{2} +\dots\,,
\end{equation}
where $\dots$ denotes the inverse radius corrections. In the main text, we shall rewrite 
\begin{equation}
\tilde{X}^4=r\,,\quad \tilde{X}^5=\psi_\perp\,,\quad \tilde{X}^6=\theta_\perp\,,\quad \tilde{X}^7=\phi_\perp\,,\quad\tilde{X}^8=\theta_{\|}\,,\quad\tilde{X}^9=\phi_{\|}\,.
\end{equation}

In the large volume coordinates, the one forms $g_i$ for $i=1,\dots,5$ are written as
\begin{align}
&g_1=\sqrt{2}\tilde{R}_{S^2}^{-1} d\phi_{\|}+\dots\,,\\
&g_2=\sqrt{2}\tilde{R}_{S^2}^{-1} d\theta_{\|}+\dots\,\\
&g_3= \tilde{R}_{S^3}^{-1}d\phi_\perp+\dots\,\\
&g_4=\tilde{R}_{S^3}^{-1}d\theta_{\perp}+\dots\,\\
&g_5=\sqrt{2} \tilde{R}_{S^3}^{-1}d\psi_\perp+\dots\,,
\end{align}
where $\dots$ denotes the inverse radius corrections.
\section{Closed-string second order background solution}\label{sec:sec back app}
In this appendix, we shall expand on the details of the closed string second-order background solution. 

Let us recall the second-order background equation
\begin{align}
Q_B\left( \tilde{R}_{S^3}^{-2}\Bbb{P}\Psi_2^{-1,-1}\right)=&\mathcal{A}_{AB} (\partial c+\bar{\partial}\bar{c})c\bar{c} e^{-\phi}\tilde{\psi}^A e^{-\bar{\phi}}\bar{\tilde{\psi}}^B+\mathcal{B}_Ac\bar{c}( \eta e^{-\bar{\phi}}\bar{\tilde{\psi}}^A+c.c.)\nonumber\\
&+\mathcal{C} (\partial c+\bar{\partial}\bar{c}) c\bar{c}\left(\eta e^{-2\phi}\bar{\partial}\bar{\xi}-e^{-2\phi}\partial\xi\bar{\eta}\right)\,,\\
=& \pi g_c^2 \left(T_{AB}-\frac{1}{8}\eta_{AB}T\right)\,,
\end{align}
where
\begin{equation}
\mathcal{A}_{AB}=-\frac{1}{4} \partial^2 \mathcal{G}_{AB}-\frac{1}{4} (\partial_B\mathcal{F}_A+\partial_A\mathcal{F}_B)\,, 
\end{equation}
\begin{equation}
\mathcal{B}_A=i\frac{1}{\sqrt{2}} \partial^B\mathcal{G}_{BA}+\frac{i}{\sqrt{2}} \mathcal{F}_A+i\frac{1}{\sqrt{2}} \partial_A\mathcal{D}\,,
\end{equation}
\begin{equation}
\mathcal{C}=\frac{1}{4}\partial^A\mathcal{F}_A-\frac{1}{4}\partial^2\mathcal{D}\,.
\end{equation}
Note again $T_{ab}=0$ due to the ISD condition.

We can rewrite the background equations as
\begin{equation}\label{eqn:F0}
    \mathcal{F}_A=-\partial_A\mathcal{D}-\partial^B\mathcal{G}_{AB}\,,
\end{equation}
\begin{equation}\label{eqn:D0}
    \partial^A\mathcal{F}_A=\partial^2\mathcal{D}\,,
\end{equation}
and
\begin{equation}\label{eqn:G0}
    \partial^2\mathcal{G}_{AB}-\partial_A\partial^C\mathcal{G}_{CB}-\partial_B\partial^C\mathcal{G}_{CA}-2\partial_A\partial_B \mathcal{D}=-4\pi g_c^2\left(T_{AB}-\frac{1}{8}\eta_{AB}T\right)\,.
\end{equation}

By combining \eqref{eqn:F0} and \eqref{eqn:D0}, we find
\begin{equation}\label{eqn:FD}
    \partial^2\mathcal{D}=-\frac{1}{2}\partial^A\partial^B\mathcal{G}_{AB}\,.
\end{equation}
By taking the trace of \eqref{eqn:G0}, we find
\begin{equation}\label{eqn:tr G0}
    \partial^2 \mathcal{G}_{AB}\eta^{AB}-\partial^A\partial^B\mathcal{G}_{AB} =\pi g_c^2 T\,.
\end{equation}
The non-compact components of \eqref{eqn:G0} are
\begin{equation}
    \partial^2\mathcal{G}_{\mu\nu}-\partial_\mu\partial^C \mathcal{G}_{C\nu}-\partial_\nu\partial^C\mathcal{G}_{C\mu}-2\partial_\mu\partial_\nu\mathcal{D}=-\frac{\pi g_c^2}{2}\eta_{\mu\nu}T\,,
\end{equation}
and the compact components are
\begin{equation}
    \partial^2\mathcal{G}_{ab}-\partial_a\partial^c \mathcal{G}_{cb}-\partial_b\partial^c\mathcal{G}_{ca}-2\partial_a\partial_b \mathcal{D}=\frac{\pi g_c^2}{2}\eta_{ab}T\,.
\end{equation}

We shall assume 
\begin{equation}
    \partial_\mu \mathcal{D}=\partial_\mu \mathcal{F}_A=\partial_\mu\mathcal{G}_{AB}=0\,,
\end{equation}
and use an ansatz
\begin{equation}
    \mathcal{D}=\alpha \mathcal{G}_{AB}\eta^{AB}+\delta\mathcal{D}\,.
\end{equation}
We then find
\begin{equation}
    \partial^2\mathcal{G}_{\mu\nu}=-\frac{\pi g_c^2}{2}\eta_{\mu\nu} T\,,\quad     \partial^2 \mathcal{G}_{ab}\eta^{ab} - \partial^a\partial^b\mathcal{G}_{ab}=3\pi g_c^2T\,.
\end{equation}
\begin{equation}
    \partial^2 \mathcal{G}_{ab} -\partial_a\partial^c\mathcal{G}_{cb} -\partial_b\partial^c\mathcal{G}_{ca}-2\alpha\partial_a\partial_b\mathcal{G}_{CD}\eta^{CD} -2\partial_a\partial_b\delta\mathcal{D}=\frac{\pi g_c^2}{2}\eta_{ab}T\,,\
\end{equation}
and
\begin{equation}
\frac{1}{2}\partial^a\partial^b\mathcal{G}_{ab} +\partial^2\delta\mathcal{D}=-\alpha \partial^2 \mathcal{G}_{AB}\eta^{AB}\,.
\end{equation}

We shall first solve $\mathcal{G}_{\mu\nu}$
\begin{equation}
    \mathcal{G}_{\mu\nu}=-\frac{\pi g_c^2}{2}\eta_{\mu\nu} T G_1 (\tilde{X})\,,
\end{equation}
where $\partial^2 G_1(\tilde{X})=1.$ We shall choose $G_1(\tilde{X})=(\tilde{X}^4)^2/2.$ Then, we find
\begin{equation}
    \partial^2 \mathcal{G}_{ab} -\partial_a\partial^c\mathcal{G}_{cb}-\partial_b\partial^c\mathcal{G}_{ca}-2\alpha\partial_a\partial_b \mathcal{G}_{cd}\eta^{cd} +\left( 4\alpha\pi g_c^2 T \partial_a\partial_b G_1-2\partial_a\partial_b\delta\mathcal{D}\right) =\frac{\pi g_c^2}{2}\eta_{ab}T\,,
\end{equation}
\begin{equation}
    \frac{1}{2} \partial^a\partial^b\mathcal{G}_{ab} +\partial^2\delta\mathcal{D}=-\alpha\partial^2\mathcal{G}_{ab}\eta^{ab} +2\alpha\pi g_c^2 T\,.
\end{equation}

We find that the following ansatz is useful
\begin{equation}
 \mathcal{G}_{ab}= A(\tilde{X}^4) \eta_{ab}-\frac{1}{12\pi} R_{acbd}\tilde{X}^c\tilde{X}^d\,.
\end{equation}
Note that the curvature term was added to take into account that the internal manifold is curved. We then find,
\begin{equation}
    \partial^2 A \eta_{ab} -(12\alpha +2)\delta_a^4\delta_b^4\partial_4^2 A +4\alpha\pi g_c^2 T \delta_a^4 \delta_b^4 -2\partial_a\partial_b\delta D =\frac{\pi g_c^2}{2}\eta_{ab}T\,,
\end{equation}
and
\begin{equation}
   \partial^2\delta\mathcal{D}=-6\alpha\partial^2 A -\frac{1}{2} \partial^2 A +2\alpha\pi g_c^2 T\,.
\end{equation}

Let us set $\alpha=-1/4.$ We then find
\begin{equation}
    \partial^2 A\eta_{ab}=\frac{3\pi g_c^2}{5}\eta_{ab} T\,,
\end{equation}
which is equivalent to
\begin{equation}
    \mathcal{G}_{ab}=\frac{3\pi g_c^2}{10} \eta_{ab} (\tilde{X}^4)^2T\,.
\end{equation}
Similarly, we find
\begin{equation}
    \partial_a\partial_b\delta\mathcal{D}=\frac{1}{20}\pi g_c^2T\eta_{ab}-\frac{1}{5} \pi g_c^2 T\delta_a^4\delta_b^4\,,
\end{equation}
and
\begin{equation}
\delta\mathcal{D}= -\frac{1}{10}\pi g_c^2 T(\tilde{X}^4)^2 +\frac{1}{40}\pi g_c^2T \tilde{X}^a\tilde{X}_a\,.
\end{equation}

To summarize, the background solution we found is given by
\begin{equation}
\mathcal{G}_{\mu\nu}=-\frac{\pi g_c^2}{4}\eta_{\mu\nu} (\tilde{X}^4)^2 T\,,\quad \mathcal{G}_{ab}=\frac{3\pi g_c^2}{10} \eta_{ab} (\tilde{X}^4)^2 T\,,
\end{equation}
\begin{equation}
    \mathcal{F}_\mu=0\,,\quad \mathcal{F}_a= -\frac{\pi g_c^2}{20} T \tilde{X}_a\,,  \quad     \mathcal{D}= -\frac{3\pi g_c^2}{10} T (\tilde{X}^4)^2+\frac{\pi g_c^2}{40}T \tilde{X}_a\tilde{X}^a\,.
\end{equation}
Note that the form of the solution has a similar feature as the supergravity solution of \cite{Klebanov:2000hb,Giddings:2001yu}. Also, the final answer does not depend on the value of $\alpha.$
\section{String vertices and the moduli space}\label{sec:vertices}
To compute amplitudes in string field theory in a consistent manner, we need to construct string vertices such that string vertices and Feynman diagrams made of string vertices cover the entire moduli space of Riemann surfaces with and without boundaries. Although we are studying many diagrams with high-dimensional moduli spaces of up to four, we won't need the details of many of them. We shall only spell out the details of the string vertices that we will actually need. We shall use the $SL(2;R)$ vertices following \cite{Sen:2019jpm,Sen:2020eck}, as we find they are most practical for the computations in the large stub limit. Following \cite{Sen:2020eck}, we will denote a closed string puncture by C and an open string puncture by O.

\subsection{Sphere with C-C-C}
The moduli space for a three-punctured sphere is a point. To study the local coordinates, we shall use a conformal map from a flat space parametrized by $z$ to a sphere. By using the $SL(2;C),$ we shall place the punctures at $z_1=0,$ $z_2=1,$ and $z_3=\infty.$ When necessary, we shall permute over the insertion of vertex operators. To each puncture around $z_i,$ we attach a unit disk parametrized by a coordinate $w_i$ such that
\begin{equation}
w_i=\lambda f_i(z)\,,
\end{equation}
where $\lambda$ is the closed string stub parameter that will be sent to infinite at the end of the computations. We choose the local coordinates as
\begin{equation}
f_1(z) =\frac{2z}{z-2} \,,\quad f_2(z)=-2\frac{1-z}{1+z} \,,\quad f_3(z)=\frac{2}{1-2z}\,.
\end{equation}
Following \cite{Sen:2019jpm}, we shall place PCOs at symmetric locations that are invariant under the $SL(2;C)$ maps that permute over the closed string punctures
\begin{equation}
p_\pm=\frac{1}{2}\pm i\frac{\sqrt{3}}{2}\,,
\end{equation}
and average over the two choices.

\subsection{Sphere with C-C-C-C}
The moduli space of a four-punctured sphere is two-dimensional. Although we will not need to use this for our purposes, we decided to include this section for future use. We shall split the moduli space into the fundamental four vertex region and the Feynman regions, which are constructed via the plumbing fixtures of two three-punctured spheres.  We shall first construct the Feynman regions with the plumbing fixtures and fill in the regions. The detailed form of the local coordinates in the interior of the moduli space won't be needed for us, as we will be only concerned with on-shell primary operators with four-punctured sphere diagrams. We shall denote the global coordinate of the four-punctured sphere by $z,$ the coordinates of the three-punctured spheres by $x_i,$ for $i=1,2.$

We shall first identify punctures at $x_1=x_2=0$ via the plumbing fixture
\begin{equation}
\lambda^2 f_1(x_1)f_1(x_2)=q\,,
\end{equation}
where $q=e^{-s+i\theta}$ is the modulus in the Schwinger parametrization. Under this plumbing fixture, a point at $x_2$ is mapped to
\begin{equation}
x_1=\frac{2q\lambda^{-2} f_1(x_2)^{-1}}{2+q\lambda^{-2}f_1(x_2)^{-1}}\,.
\end{equation}
We find punctures at $x_2=0,~1,$ and $\infty$ map to
\begin{equation}
x_1=2\,,\quad \frac{-2q\lambda^{-2}}{4-q\lambda^{-2}}\,,\quad \frac{2q\lambda^{-2}}{4+q\lambda^{-2}}\,,
\end{equation}
respectively. Note that the symmetric points $x_2=p_\pm$ PCOs can be inserted are mapped to
\begin{equation}
x_1= \frac{-2q\lambda^{-2}(-2+e^{\pm i\pi/3})}{-4e^{\pm i\pi/3}-(-2 +e^{\pm i\pi/3})q\lambda^{-2}}\,.
\end{equation}

We now define the global coordinate $z$ as
\begin{equation}
z:=\frac{ax_1+b}{cx_1+d}\,,
\end{equation}
where
\begin{equation}
a=1\,,\quad b=\frac{2q\lambda^{-2}}{4-q\lambda^{-2}}\,,\quad c=0\,,\quad d=1+\frac{2q\lambda^{-2}}{4-q\lambda^{-2}}\,.
\end{equation}
We find $x_1=1,~\infty,$ and $x_2=1,~\infty$ map to
\begin{equation}
z=1\,,\quad z=\infty\,,\quad z=0\,,\quad z=\frac{16q\lambda^{-2}}{(4+q\lambda^{-2})^2}\,,
\end{equation}
respectively. Similarly, the symmetric points $x_2=p_\pm$ map to
\begin{equation}
y=e^{\pm i\pi/3} q\lambda^{-2}+\dots\,.
\end{equation}

We constructed the Feynman region by the plumbing fixture when a movable vertex comes close to $z=0.$ To construct the Feynman region for other degeneration channels, one can simply use the $SL(2;C)$ map to permute over vertices.

\subsection{Disk with O-O-O}
The moduli space of a disk with three boundary punctures is zero-dimensional. Therefore, we shall declare that the entirety of it is the open string three-point vertex. To study disk amplitudes, we shall use a conformal map from a half upper plane parametrized by $z$ to a unit disk. Using $SL(2;R),$ we shall place three open string punctures at $z_1=0,\,z_2=1,$ and $z_3=\infty.$ We shall also permute over the vertices. To each open string puncture, we attach a unit half-disk
\begin{equation}
w_i=\mu g_i(z)\,,
\end{equation}
where
\begin{equation}
g_1(z)=\frac{2z}{2-z}\,,\quad g_2=-2\frac{1-z}{1+z}\,,\quad g_3=\frac{2}{1-2z}\,,
\end{equation}
and $\mu$ is the open string stub parameter that will be sent to infinity at the end of the computation. We shall place a PCO at 
\begin{equation}
p_\pm=\frac{1}{2}\pm i\frac{\sqrt{3}}{2}\,,
\end{equation}
which is symmetric under $SL(2;R).$ Note that the PCO located at $p_-$ can also be viewed as an anti-holomorphic PCO at $p_+.$ When we can show that the PCO location does not affect the answer, we may move PCOs as we prefer. 

\subsection{Disk with C-O}
The moduli space of a disk with one closed string puncture and one open string puncture is zero-dimensional. Again, we declare that the entirety of the moduli space belongs to the two-point string vertex. We shall denote the global coordinate on the unit disk by $y,$ such that $|y|\leq1,$ and the global coordinate on the upper-half-plane by $z.$ We shall place the closed string puncture at $y=0,$ and the open string puncture at $y=1.$ These two global coordinates are related by
\begin{equation}
z=i\frac{1-y}{1+y}\,.
\end{equation}
The closed puncture at $y=0$ is mapped to $z=i,$ and the open string puncture at $y=1$ is mapped to $z=0.$ 

We choose the local coordinates around the closed and open string punctures as
\begin{equation}
w=\lambda y=\lambda \frac{i-z}{i+z}\,,
\end{equation}
and
\begin{equation}
w=\mu z\,.
\end{equation}

If both the closed and the open string punctured are inserted with NS states, we need to insert a PCO. Ideally, we would choose to insert a PCO at a symmetric location. Treating the anti-holomorphic part of the closed string vertex at $i$ as a holomorphic vertex located at $-i,$ we can attempt to place a PCO at the origin where the open string puncture is located. This is a valid choice only if there is no singularity, as PCO comes close to the open string puncture. We can average over PCO locations for generic cases, e.g., $1/2$ and $-1/2.$ 

\subsection{Disk with C-O-O}
The moduli space of a disk with one closed string puncture and two open string punctures is one-dimensional. We shall denote the modulus by $t,$ and place the closed string puncture at $i,$ and the open string punctures at $\pm t.$ Note that the range of $t$ is $0\leq t\leq \infty.$ There are two degeneration channels, first at $t\simeq0,$ and the other at $t\simeq\infty.$ We shall first determine the local coordinates and the location of the PCOs for a Ramond-Ramond closed string puncture and NS open string punctures of the Feynman regions. We shall construct the local coordinates and the PCO location of the vertex region by filling in the gaps.

The Feynman region of the disk with one closed string puncture and two open punctures is obtained by joining one disk with C-O and one disk with O-O-O via the plumbing fixture. We shall denote the global coordinate in the first upper half plane by $x_1$ and the global coordinate in the second upper half plane by $x_2.$ We shall denote the global coordinate of the disk with C-O-O by $z.$ We then glue open string punctures at $x_1=x_2=0$ by
\begin{equation}
\mu^2 x_1 g_1(x_2)=-q\,,
\end{equation}
where $q=e^{-s}$ is the Schwinger parameter. As a result, we find 
\begin{equation}
x_1=-q\mu^{-2} \frac{2-x_2}{2x_2}\,,
\end{equation}
and equivalently,
\begin{equation}
x_2=\frac{2q\mu^{-2}}{q\mu^{-2} -2x_1}\,.
\end{equation}
Open string punctures at $x_2=1$ and $x_2=\infty$ are mapped to
\begin{equation}
x_1= -\frac{1}{2} q\mu^{-2}\,,
\end{equation}
and
\begin{equation}
x_1= \frac{1}{2} q\mu^{-2}\,,
\end{equation}
respectively. We can therefore identify $z$ with $x_1,$ and $t$ with $q\mu^{-2}/2.$  The PCOs at, with $q=1,$ $x_2=p_\pm$ are mapped to
\begin{equation}
    x_1= \pm i\frac{\sqrt{3}}{2}\mu^{-2}\,.
\end{equation}
The local coordinates around the open string punctures in the Feynman region are given as
\begin{equation}
w_1= \mu g_2(x_2)=\mu \frac{2(q\mu^{-2}+2x_1)}{3q\mu^{-2}-2x_1}\,,
\end{equation}
\begin{equation}
w_2=\mu g_3(x_2)=-\mu \frac{2(q\mu^{-2}-2x_1)}{3q\mu^{-2}+2x_1}\,.
\end{equation}

The other Feynman region can be simply constructed by replacing $x_1$ with $-1/x_1.$ It is crucial to note that for the evaluation of amplitudes, we shall not use the map $x_1\mapsto-1/x_1.$ We are only using the map to construct the local coordinates. We find, at $q=1,$
\begin{equation}
    w_1= \mu\frac{2(2\mu^2+x_1)}{2\mu^2-3x_1}\,,
\end{equation}
\begin{equation}
    w_2=\mu \frac{2(-2\mu^2 +2x_1)}{2\mu^2+3x_1}\,.
\end{equation}
The PCOs at $x_2=p_\pm$ are mapped to
\begin{equation}
    x_1=\pm i \frac{2}{\sqrt{3}}\mu^2\,.
\end{equation}

We shall define the fundamental string vertex to cover the moduli space of $t\geq \mu^{-2}/2.$ We shall require that the local coordinates of the open string punctures at $t=\mu^{-2}/2$ coincide with that of the Feynman region at $q=1.$ Following \cite{Sen:2020eck}, we choose the local coordinates of the open string punctures as
\begin{equation}
w_i= \mu^3 \frac{4\mu^4+1}{4\mu^4} \frac{z-z_i}{(1+z_iz)+\mu^2 h(t) (z-z_i)}\,,
\end{equation}
where $z_1=-t,$ and $z_2=t.$ We shall choose the holomorphic function $h(z)$ such that $h(z)$ interpolates
\begin{equation}
h\left(\pm\frac{1}{2\mu^2}\right) =\pm \frac{4\mu^4-3}{8\mu^4}\,, 
\end{equation}
and 
\begin{equation}
h(\pm1)=0\,.
\end{equation}
We choose
\begin{equation}
h(z):=-\frac{3-4\mu^4}{4\mu^2-16\mu^6} \left(z-\frac{1}{z}\right)\,.
\end{equation}
We implicitly defined $h(z)$ over the range of $ (2\mu^{2})^{-1}\leq|z|\leq1.$ We will extend the range by using the following definition for $ 1\leq |z|\leq 2\mu^2$
\begin{equation}
h(z)=h(-1/z)=-h(1/z)\,.
\end{equation}
We choose to fill in the gaps for the PCOs
\begin{equation}
p(t)=\pm i\frac{2\mu^2/\sqrt{3}-\sqrt{3}\mu^{-2}/2}{2\mu^2-(2\mu^2)^{-1}}\left(t-\frac{1}{2\mu^2}\right) \pm i\frac{\sqrt{3}}{2}\mu^{-2}\,.
\end{equation}

It will be often convenient to work with a new coordinate $y$ that is defined as
\begin{equation}
y= \frac{z+t}{-t z+1}\,.
\end{equation}
Such that the open string punctures at $z=-t$ and $z=t$ are mapped to
\begin{equation}
y=0\,,\quad y=\frac{2t }{1-t^2}\,,
\end{equation}
respectively. Note that the closed string puncture at $z=i$ is mapped to $y=i.$

\subsection{Disk with O-O-O-O}
The moduli space of a disk with four open string punctures is one-dimensional. The Feymann regions of the moduli space are constructed by gluing two disks with three open string punctures. We shall denote the global coordinate of the disk with four open string punctures by $z,$ the global coordinates of disks with three open string punctures by $x_1$ and $x_2.$ 

We shall glue two open string punctures at $x_1=0$ and $x_2=0$ by the plumbing fixture
\begin{equation}
\mu^2 g_1(x_1)g_1(x_2)=-q\,,
\end{equation}
where $q=e^{-s}$ is the Schwinger parameter. Under the gluing, a point at $x_2$ is mapped to
\begin{equation}
x_1=-\frac{2q\mu^{-2}g_1(x_2)^{-1}}{2-q\mu^{-2}g_1(x_2)^{-1}}\,.
\end{equation}
The open string punctures at $x_2=1\,, \infty$ are mapped to
\begin{equation}
x_1=-\frac{2q\mu^{-2}}{4-q\mu^{-2}}\,,\quad x_1=\frac{2q\mu^{-2}}{4+q\mu^{-2}}\,,
\end{equation}
respectively.

We can now either declare that $x_1$ is the global coordinate $z,$ or we can redefine $z.$ We shall do the latter by defining $z$ as
\begin{equation}
z=\mathcal{M}_4^+:=\frac{a_+x_1+b_+}{c_+x_1+d_+}\,,
\end{equation}
where
\begin{equation}
a_+=1\,,\quad b_+=\frac{2q\mu^{-2}}{4-q\mu^{-2}}\,,\quad c_+=0\,,\quad d_+=1+\frac{2q\mu^{-2}}{4-q\mu^{-2}}\,.
\end{equation}
The open string punctures at $x_1=1,~\infty,$ $x_2=1,$ and $x_2=\infty$ are mapped to
\begin{equation}
z=1\,,\quad z=\infty\,,\quad z=0\,,\quad z=16\frac{q\mu^{-2}}{(4+q\mu^{-2})^2}= q\mu^{-2}-\frac{q^2\mu^{-4}}{2}+\dots\,.
\end{equation}
Also, we find that PCOs located at $x_1=p_\pm$ and $x_2=p_\pm$ are mapped to
\begin{equation}
z=p_\pm +\frac{p_\mp}{2\mu^2}q-\frac{p_\mp}{8\mu^4}q^2+\dots\,,
\end{equation}
and
\begin{equation}
z=\frac{16q\mu^{-2} (p_\pm-1)}{(4+q\mu^{-2})(q\mu^{-2}(p_\pm-2)+4p_\pm)}= p_\pm q\mu^{-2}+\frac{1}{2}p_\mp q\mu^{-2}+\dots\,,
\end{equation}
respectively.

By replacing $a_+,~b_+,~c_+,$ and $d_+$ with
\begin{equation}
a_-=1\,,\quad b_-=-\frac{2q\mu^{-2}}{4+q\mu^{-2}}\,,\quad c_-=0\,,\quad d_-=1-\frac{2q\mu^{-2}}{4+q\mu^{-2}}\,,
\end{equation}
to define a new Mobius transformation $\mathcal{M}_4^-,$ we can cover a different region of the moduli space as the open string punctures at $x_1=1\,,\infty,$ and $x_2=1\,,\infty$ are mapped to
\begin{equation}
z=1\,,\quad z=\infty \,,\quad z=-16\frac{q\mu^{-2}}{(-4+q\mu^{-2})^2}=-q\mu^{-2}-\frac{q^2\mu^{-4}}{2}+\dots\,, \quad z=0\,,
\end{equation}
respectively. Also, we find that PCOs located at $x_1=p_\pm$ and $x_2=p_\pm$ are mapped to
\begin{equation}
    z= p_\pm-\frac{p_\pm}{2\mu^2}q-\frac{p_\pm}{8\mu^4}q^2+\dots\,,
\end{equation}
\begin{equation}
z=-\frac{16q\mu^{-2} (p_\pm-1)}{(-4+q\mu^{-2})(q\mu^{-2}(p_\pm-2)-4p_\pm)}=- p_\mp q\mu^{-2}+\frac{1}{2}p_\pm q\mu^{-2}+\dots\,.
\end{equation}

We define a function $\mathcal{L}_4^\pm$
\begin{equation}
x_2=\mathcal{L}^\pm_4(z):= g_1^{-1}\left( -q\mu^{-2} g_1\left( (\mathcal{M}_4^\pm)^{-1}(z)\right)^{-1}\right)\,,
\end{equation}
which relates $z$ to $x_2.$  The local coordinates around the open string punctures are given as
\begin{equation}
w_1= \mu g_2( (\mathcal{M}_4^\pm)^{-1}(z))\,,\quad w_2= \mu g_3( (\mathcal{M}_4^\pm)^{-1}(z))\,,
\end{equation}
\begin{equation}
w_3=\mu g_2\left(\mathcal{L}_4^\pm(z) \right)\,,\quad w_4=\mu g_3\left(\mathcal{L}_4^\pm(z)\right)\,.
\end{equation}

By using $SL(2;R)$ maps
\begin{equation}
z\mapsto 1-z\,,\quad z\mapsto \frac{1}{z}\,,
\end{equation}
we can construct the moduli space of the remaining Feynman regions. It is important to note that the above $SL(2;C)$ maps invert the orientation. To take into account the change of orientation, we shall include an additional sign to define the local coordinates.

In the large stub limit, the fundamental vertex region is therefore determined to be
\begin{equation}
\mu^{-2}-\mu^{-4}/2<z<1-\mu^{-2}+\mu^{-4}/2\,,\quad 1+\mu^{-2}+\mu^{-4}/2<z<\mu^2+1/2\,,
\end{equation}
\begin{equation}
-\mu^2+1/2<z<-\mu^{-2}-\mu^{-4}/2\,.
\end{equation}
Note that we fixed the position of three vertices at $0,~1,$ and $\infty.$ To determine the local coordinates of the vertex region, we shall first determine the boundary conditions for the local coordinates.

The local coordinates at $z= - (4-\mu^{-2})^2/16\mu^{-2}$ are
\begin{equation}
w_1= \frac{2\mu(-1+4\mu^2)(z-1)}{(4\mu^2-1)z+4\mu^2-1}\,,\quad w_2= -\frac{32\mu^5-2\mu}{3+16\mu^2(-1+\mu^2+2z)}\,,
\end{equation}
\begin{equation}
w_3=-\frac{2\mu(1+4\mu^2)z}{(4\mu^2-3)z-8\mu^2+2}\,,\quad w_4=\frac{ 2\mu(1+16\mu^4+8\mu^2(-1+2z))}{-3+16\mu^4+\mu^2(8-16z)}\,.
\end{equation}
The local coordiantes at $z=-16\mu^{-2}/(4-\mu^{-2})^2$ are 
\begin{equation}
w_1= \frac{2\mu (-1+4\mu^2)(z-1)}{(-1+4\mu^2)z +4\mu^2+3}\,,\quad w_2=-\frac{2\mu(1+4\mu^2)}{(8\mu^2-2)z-4\mu^2+3}\,,
\end{equation}
\begin{equation}
w_3=\frac{2\mu(-1+16\mu^4)z}{(3-16\mu^2+16\mu^4)z+32\mu^2}\,,\quad w_4= -\frac{2\mu(-8\mu^2(z-2)+z+16\mu^4z)}{8\mu^2(z-2)-3z+16\mu^4z}\,.
\end{equation}
The local coordinates at $z=16\mu^{-2}/(4+\mu^{-2})^2$ are determined to be
\begin{equation}
w_1=2\mu (4+\mu^{-2}) \frac{z-1}{(4+\mu^{-2})z +4-3\mu^{-2}}\,,\quad w_2=\mu\frac{2(-4+\mu^{-2})}{(8+2\mu^{-2})z-4-3\mu^{-2}}\,,
\end{equation}
\begin{equation}
w_3=\frac{2\mu (-16+\mu^{-4}) z}{ (16+16\mu^{-2}+3\mu^{-4})z -32\mu^{-2}}\,,\quad w_4=\mu \frac{-2(4+\mu^{-2})^2z +32\mu^{-2}}{(4+\mu^{-2})(-4+3\mu^{-2})z-16\mu^{-2}}\,.
\end{equation}
The local coordinates at $z= 1-16\mu^{-2}/(4+\mu^{-2})^2$ are
\begin{equation}
w_1=\frac{2\mu (16\mu^4-1)(z-1)}{3(z-1)+16\mu^2(1+\mu^2(z-1)+z)}\,,\quad w_2=-\frac{ 8\mu^3-2\mu}{(8\mu^2+2)z-4\mu^2+1}\,,
\end{equation}
\begin{equation}
w_3=- \frac{2\mu(4\mu^2+1)z}{(4\mu^2+1)z-8\mu^2+2}\,,\quad w_4=-\frac{2\mu (-1+z+8\mu^2(1+2\mu^2(z-1)+z))}{3+16\mu^4(z-1)-3z-8\mu^2(z+1)}\,.
\end{equation}
The local coordinates at $z= 1+16\mu^{-2}/(4-\mu^{-2})^2$ are
\begin{equation}
w_1=-\frac{2\mu(16\mu^4-1)(z-1)}{(3-16\mu^2+16\mu^4)z-3-16\mu^2-16\mu^4}\,,\quad w_2=-\frac{2\mu (1+4\mu^2)}{(8\mu^2-2)z-4\mu^2-1}\,,
\end{equation}
\begin{equation}
w_3=\frac{2\mu(4\mu^2-1)z}{-(4\mu^2-1)z+8\mu^2+2}\,,\quad w_4= -\frac{2\mu (-1+z-8\mu^2(1+z)+16\mu^4(z-1))}{3-3z+8\mu^2(1+z)+16\mu^4(z-1)}\,.
\end{equation}
The local coordinates at $z=(4+\mu^{-2})^2/16\mu^{-2}$ are
\begin{equation}
w_1=\frac{2\mu(4\mu^2+1)(z-1)}{(4\mu^2-3)z+4\mu^2+1}\,,\quad w_2=-\frac{2\mu(1-32\mu^4)}{-32\mu^2z+3+16\mu^2+16\mu^4}\,,
\end{equation}
\begin{equation}
w_3=-\frac{2\mu(4\mu^2-1)z}{(4\mu^2+3)z-8\mu^2-2}\,,\quad w_4=-\frac{2\mu(1+16\mu^4+\mu^2(8-16z))}{-3+16\mu^4+8\mu^2(2z-1)}\,.
\end{equation}

We will only construct the local coordinates of the vertical region up to the few leading orders in the stub parameter. Also, the local coordinates we shall choose do not strictly speaking respect $SL(2;R).$ However, this is not a problem as we shall permute over vertices to compute amplitudes. Also, we only need the local coordinate for the open string puncture at $0,$ as we shall use this puncture to construct the Feynman region of higher vertices. Other local coordinates can be similarly constructed. We shall use the following ansatz 
\begin{equation}
w_3=\mu \frac{\alpha_1(t) z}{\alpha_2(t)z+\alpha_3(t)}\,.
\end{equation}
The boundary conditions for $\alpha_i(t)$ are 
\begin{equation}
\alpha_1(-\mu^2+\dots)=\alpha_1(-\mu^{-2}+\dots)=\alpha_1(\mu^{-2}+\dots)=2
\end{equation}
\begin{equation}
\alpha_1(1-\mu^{-2}+\dots)=\alpha_1(1+\mu^{-2}+\dots)=\alpha_1(\mu^2+\dots)=2\,,
\end{equation}
\begin{equation}
\alpha_2(-\mu^2+\dots)=-\frac{4\mu^2-3}{4\mu^2+1}\,, \quad\alpha_2(-\mu^{-2}+\dots)=\frac{16\mu^4-16\mu^2+3}{16\mu^4-1}\,,
\end{equation}
\begin{equation}
\alpha_2(\mu^{-2}+\dots)= - \frac{16\mu^4+16\mu^2+3}{16\mu^4-1}\,,\quad \alpha_2(1-\mu^{-2}+\dots)=-1\,,
\end{equation}
\begin{equation}
\alpha_2(1+\mu^{-2}+\dots)= -1\,,\quad \alpha_2(\mu^2+\dots)= -\frac{4\mu^2+3}{4\mu^2-1}\,,
\end{equation}
\begin{equation}
\alpha_3(-\mu^2+\dots)=\frac{8\mu^2-2}{4\mu^2+1}\,,\quad \alpha_3(-\mu^{-2}+\dots)=\frac{32\mu^2}{16\mu^4-1}\,,
\end{equation}
\begin{equation}
\alpha_3(\mu^{-2}+\dots)=\frac{32\mu^2}{16\mu^4-1}\,,\quad \alpha_3(1-\mu^{-2}+\dots)=\frac{8\mu^2-2}{4\mu^2+1}\,,
\end{equation}
\begin{equation}
\alpha_3(1+\mu^{-2}+\dots)=\frac{8\mu^2+2}{4\mu^2-1}\,,\quad \alpha_3(\mu^2+\dots)=\frac{8\mu^2+2}{4\mu^2-1}\,.
\end{equation}

We choose
\begin{equation}
\alpha_1(t)=2\,,
\end{equation}
\begin{equation}
\alpha_2(t):= \begin{cases}
(2\mu^{-2}-\mu^{-4})t+1-\mu^{-2}+\frac{9}{4}\mu^{-4}\,,& \text{for}\,-\mu^2+\dots<t<-\mu^{-2}+\dots\\
\left(\mu^{-2}+\frac{9}{4}\mu^{-4}\right)t -\left(1+\mu^{-2}+\frac{5}{4}\mu^{-4}\right) \,,&\text{for}\, \mu^{-2}+\dots<t<1-\mu^{-2}+\dots\\
-\mu^{-4}t -1+\mu^{-4}\,,&\text{for}\, 1+\mu^{-2}+\dots<t<\mu^2+\dots\,
\end{cases}\,,
\end{equation}
\begin{equation}
\alpha_3(t):= \begin{cases}
-2(\mu^{-2}-\mu^{-4})t+2(\mu^{-2}-\mu^{-4})\,,& \text{for}\,-\mu^2+\dots<t<-\mu^{-2}+\dots\\
\left(2+\mu^{-2}+\frac{1}{4}\mu^{-4}\right)t \,,&\text{for}\, \mu^{-2}+\dots<t<1-\mu^{-2}+\dots\\
2+\mu^{-2}+\frac{1}{4}\mu^{-4}\,,&\text{for}\, 1+\mu^{-2}+\dots<t<\mu^2+\dots\,
\end{cases}\,.
\end{equation}

\subsection{Disk with C-O-O-O}\label{sec:loc cord COOO}
The moduli space of this final case is two-dimensional. To make the comparison with the conventional worldsheet normalization, e.g., that of \cite{Polchinski:1998rq,Polchinski:1998rr}, we shall use $SL(2;R)$ invariance to fix the location of the closed string puncture to be $i,$ and one open string puncture at $0,$ and let the position of the two other open string punctures to be moduli. 

The Feynman regions of the disk amplitude with C-O-O-O consist of three contributions. First, a disk diagram with C-O is glued to a disk diagram with O-O-O-O. Second, a disk diagram with C-O-O is glued to a disk diagram with O-O-O. Third, a disk diagram with C-O is glued to a disk diagram with O-O-O, which is again glued to a disk diagram with O-O-O. By construction, if the boundary region of the fundamental vertex meets with the boundary regions of the first two Feynman regions, the whole moduli space is entirely covered. 

Let us first construct the moduli space covered by joining a disk diagram with C-O-O to a disk diagram with O-O-O. We shall denote the global coordinate of the first disk by $x_1,$ and the global coordinate of the second disk by $x_2.$ We shall glue a vertex at $x_1=t$ to a vertex at $x_2=0$ by the plumbing fixture
\begin{equation}
2\mu^4 \frac{4\mu^4+1}{4\mu^4} \frac{x_1-t}{(1+tx_1)+\mu^2 h(t) (x_1-t)} \frac{x_2}{2-x_2}=-q_1\,.
\end{equation}
The global coordinate $x_2$ is therefore mapped to
\begin{equation}\label{eqn:plumbing1 COOO}
x_1= \frac{2q_1(x_2-2)+(1+4\mu^4)tx_2-2\mu^2 qt (-2+x_2)h(t)}{-2qt(-2+x_2)+x_2+4\mu^4 x_2-2\mu^2 q(x_2-2)h(t)}\,.
\end{equation}
The open string punctures at $x_2=1$ and $x_2=\infty$ are mapped to
\begin{equation}\label{eqn:sec punct COOO deg1}
x_1= \frac{-2q_1 +t+4\mu^4 t+2\mu^2 q_1th(t)}{1+4\mu^4+2q_1t+2\mu^2q_1h(t)}\,,
\end{equation}
and
\begin{equation}\label{eqn:third punct COOO deg1}
x_1=\frac{2q_1+t+4\mu^4t -2\mu^2 q_1t h(t)}{1+4\mu^4 -2q_1t-2\mu^2q_1h(t)}\,,
\end{equation}
respectively. The remaining open string puncture is located at $-t.$ For the disk diagram with an NSNS closed string puncture, and NS open string punctures, we shall place PCOs at
\begin{equation}
    p_1(t)=\pm i\frac{2\mu^2/\sqrt{3}-\sqrt{3}\mu^{-2}/2}{2\mu^2-(2\mu^2)^{-1}}\left(t-\frac{1}{2\mu^2}\right) \pm i\frac{\sqrt{3}}{2}\mu^{-2}\,,
\end{equation}
\begin{equation}
    p_2(t,q_1)=t\pm\frac{i\sqrt{3}(1+t^2)q_1}{2\mu^4}+\dots\,.
\end{equation}
The PCO location $p_1(t)$ was chosen to match the PCO location of the disk diagram with C-O-O, and the PCO location $p_2(t)$ was chosen to match the PCO location of the disk diagram with O-O-O. Note that given the plumbing fixture \eqref{eqn:plumbing1 COOO}, we can use various $SL(2;R)$ transformations to cover different regions of the moduli space. 

Let us first cover the range $z_1<0<z_3.$ As we are interested in the range of the fundamental vertex, we shall set $q_2=1.$ Note that the range of $t$ is defined to be $(2\mu^{2})^{-1}\leq t\leq1.$ For $t>1,$ we can use $t'=1/t$ as a modulus. Expanding \eqref{eqn:sec punct COOO deg1} and \eqref{eqn:third punct COOO deg1} in large $\mu,$ we find that the open string punctures at $x_2=1$ and $x_2=\infty$ are mapped to
\begin{equation}
x_1= t-\frac{1+t^2}{2\mu^4}+\frac{(1+t^2)(1+2t+3t^2)}{16t\mu^8}+\dots\,,
\end{equation}
and
\begin{equation}
x_1= t+\frac{1+t^2}{2\mu^4}+\frac{(1+t^2)(1-2t+3t^2)}{16t\mu^8}+\dots\,.
\end{equation}
Similarly, the points $x_2=p_\pm$ are mapped to
\begin{equation}
    x_1=t\pm \frac{i\sqrt{3}(1+t^2)}{2\mu^4}+\dots\,.
\end{equation}
We can then use the following $SL(2;R)$ transformation to define the global coordinate of the disk with C-O-O-O as
\begin{equation}
z:= \frac{ax_1+b}{cx_1+d}\,,
\end{equation}
where 
\begin{equation}
a=-3+4\mu^4-2t+32\mu^8t-t^2+12\mu^4t^2\,,~ b=-t+12\mu^4t+2t^2-32\mu^8 t^2-3t^3+4\mu^4t^3\,,
\end{equation}
\begin{equation}
c=t-12\mu^4t-2t^2+32\mu^8t^2+3t^3-4\mu^4t^3\,,~d=-3+4\mu^4-2t+32\mu^8t-t^2+12\mu^4t^2\,.
\end{equation}
Under the global coordinate, we find that the open string punctures are located at
\begin{align}
z_1=&\frac{2 \left(4 \mu ^4+1\right) t \left(t \left(8 \mu ^4+t-2\right)-1\right)}{32 \mu ^8 t \left(t^2-1\right)-4 \mu ^4 \left(t^4+6 t^2+1\right)+t (t (t (3 t-2)+2)+2)+3}\,,\\
=&\frac{2 t}{t^2-1}+\frac{\left(t^2+1\right)^2}{2 \mu ^4 \left(t^2-1\right)^2}+\frac{\left(t^4-2 t^3+6 t^2+2 t+1\right) \left(t^2+1\right)^2}{16 \mu ^8 t \left(t^2-1\right)^3}+\dots\,,
\end{align}
\begin{equation}
z_2=0\,,
\end{equation}
\begin{align}
z_3=&-\frac{16 \left(4 \mu ^4+1\right) \left(t-4 \mu ^4 t\right)^2}{\left(3-4 \mu ^4\right)^2+\left(3-4 \mu ^4\right)^2 t^4-2 \left(8 \left(64 \mu ^{12}-22 \mu ^4+5\right) \mu ^4+3\right) t^2}\,,\\
=&\mu^{-4}-\frac{1}{4}\mu^{-8}+\dots\,.
\end{align}
We find that the PCOs at the symmetric point in the disk with O-O-O are mapped to
\begin{equation}
p_2= p_\pm \mu^{-4}+\dots\,,
\end{equation}
and the PCOs in the disk with C-O-O are mapped to
\begin{equation}
    p_1=\frac{(\pm 3i +\sqrt{3})t}{\mp3i+\sqrt{3}t^2}\mp\frac{3i\sqrt{3}(1+t^2)}{(\mp3i+\sqrt{3}t^2)^2\mu^2}+\dots\,.
\end{equation}
Therefore, we obtained one boundary of the fundamental vertex regions
\begin{equation}
-\infty<z_1<-\mu^{-2}+\frac{1}{2}\mu^{-4}-\frac{3}{8}\mu^{-6}+\frac{13}{32}\mu^{-8}+\dots\,,
\end{equation}
and
\begin{equation}
z_3=\mu^{-4}-\frac{1}{4}\mu^{-8}+\dots\,.
\end{equation}

We can similarly obtain the other boundary region of the moduli space
\begin{equation}
z_1=-\mu^{-4}+\frac{1}{4}\mu^{-8}+\dots\,,
\end{equation}
\begin{equation}
z_2=0\,,
\end{equation}
\begin{equation}
\mu^{-2}-\frac{1}{2}\mu^{-4}+\frac{3}{8}\mu^6-\frac{13}{32}\mu^{-8}+\dots<z_3<\infty\,.
\end{equation}
The PCOs are located at
\begin{equation}
    p_2=-p_\pm \mu^{-4}+\dots\,,
\end{equation}
\begin{equation}
    p_1=\frac{(\pm3i+\sqrt{3})t}{\mp3i+\sqrt{3}t^2}\mp \frac{3i\sqrt{3}(1+t^2)}{(\mp3i +\sqrt{3}t^2)^2\mu^2}+\dots\,.
\end{equation}

Second, we shall now cover the region $0<z_3<z_1.$ One of the two boundaries of this fundamental region can be easily obtained by extending $t\leq1$ to $1\leq t\leq 2\mu^2$ for the following expressions we already obtained
\begin{align}
z_1=&\frac{2 \left(4 \mu ^4+1\right) t \left(t \left(8 \mu ^4+t-2\right)-1\right)}{32 \mu ^8 t \left(t^2-1\right)-4 \mu ^4 \left(t^4+6 t^2+1\right)+t (t (t (3 t-2)+2)+2)+3}\,,\\
=&\frac{2 t}{t^2-1}+\frac{\left(t^2+1\right)^2}{2 \mu ^4 \left(t^2-1\right)^2}+\frac{\left(t^4-2 t^3+6 t^2+2 t+1\right) \left(t^2+1\right)^2}{16 \mu ^8 t \left(t^2-1\right)^3}+\dots\,,
\end{align}
\begin{equation}
z_2=0\,,
\end{equation}
\begin{align}
z_3=&-\frac{16 \left(4 \mu ^4+1\right) \left(t-4 \mu ^4 t\right)^2}{\left(3-4 \mu ^4\right)^2+\left(3-4 \mu ^4\right)^2 t^4-2 \left(8 \left(64 \mu ^{12}-22 \mu ^4+5\right) \mu ^4+3\right) t^2}\,,\\
=&\mu^{-4}-\frac{1}{4}\mu^{-8}+\dots\,.
\end{align}
The PCOs are located at
\begin{equation}
    p_2= p_\pm \mu^{-4}+\dots\,,
\end{equation}
\begin{equation}
    p_1=\frac{(\pm 3i +\sqrt{3})t}{\mp3i+\sqrt{3}t^2}\mp\frac{3i\sqrt{3}(1+t^2)}{(\mp3i+\sqrt{3}t^2)^2\mu^2}+\dots\,.
\end{equation}

The other boundary region can be obtained by defining the global coordinate $z$ as
\begin{equation}
    z=\frac{x_1+t}{-tx_1+1}\,.
\end{equation}
In this patch, we find
\begin{equation}
    z_2=0\,,
\end{equation}
\begin{align}
    z_3=&-\frac{2 \left(4 \mu ^4+1\right) t \left(t \left(8 \mu ^4+t-2\right)-1\right)}{32 \mu ^8 t \left(t^2-1\right)-4 \mu ^4 \left(t^4+6 t^2+1\right)+t (t (t (3 t-2)+2)+2)+3}\,\\
    =&-\frac{2 t}{t^2-1}-\frac{(t^2+1)^2}{2 \mu ^4 \left(t^2-1\right)^2}+\dots\,,
\end{align}
\begin{align}
    z_1=&-\frac{2 \left(4 \mu ^4+1\right) t \left(t \left(8 \mu ^4-t-2\right)+1\right)}{32 \mu ^8 t \left(t^2-1\right)+4 \mu ^4 \left(t^4+6 t^2+1\right)-t (t (t (3 t+2)+2)-2)-3}\,,\\
    =&-\frac{2 t}{t^2-1}+\frac{(t^2+1)^2}{2 \mu ^4 \left(t^2-1\right)^2}+\dots\,,
\end{align}
for $(2\mu^2)^{-1}\leq t\leq1.$ The PCOs are located at
\begin{equation}
    p_2=-\frac{2t}{t^2-1}\pm \frac{i\sqrt{3}(1+t^2)^2}{2(-1+t^2)^2\mu^4}+\dots\,,
\end{equation}
\begin{equation}
    p_1=-\frac{(\mp 3i +\sqrt{3})t}{\pm3i+\sqrt{3}t^2}\mp \frac{3i\sqrt{3}(1+t^2)}{(\pm3i+\sqrt{3}t^2)^2\mu^2}+\dots\,.
\end{equation}

Finally, we shall cover the region $z_3<z_1<0.$ In one of the Feynman regions, we find
\begin{equation}
    z_2=0\,,
\end{equation}
\begin{align}
    z_3=&-\frac{2 \left(4 \mu ^4+1\right) t \left(t \left(8 \mu ^4+t-2\right)-1\right)}{32 \mu ^8 t \left(t^2-1\right)-4 \mu ^4 \left(t^4+6 t^2+1\right)+t (t (t (3 t-2)+2)+2)+3}\,\\
    =&-\frac{2 t}{t^2-1}-\frac{(t^2+1)^2}{2 \mu ^4 \left(t^2-1\right)^2}+\dots\,,
\end{align}
\begin{align}
    z_1=&-\frac{2 \left(4 \mu ^4+1\right) t \left(t \left(8 \mu ^4-t-2\right)+1\right)}{32 \mu ^8 t \left(t^2-1\right)+4 \mu ^4 \left(t^4+6 t^2+1\right)-t (t (t (3 t+2)+2)-2)-3}\,,\\
    =&-\frac{2 t}{t^2-1}+\frac{(t^2+1)^2}{2 \mu ^4 \left(t^2-1\right)^2}+\dots\,,
\end{align}
for $1\leq t\leq 2\mu^2.$ The PCOs are located at
\begin{equation}
    p_2=-\frac{2t}{t^2-1}\pm \frac{i\sqrt{3}(1+t^2)^2}{2(-1+t^2)^2\mu^4}+\dots\,,
\end{equation}
\begin{equation}
    p_1=-\frac{(\mp 3i +\sqrt{3})t}{\pm3i+\sqrt{3}t^2}\mp \frac{3i\sqrt{3}(1+t^2)}{(\pm3i+\sqrt{3}t^2)^2\mu^2}+\dots\,.
\end{equation}

In the other Feynman region, we find
\begin{align}
    z_3=&\frac{2 \left(4 \mu ^4+1\right) t \left(-t^2+\left(8 \mu ^4-2\right) t+1\right)}{-3 t^4-2 t^3+32 \mu ^8 t \left(t^2-1\right)-2 t^2+4 \mu ^4 \left(t^4+6 t^2+1\right)+2 t-3}\,,\\
    =&\frac{2 t}{t^2-1}+\frac{-t^4-2 t^2-1}{2 \mu ^4 \left(t^2-1\right)^2}+\dots\,,
\end{align}
\begin{align}
    z_1=&-\frac{16 \left(4 \mu ^4+1\right) \left(t-4 \mu ^4 t\right)^2}{-9 t^4+1024 \mu ^{16} t^2+6 t^2-16 \mu ^8 \left(t^4+22 t^2+1\right)+8 \mu ^4 \left(3 t^4+10 t^2+3\right)-9}\,,\\
    =&-\mu^{-4}+\frac{1}{4}\mu^{-8}\dots\,,
\end{align}
for $(2\mu^2)^{-1}\leq t\leq1.$ The PCOs are located at
\begin{equation}
    p_2=-p_\pm\mu^{-4}+\dots\,,
\end{equation}
\begin{equation}
    p_1=\frac{(\pm3i+\sqrt{3})t}{\mp3i+\sqrt{3}t^2}\mp\frac{3i\sqrt{3}(1+t^2)}{(\mp3i+\sqrt{3}t^2)^2\mu^2}+\dots\,.    
\end{equation}

Let us now construct the moduli space covered by joining a disk diagram with C-O to a disk diagram with O-O-O-O. We shall denote the global coordinate of the disk with C-O-O-O by $z,$ the global coordinate of the first disk in the gluing by $x_1,$ and the global coordinate of the second disk in the gluing by $x_2.$  We identify the open string puncture at $x_1=0$ with the open string puncture at $x_2=0$ via the plumbing fixture
\begin{equation}
\mu^2 x_1 \frac{2 x_2}{\alpha_2(t) x_2+\alpha_3(t)}=-q_2\,, 
\end{equation}
or equivalently,
\begin{equation}
x_1= -q_2\mu^{-2}  \frac{\alpha_3(t)+\alpha_2(t)x_2}{2x_2}\,,
\end{equation}
where $t$ stands for the modulus of the disk with O-O-O-O. The open string punctures at $x_2=1\,,~ \infty\,,$ and $x_2=t$ are mapped to
\begin{equation}
x_1= -q_2\mu^{-2} \frac{\alpha_2(t)+\alpha_3(t)}{2}\,,\quad x_1=-\frac{1}{2}\alpha_2(t)q_2\mu^{-2}\,,\quad x_1= -q_2\mu^{-2} \frac{\alpha_2(t)t+\alpha_3(t)}{2t}\,.
\end{equation}

We shall divide the moduli space $t$ into three to study the location of the open string punctures. We shall set $q_2=1$ to study the boundary conditions. When $\mu^{-2}<t<1-\mu^{-2}+\dots,$ we have
\begin{equation}
\alpha_2(t)=\frac{4 \left(16 \mu ^4+8 \mu ^2+1\right)}{64 \mu ^6-112 \mu ^4+28 \mu ^2-1}t+-\frac{16 \mu ^4-8 \mu ^2-3}{16 \mu ^4-24 \mu ^2+1}\,,
\end{equation}
and
\begin{equation}
\alpha_3(t)=-\frac{2 \left(4 \mu ^2+1\right)}{4 \mu ^2-1} t\,.
\end{equation}
The location of the open string punctures is, therefore, determined to be
\begin{equation}
x_1=\frac{1-2 t}{2 \mu ^2}+\frac{1-2 t}{2 \mu ^4}-\frac{5 (2 t-1)}{8 \mu ^6}+\dots\,,
\end{equation}
\begin{equation}
x_1=\frac{1}{2 \mu ^2}+\frac{1-t}{2 \mu ^4}+\frac{5-9 t}{8 \mu ^6}+\dots\,,
\end{equation}
\begin{equation}
x_1=-\frac{1}{2 \mu ^2}-\frac{t}{2 \mu ^4}+\frac{4-9 t}{8 \mu ^6}+\dots\,.
\end{equation}
We define the global coordinate $z$ as
\begin{equation}
z= \frac{2x_1+\mu^{-2}(\alpha_2(t)+\alpha_3(t))}{ -\mu^{-2}(\alpha_2(t)+\alpha_3(t))x_1+2}\,,
\end{equation}
so that the open string punctures are mapped to
\begin{equation}
z_1=0\,,
\end{equation}
\begin{equation}
z_2=\frac{t-1}{\mu ^2}+\frac{t-1}{2 \mu ^4}+\frac{-4 t^2+7 t-3}{8 \mu ^6}+\frac{-16 t^3+29 t-13}{32 \mu ^8}+\dots\,,
\end{equation}
and
\begin{equation}
z_3=\frac{t}{\mu ^2}+\frac{t}{2 \mu ^4}+\frac{t (4 t-1)}{8 \mu ^6}-\frac{t \left(16 t^2-48 t+19\right)}{32 \mu ^8}+\dots\,.
\end{equation}
It is straightforward to confirm that the boundary region of this degeneration channel coincides with that of the degeneration channel into a disk with C-O-O and a disk with O-O-O.

Lastly, for completeness, we shall determine the region of the moduli space covered by three disk diagrams, one with C-O, two with O-O-O. Let us denote the coordinates of the disk with C-O by $x_1,$ and those of the disks with O-O-O by $x_2$ and $x_3.$ 

First, let us glue open string punctures at $x_1=0$ to $x_2=0,$ and $x_2=1$ to $x_3=0$
\begin{equation}
    \mu^2 x_1 g_1(x_2)=-q_1\,,
\end{equation}
\begin{equation}
    \mu^2 g_2(x_2) g_1(x_3)=-q_2\,.
\end{equation}
We find that the open string punctures are located at, in $x_1$ coordinate, 
\begin{equation}
    \left(\frac{q_1}{2\mu^2},\frac{3q_1}{2\mu^2}-\frac{8q_1}{4\mu^2+q_2},\frac{3q_1}{2\mu^2}+\frac{8q_1}{-4\mu^2+q_2}\right)\,.
\end{equation}
We shall refer to this plumbing fixture as plumbing A. Similarly, we can glue open string punctures at $x_1=0$ to $x_2=0,$ and $x_2=\infty$ to $x_3=0$
\begin{equation}
    \mu^2 x_1g_1(x_2)=-q_1\,,
\end{equation}
\begin{equation}
    \mu^2 g_3(x_2)g_1(x_3)=-q_2\,.
\end{equation}
We find that the open string punctures are located at
\begin{equation}
    \left(\frac{q_1}{2\mu^2},-\frac{3q_1}{2\mu^2}-\frac{8q_1}{-4\mu^2+q_2},-\frac{3q_1}{2\mu^2}+\frac{8q_1}{4\mu^2+q_2}\right)\,.
\end{equation}
We shall refer to this plumbing fixture as plumbing B.

Now, we shall use the conformal killing group to fix the location of one of the open string punctures at $0.$ This will generate six different Feynman regions. As we are interested in determining the boundary conditions of this Feynman region, we will set $q_1=q_2=1$ from now on.

We shall use the Mobius map
\begin{equation}
    z:=\frac{x_1-\mu^{-2}/2}{x_1\mu^{-2}/2+1}\,,
\end{equation}
to move the location of punctures at $q_1/2\mu^2$ to $0$ in both of the plumbing fixtures. We find that for plumbing A open string punctures are located at 
\begin{align}
    (z_1,z_2,z_3)=&\left(0,\frac{4 \mu ^2-16 \mu ^4}{16 \mu ^6+4 \mu ^4-4 \mu ^2+3},\frac{4 \left(4 \mu ^4+\mu ^2\right)}{-16 \mu ^6+4 \mu ^4+4 \mu ^2+3}\right)\,,\\
    =&\left(0,-\mu^{-2}+\frac{1}{2}\mu^{-4}-\frac{3}{8}\mu^{-6}+\dots,-\mu^{-2}-\frac{1}{2}\mu^{-4}-\frac{3}{8}\mu^{-6}+\dots\right)\,,
\end{align}
and the PCOs are located at
\begin{equation}
    p_1= -p_\pm\mu^{-2}+\dots\,,
\end{equation}
\begin{equation}
    p_2=-\mu^{-2} \pm\frac{i\sqrt{3}}{2}\mu^{-4}+\dots\,.
\end{equation}
For plumbing B, we find that open string punctures are located at 
\begin{align}
    (z_1,z_2,z_3)=&\left\{0,\frac{4 \mu ^2 \left(4 \mu ^2-1\right)}{16 \mu ^6+4 \mu ^4-4 \mu ^2+3},\frac{4 \mu ^2 \left(4 \mu ^2+1\right)}{16 \mu ^6-4 \mu ^4-4 \mu ^2-3}\right\}\,,\\
    =&\left(0,\mu^{-2}-\frac{1}{2}\mu^{-4}+\frac{3}{8}\mu^{-6}+\dots,\mu^{-2}+\frac{1}{2}\mu^{-4}+\frac{3}{8}\mu^{-6}+\dots\right)\,,
\end{align}
and PCOs are located at
\begin{equation}
    p_1=p_\pm \mu^{-2}+\dots\,,
\end{equation}
\begin{equation}
    p_2=\mu^{-2}\pm\frac{i\sqrt{3}}{2}\mu^{-4}+\dots\,.
\end{equation}

Let us now bring the second punctures to $0.$ For plumbing A, we have the following Mobius map
\begin{equation}
    z=\frac{-16 \mu ^4+16 \mu ^2+\left(2 \mu ^2-32 \mu ^6\right) x_1-3}{-32 \mu ^6+2 \mu ^2+\left(16 \mu ^4-16 \mu ^2+3\right) x_1}\,,
\end{equation}
which maps the open string punctures to
\begin{align}
    (z_1,z_2,z_3)=&\left(\frac{4 \mu ^2 \left(4 \mu ^2-1\right)}{16 \mu ^6+4 \mu ^4-4 \mu ^2+3},0,\frac{64 \mu ^4}{-64 \mu ^8-12 \mu ^4+9}\right)\,,\\
    =&\left(\mu^{-2}-\frac{1}{2}\mu^{-4}+\frac{3}{8}\mu^{-6}+\dots,0,-\mu^{-4}+\dots\right)\,,
\end{align}
and PCOs to
\begin{equation}
    p_1=p_\pm \mu^{-2}-\frac{1}{2}\mu^{-4}+\dots\,,
\end{equation}
\begin{equation}
    p_2=-p_\pm \mu^{-4}+\dots\,.
\end{equation}
For plumbing B, we have the following Mobius map
\begin{equation}
    z=\frac{4 \mu ^2+\left(-8 \mu ^4-2 \mu ^2\right) x_1-3}{-8 \mu ^4-2 \mu ^2+\left(3-4 \mu ^2\right) x_1}\,,
\end{equation}
which maps the open string punctures to
\begin{align}
    (z_1,z_2,z_3)=&\left(\frac{4 \mu ^2-16 \mu ^4}{16 \mu ^6+4 \mu ^4-4 \mu ^2+3},0,\frac{64 \mu ^4}{64 \mu ^8+12 \mu ^4-9}\right)\,,\\
    =&\left(-\mu^{-2}+\frac{1}{2}\mu^{-4}-\frac{3}{8}\mu^{-6}+\dots,0,\mu^{-4}+\dots\right)\,,
\end{align}
and PCOs to
\begin{equation}
    p_1=-p_\pm \mu^{-2}+\frac{1}{2}\mu^{-4}+\dots\,,
\end{equation}
\begin{equation}
    p_2=p_\pm \mu^{-4}+\dots\,.
\end{equation}

Finally, let us bring the third puncture to $0.$ For plumbing A, we have the following Mobius map
\begin{equation}
    z=\frac{4 \mu ^2+\left(8 \mu ^4-2 \mu ^2\right) x_1+3}{8 \mu ^4-2 \mu ^2-\left(4 \mu ^2+3\right) x_1}\,,
\end{equation}
which maps the open string punctures to
\begin{align}
    (z_1,z_2,z_3)=&\left(\frac{4 \mu ^2 \left(4 \mu ^2+1\right)}{16 \mu ^6-4 \mu ^4-4 \mu ^2-3},\frac{64 \mu ^4}{64 \mu ^8+12 \mu ^4-9},0\right)\,,\\
    =&\left(\mu^{-2}+\frac{1}{2}\mu^{-4}+\frac{3}{8}\mu^{-6}+\dots,\mu^{-4}+\dots,0\right)\,,
\end{align}
and PCOs to
\begin{equation}
    p_1=p_\pm \mu^{-2}+\frac{1}{2}\mu^{-4}+\dots\,,
\end{equation}
\begin{equation}
    p_2=p_\pm\mu^{-4}+\dots\,.
\end{equation}
For plumbing B, we have the following Mobius map
\begin{equation}
    z=\frac{4 \mu ^2+\left(2 \mu ^2-8 \mu ^4\right) x_1+3}{-8 \mu ^4+2 \mu ^2-\left(4 \mu ^2+3\right) x_1}\,,
\end{equation}
which maps the open string punctures to
\begin{align}
    (z_1,z_2,z_3)=&\left(\frac{4 \left(4 \mu ^4+\mu ^2\right)}{-16 \mu ^6+4 \mu ^4+4 \mu ^2+3},\frac{64 \mu ^4}{-64 \mu ^8-12 \mu ^4+9},0\right)\,,\\
    =&\left( -\mu^{-2}-\frac{1}{2}\mu^{-4}-\frac{3}{8}\mu^{-6}+\dots,-\mu^{-4}+\dots,0\right)\,,
\end{align}
and PCOs to
\begin{equation}
    p_1=-p_\pm\mu^{-2}-\frac{1}{2}\mu^{-4}+\dots\,,
\end{equation}
\begin{equation}
    p_2=-p_\pm\mu^{-4}+\dots\,.
\end{equation}

\section{Tadpole for the off-diagonal modes}\label{app:off diag}

The source term $\mathcal{S}_1$ provides the tadpole for the off-diagonal modes, which will determine if we can indeed find a puffed-up NS5-brane as a perturbative solution to string field theory. Because we are taking the large stub limit, and there is no tachyonic spectrum in the original CFT we started with, only the following terms will be non-trivial
\begin{equation}
\mathcal{S}_{1,1}:= -\tilde{R}_{S^3}^{-3}\mathcal{G}\Bbb{P}\left[\frac{1}{3!}(\Bbb{P}\Psi_{1,0}^o)^3\right]_{D^2}^o\,,
\end{equation}
\begin{equation}
\mathcal{S}_{1,2}:=-\tilde{R}_{S^3}^{-3}\mathcal{G}\Bbb{P}\left[ \frac{1}{2}\Psi_{1,0}^c \otimes(\Bbb{P}\Psi_{1,0}^o)^2\right]_{D^2}^o\,.
\end{equation}

Let us recall that the first order open string background solution in $-1$ picture takes the following form
\begin{equation}
(\tilde{R}_{S^3}^{-1}\Bbb{P}\Psi_{1,0}^o)^{-1}=f_i  c e^{-\phi}\tilde{\psi}^i\,.
\end{equation}
Similarly, the test fields in $-1$ picture take the following forms
\begin{equation}
V_{t,1}^o=g_A c e^{-\phi}\tilde{\psi}^A\,,\quad V_{t,2}^o=h c\partial c e^{-2\phi}\partial\xi\,.
\end{equation}
We shall now find $0$ picture forms of the above vertices
\begin{equation}
\left( \tilde{R}_{S^3}^{-1}\Bbb{P}\Psi_{1,0}^o\right)^{0}=-i\sqrt{2} f_i c\partial \tilde{X}^i-f_i\eta e^\phi\tilde{\psi}^i\,,
\end{equation}
\begin{equation}
(V_{t,1}^o)^0=-i\sqrt{2} g_A c\partial\tilde{X}^A-g_A\eta e^\phi\tilde{\psi}^A\,,
\end{equation}
\begin{equation}
(V_{t,2}^o)^0=-h\partial c\,.
\end{equation}
Note that we assumed $g_A$ and $h$ are constant $p\times p$ matrices. Note that we defined the zero picture vertices as
\begin{equation}
\mathcal{X}_0 V^{-1}\,.
\end{equation}
For the first order background solution and $V_{t,1}^o,$ since there is no singularity as the PCO comes close to the vertex, we can instead just treat the zero picture vertex as a PCO on the $-1$ picture vertex.

To compute $\mathcal{S},$ we shall proceed as follows. Open string amplitudes are ordered in the position of the open string punctures. Let $z_1,$ $z_2,$ and $z_3$ denote the positions of three first-order background solutions, and let $z_4$ denote the position of the test field. Then, we shall sum over the six distinct orderings to compute the correlator. We shall then fix the position of three vertices to be $0,~1,~\infty,$ and use the rest of the position as the modulus to be integrated over. This cumbersome procedure ensures that despite the string vertices we chose for the four-point function do not respect the $SL(2;R),$ the final on-shell amplitude is gauge invariant. Also, we haven't yet completely determined the orientation of the moduli integral. Although one can, in principle, determine the orientation purely within string field theory \cite{Sen:2024npu}, in this work, we shall use the orientation provided by the conventional worldsheet approach, which fixes the orientation of the moduli integral of all the permutations.

\subsection{Computation of $\mathcal{S}_{1,1}$}
Let us first study $\mathcal{S}_{1,1}.$

We write the amplitude for $z_1<z_2<z_3<z_4$ as with the test field $V_{t,1}^o$, in the interior of the moduli space,
\begin{align}\label{eqn:amp gen}
\mathcal{A}_{1,1,1}:=& -\left\{ V_{t,1}^o\otimes \frac{1}{3!} (\tilde{R}_{S^3}^{-3}\Bbb{P}\Psi_{1,0}^o)^4\right\}\,,\\
=&\frac{\Omega}{3!}\int dt \oint dw  \text{Tr}\biggr\langle \mathfrak{B} \otimes f_i  ce^{-\phi}\tilde{\psi}^i(z_1)\otimes f_j ce^{-\phi}\tilde{\psi}^j(z_2)\otimes f_k ce^{-\phi}\tilde{\psi}^k(z_3)\nonumber\\&\otimes g_A ce^{-\phi}\tilde{\psi}^A  (z_4)\otimes \mathcal{X}(p_1)\otimes\mathcal{X}(p_2) \biggr\rangle\,,
\end{align}
where 
\begin{equation}
\mathfrak{B}:=\left( \sum_{i} \frac{\partial F_i}{\partial t} b(w)-\sum_i \frac{1}{\mathcal{X}(p_i)} \frac{\partial p_i}{\partial t}\partial\xi(p_i)\right)\,,
\end{equation}
and $\Omega$ is an orientation of the open-string diagram we shall determine momentarily. For the amplitude involving the test field $V_{t,2}^o,$ we can just simply replace $V_{t,1}^o$ with $V_{t,2}^o.$ We define the modulus $t$ as the cross ratio
\begin{equation}
t:= \frac{(z_1-z_3)(z_2-z_4)}{(z_1-z_2)(z_3-z_4)}\,.
\end{equation}
To determine the orientation $\Omega,$ we shall compare our amplitude to that of \cite{Polchinski:1998rq}. Let us first fix $z_1=0,~z_2=1,~z_4=\infty.$ Then the modulus $t$ is related to $z_3$ as
\begin{equation}
t=z_3\,.
\end{equation}
The amplitude \eqref{eqn:amp gen} is then written as
\begin{align}
\frac{\Omega}{3!}\int dz_3 \text{Tr}(f_i f_j f_k g_A)\biggr\langle & ce^{-\phi}\tilde{\psi}^i(0)\otimes ce^{-\phi}\tilde{\psi}^j(1)\otimes  g_A ce^{-\phi}\tilde{\psi}^A  (\infty) \otimes  e^{-\phi}\tilde{\psi}^k(z_3)\nonumber\\&\otimes \mathcal{X}(p_1)\otimes\mathcal{X}(p_2) \biggr\rangle\,.
\end{align}
The amplitude above therefore agrees with the convention of \cite{Polchinski:1998rq} provided that $\Omega=1$ and
\begin{equation}
\langle c(z_1)c(z_2)c(z_3)\rangle= - C_{D^2}\int d^{p+1}\tilde{X} (z_1-z_2)(z_2-z_3)(z_1-z_3)\,.
\end{equation}
Therefore, we shall fix $\Omega=1.$ 

\eqref{eqn:amp gen} is not the complete answer, as we need to move PCOs on the boundary of the moduli space correspondingly to match the location of the PCOs to that of the boundary regions of the Feynman regions. The effect of the vertical integration can be obtained by replacing
\begin{equation}
\mathcal{X}(p_1)\mathcal{X}(p_2) dt \oint \left(\sum_{i=1}^4 \frac{\partial F_i}{\partial t} b(w) \right)\,,
\end{equation}
with
\begin{equation}
(\xi(p_1)-\xi(W_1))\mathcal{X}(p_2)+ \mathcal{X}(W_1)(\xi(p_2)-\xi(W_2))\,,
\end{equation}
where we assumed that we moved the PCOs from $p_1$ and $p_2$ to $W_1$ and $W_2$ sequentially.

Let us compute the correlator in $\mathcal{A}_{1,1}.$ As all of the vertices in the computation of $\mathcal{A}_{1,1},$ are on-shell primaries we can move PCOs to some of the vertices to simplify the computations. Note that once this choice is made, we need to use the same PCO configurations in the interior of the moduli space to ensure that the amplitudes are well-defined. We shall bring the PCO at $p_1$ to $z_1$ and the PCO at $p_2$ to $z_4.$ In the interior of the moduli space, we find
\begin{align}
\mathcal{A}_{1,1}^i=&\frac{1}{3!} C_{D^2}\text{Tr}( f_i f_j f_k g_l) \int_{\mathcal{M}_1} dt \oint dw \left(\sum_{i=1}^4 \frac{\partial F_i}{\partial t} \right) \mathcal{A}_{1,1}^{bc}(\mathcal{A}_{1,1}^m)^{ijkl}\,, 
\end{align}
for $z_1<z_2<z_3<z_4,$  where
\begin{equation}
\mathcal{A}_{1,1}^{bc}(z_1,z_2,z_3,z_4)= \left( \frac{1}{w-z_1}z_{23}z_{24}z_{34}-\frac{1}{w-z_2} z_{13}z_{14}z_{34} +\frac{1}{w-z_3} z_{12}z_{14}z_{24}-\frac{1}{w-z_4}z_{12}z_{13}z_{23}\right)\,,
\end{equation}
\begin{equation}
(\mathcal{A}_{1,1}^m)^{ijkl}(z_1,z_2,z_3,z_4)= -\frac{\eta^{il}\eta^{jk}}{(z_1-z_4)^2(z_2-z_3)^2}\,,
\end{equation}
and $\mathcal{M}_1$ denotes the moduli space for the ordering $z_1<z_2<z_3<z_4.$ Note that the ordering $<$ we are using is the circular ordering on the boundary of the disk, not the numerical ordering. In particular, we have $ 0<1<\infty<0.$ Also, we suppressed $\int d^4\tilde{X}.$

There are, in total, five more contributions coming from 5 additional orderings. We summarize those contributions below
\begin{equation}
\mathcal{A}_{1,1,2}^i=\frac{1}{3!}C_{D^2} \text{Tr}(f_i f_kf_j g_l) \int dt\oint dw \left(\sum_{i=1}^4 \frac{\partial F_i}{\partial t} \right)  \mathcal{A}_{1,1}^{bc}(z_1,z_3,z_2,z_4) (\mathcal{A}_{1,1}^m)^{ikjl} (z_1,z_3,z_2,z_4)\,,
\end{equation}
\begin{equation}
\mathcal{A}_{1,1,3}^i =\frac{1}{3!}C_{D^2} \text{Tr}(f_if_jg_lf_k) \int dt \oint dw \left(\sum_{i=1}^4 \frac{\partial F_i}{\partial t} \right)\mathcal{A}_{1,1}^{bc} (z_1,z_2,z_4,z_3) (\mathcal{A}_{1,1}^m)^{ijlk} (z_1,z_2,z_4,z_3)\,,
\end{equation}
\begin{equation}
\mathcal{A}_{1,1,4}^i=\frac{1}{3!}C_{D^2}\text{Tr}(f_i f_kg_lf_j) \int dt\oint dw \left(\sum_{i=1}^4 \frac{\partial F_i}{\partial t} \right) \mathcal{A}_{1,1}^{bc}(z_1,z_3,z_4,z_2)(\mathcal{A}_{1,1}^m)^{iklj}(z_1,z_3,z_4,z_2)\,,
\end{equation}
\begin{equation}
\mathcal{A}_{1,1,5}^i=\frac{1}{3!}C_{D^2}\text{Tr}(f_ig_lf_jf_k)\int dt\oint dw \left(\sum_{i=1}^4 \frac{\partial F_i}{\partial t} \right) \mathcal{A}_{1,1}^{bc} (z_1,z_4,z_2,z_3) (\mathcal{A}_{1,1}^m)^{iljk} (z_1,z_4,z_2,z_3)\,,
\end{equation}
\begin{equation}
\mathcal{A}_{1,1,6}^i=\frac{1}{3!}C_{D^2}\text{Tr}(f_ig_lf_kf_j)\int dt\oint dw \left(\sum_{i=1}^4 \frac{\partial F_i}{\partial t} \right)\mathcal{A}_{1,1}^{bc}(z_1,z_4,z_3,z_2)(\mathcal{A}_{1,1}^m)^{ilkj}(z_1,z_4,z_3,z_2)\,.
\end{equation}
We can collect $\mathcal{A}_{1,2n-1}$  into a group, and similarly collect $\mathcal{A}_{1,2n}$ into one group.

To illustrate how to compute $\mathcal{A}_{1,1,n},$ we shall evaluate $\mathcal{A}_{1,1,1}$ in detail and quote the results for the other contributions. We shall first fix $z_1=0,~z_2=1,~z_3=\infty,$ and let $t=-z_4.$ The moduli integral measure is therefore given as
\begin{equation}
\int_{\mathcal{M}_1}dt= -\int_{-\mu^2+1/2}^{-\mu^{-2}-\mu^{-4}/2} dz_4\,.
\end{equation}
We therefore find
\begin{align}
\mathcal{A}_{1,1,1}= &\frac{1}{3!} C_{D^2} \text{Tr}(f_if_jf_kg_l) \int_{-\mu^2+1/2}^{-\mu^{-2}-\mu^{-4}/2} \frac{dz_4}{z_4^2}\eta^{il}\eta^{jk}\,,\\
=&\frac{1}{3!} C_{D^2} \text{Tr} (f^i f_j f^j g_i) (\mu^2-1/2+\dots)\,.
\end{align}

\begin{table}
\centering
\begin{tabular}{|c|c|c|c|c|}
\hline
$p_1$&$p_2$&$W_1$&$W_2$&$\Omega$\\
\hline
$-\mu^2+1/2$&0&$ -p_\pm \mu^2 +1/2+(-2+p_\pm)\mu^{-2}/16$& $p_\pm-1/2\mu^2\mp i\sqrt{3}/8\mu^4$&$-$\\
$-\mu^{-2}-\mu^{-4}/2$&0&$-p_\mp \mu^{-2}+p_\pm\mu^{-4}/2$&$p_\pm-p_\mp/2\mu^2-p_\mp/8\mu^4$&$+$\\
$\mu^{-2}-\mu^{-4}/2$&0&$p_\pm\mu^{-2}+p_\mp\mu^{-4}/2$&$p_\pm+p_\mp/2\mu^2-p_\mp/8\mu^4$&$-$\\
$1-\mu^{-2}+\mu^{-4}/2$&0&$1-p_\pm \mu^{-2}-p_\mp\mu^{-4}/2$&$p_\pm-p_\pm/2\mu^2+p_\pm/8\mu^4$&$+$\\
$1+\mu^{-2}+\mu^{-4}/2$&0&$1+p_\mp \mu^{-2}-p_\pm\mu^{-4}/2$&$p_\pm+p_\pm/2\mu^2+p_\pm/8\mu^4$&$-$\\
$\mu^2+1/2$&0&$p_\mp\mu^2+1/2+(-2+p_\pm)/16\mu^{-2}$&$p_\pm+1/2\mu^2\pm i\sqrt{3}/8\mu^4$&$+$\\\hline
\end{tabular}
\caption{Initial and final conditions for the PCO locations for the vertical integrations for the disk amplitude with O-O-O-O. $\Omega$ is the overall sign one needs to multiply to account for the orientation of the boundary.}\label{tab:bound vert1}
\end{table}

Let us now determine the vertical integration. There are two disconnected boundaries of the moduli space. On the left-hand side, we have $t=-\mu^2+1/2,$ with $p_1=-\mu^2+1/2$ and $p_2=0.$ On the right-hand side of the boundaries, we have $t=-\mu^{-2}-\mu^{-4}/2$ with $p_1=-\mu^{-2}-\mu^{-4}/2$ and $p_2=0.$ On the first boundary of the moduli space, we shall move the PCO at $p_1=t$ to $W_1=-p_\pm\mu^2+1/2+(-2+p_\pm)\mu^{-2}/16$ and the PCO at $p_2=0$ to $W_2=p_\pm.$ On the second boundary of the moduli space, we shall move the PCO at $p_1=-\mu^{-2}-\mu^{-4}/2$ to $W_1=-p_\pm \mu^{-2}+p_{\mp}\mu^{-4}/2,$ and the PCO at $0$ to $W_2= p_\pm.$ For the boundary conditions for the rest of the boundaries of the moduli space, see Table \ref{tab:bound vert1}. The vertical integration is then determined as
\begin{align}
\mathcal{B}_{1,1,1}:=& -\frac{\Omega}{3!} \text{Tr}(f_if_jf_kg_l) \biggr\langle ((\xi(p_1)-\xi(W_1))\mathcal{X}(p_2)+\mathcal{X}(W_1)(\xi(p_2)-\xi(W_2)))\otimes  ce^{-\phi}\tilde{\psi}^l(-\mu^{-2}) \nonumber\\&\otimes  ce^{-\phi}\tilde{\psi}^i(0)\otimes  ce^{-\phi}\tilde{\psi}^j( 1)\otimes ce^{-\phi}\tilde{\psi}^k(\infty)  \biggr\rangle\,,
\end{align}
where $\Omega$ is introduced to keep track of the orientation of the boundaries. The boundary on the left shall be evaluated with $\Omega=-1,$ and the boundary on the right shall be evaluated with $\Omega=1.$ In order to saturate the background $\phi$ charge and the $c$-ghost charge, we must contract $-\partial\eta be^{2\phi}-\partial(\eta be^{2\phi})$ from $\mathcal{X}.$ We write the gauge unfixed version of the vertical integration
\begin{align}
\mathcal{B}(p,q,W,z_1,z_2,z_3,z_4)=&-\frac{\Omega}{3!} \text{Tr}(f_if_jf_kg_l) \left[\frac{1}{(p-q)^2} -\frac{1}{(W-q)^2}+ \partial_q\left( \left(\frac{1}{p-q}-\frac{1}{W-q} \right)\_ \right)  \right]\nonumber\\
& \times\mathcal{B}^{bc}(q,z_1,z_2,z_3,z_4) \mathcal{B}^\phi(q,z_1,z_2,z_3,z_4)\mathcal{B}^{\psi}(z_1,z_2,z_3,z_4)^{lijk}\frac{\partial t}{\partial z_i}\,,
\end{align}
where $z_i$ is the unfixed vertex, and
\begin{equation}
\mathcal{B}^{bc}(q,z_1,z_2,z_3,z_4)=\left(\frac{1}{q-z_1} z_{23}z_{34}z_{24}-\frac{1}{q-z_2}z_{13}z_{14}z_{34}+\frac{1}{q-z_3}z_{12}z_{14}z_{24}-\frac{1}{q-z_4}z_{12}z_{13}z_{23}\right)\,,
\end{equation}
\begin{equation}
\mathcal{B}^{\phi}(q,z_1,z_2,z_3,z_4)=(q-z_1)^2(q-z_2)^2(q-z_3)^2(q-z_4)^2z_{12}^{-1}z_{13}^{-1}z_{14}^{-1}z_{23}^{-1}z_{24}^{-1}z_{34}^{-1}\,,
\end{equation}
\begin{equation}
(\mathcal{B}^\psi)^{ijkl}(q,z_1,z_2,z_3,z_4)=\frac{\eta^{il}\eta^{jk}}{z_{12}z_{34}} -\frac{\eta^{ik}\eta^{jl}}{z_{13}z_{24}}+\frac{\eta^{lk}\eta^{ij}}{z_{14}z_{23}}\,.
\end{equation}
Then, we find
\begin{align}
\mathcal{B}_{1,1,1}=& \mathcal{B}_{1,1} (p_1,p_2,W_1,0,1,\infty,-\mu^{-2}) +\mathcal{B}_{1,1}(p_2,W_1,W_2,0,1,\infty,-\mu^{-2})\,,\\
=&\frac{1}{3!}C_{D^2} \text{Tr}(f_i f_j f_k g_l ) \left( -\mu^2 \eta^{il}\eta^{jk} +(\eta^{ik}\eta^{jl}-\eta^{ij}\eta^{lk}/2)+\dots\right)\,.
\end{align}
As a result, for the ordering $z_1<z_2<z_3<z_4,$ we find
\begin{equation}
\mathcal{A}_{1,1,1}=\frac{1}{3!}C_{D^2} \text{Tr}(f_i f_j f_k g_l ) \left( -\eta^{il}\eta^{jk}/2 +\eta^{ik}\eta^{jl}-\eta^{ij}\eta^{lk}/2\right)\,,
\end{equation}
in the large stub limit. 

For the ordering $z_1<z_4<z_2<z_3,$ we find
\begin{equation}
\mathcal{A}_{1,1,5}^{i}=\frac{1}{3!}C_{D^2}\text{Tr}(f_i g_l f_jf_k) \eta^{il}\eta^{jk}(\mu^2-1/2+\dots)\,,
\end{equation}
\begin{equation}
\mathcal{B}_{1,1,5}=\frac{1}{3!} C_{D^2}\text{Tr}(f_ig_l f_jf_k) \left( -\mu^2\eta^{il}\eta^{jk}+(\eta^{ij}\eta^{lk}-\eta^{ik}\eta^{jl}/2) \right)\,,
\end{equation}
and
\begin{equation}
\mathcal{A}_{1,1,5}=\frac{1}{3!} C_{D^2}\text{Tr}(f_ig_l f_jf_k) \left( -\eta^{il}\eta^{jk}/2+\eta^{ij}\eta^{lk}-\eta^{ik}\eta^{jl}/2 \right)\,.
\end{equation}

For the ordering $z_1<z_2<z_4<z_3,$ we find
\begin{equation}
\mathcal{A}_{1,1,3}^{i}=\frac{1}{3!}C_{D^2}\text{Tr}(f_i  f_j g_lf_k) \eta^{il}\eta^{jk}(1+\dots)\,,
\end{equation}
\begin{equation}
\mathcal{B}_{1,1,3}=\frac{1}{3!} C_{D^2}\text{Tr}(f_if_jg_lf_k) (-\eta^{ik}\eta^{jl}/2-\eta^{ij}\eta^{lk}/2)+\dots\,,
\end{equation}
and
\begin{equation}
\mathcal{A}_{1,1,3}=\frac{1}{3!} C_{D^2}\text{Tr}(f_if_jg_lf_k) (\eta^{il}\eta^{jk}-\eta^{ik}\eta^{jl}/2-\eta^{ij}\eta^{lk}/2)+\dots\,.
\end{equation}

By collecting terms obtained by exchanging $z_2$ with $z_3$ as well, we find
\begin{equation}
\mathcal{A}_{1,1}= \frac{1}{2}C_{D^2} \text{Tr}\left( [ f_i,f_j] [f^i,g^j]\right)\,.
\end{equation}
Note that the above expression was expected from the structure of the DBI action of the low-energy supergravity. For example, $\mathcal{A}_{1,1}$ should compute a term in the effective potential had we replaced $g$ with $f$.

Next, we shall now compute the overlap between $V_{t,2}^o$ and $\mathcal{S}_{1,1}.$ Because there is no way to have a non-trivial anti-commutator involving two $f_i$ and $h$ that respects the spacetime covariance, one can expect that the overlap between $V_{t,2}^o$ and $\mathcal{S}_{1,1}$ shall vanish. We will show this explicitly now.

In order to evaluate
\begin{equation}
\frac{1}{3!}\{ (\tilde{R}_{S^3}^{-1}\Bbb{P}\Psi_{1,0}^o)^3\otimes V_{t,2}^o\}\,,
\end{equation}
same as before, we need to insert two PCOs. To properly saturate the $c$-ghost charge and the background $\phi$ charge, one $e^\phi T_F$ shall be contracted from one PCO, and $-\partial \eta be^{2\phi}-\partial(\eta be^{2\phi})$ from the other PCO. Because $T_F$ contains a worldsheet boson $\partial X$ that cannot be contracted, we conclude that the above amplitude vanishes. This also implies that the vertical integration vanishes, as the vertical integration cannot saturate the background $\phi$ charge.

Therefore, we conclude
\begin{equation}
\mathcal{S}_{1,1}=  -\frac{1}{2}C_{D^2}  [ [f^j,f_i],f^i] c\partial c e^{-\phi}\tilde{\psi}^j\,.
\end{equation}

\subsection{Computation of $\mathcal{S}_{1,2}$} 
To compute $\mathcal{S}_{1,2},$ let us evaluate
\begin{equation}
\mathcal{A}_{2,i}:=\langle V_{t,i}^o|\mathcal{S}_{1,2}\rangle= -\tilde{R}_{S^3}^{-3} \left\{V_{t,2}^o\otimes \frac{1}{2}\Psi_{1,0}^c\otimes (\Bbb{P}\Psi_{1,0}^o)^2\right\}\,.
\end{equation}
$\mathcal{A}_{2,i}$ gets contributions from
\begin{equation}
\mathcal{A}_{2,i}^{B}=-\tilde{R}_{S^3}^{-2}\left\{ V_{t,i}^o\otimes \tilde{V}_{NSNS}\otimes \frac{1}{2}(\Bbb{P}\Psi_{1,0}^o)^2\right\}\,,
\end{equation}
and
\begin{equation}
\mathcal{A}_{2,i}^{F}=-\tilde{R}_{S^3}^{-2}\left\{ V_{t,i}^o\otimes \tilde{V}_{RR}\otimes \frac{1}{2}(\Bbb{P}\Psi_{1,0}^o)^2\right\}\,.
\end{equation}
Because the anti-D3-brane stack preserves the opposite of the spacetime supersymmetry, $\mathcal{A}_{2,i}^B=\mathcal{A}_{2,i}^F.$ On the other hand, for a stack of D3-brane that preserves the same supersymmetry as the closed string background, we have $\mathcal{A}_{2,i}^B=-\mathcal{A}_{2,i}^F.$ 

To evaluate $\mathcal{A}_{2,i}^B,$ we need to insert 3 PCOs, and to evaluate $\mathcal{A}_{2,i}^F,$ we need to insert 2 PCOs. To compute $\mathcal{A}_{2,1}^B,$ in the interior of the moduli space, we shall choose to insert the PCOs at the open string punctures. Similarly, to compute $\mathcal{A}_{2,1}^F,$ we shall insert the PCOs at the movable open string punctures. As we will show momentarily, this choice of PCOs makes the contribution from the interior of the moduli space completely vanishes, and the vertical integration is solely responsible for the non-trivial contribution.  

Let us first comment on $\mathcal{A}_{2,2}^B$ and $\mathcal{A}_{2,2}^F.$ Because of the ghost structure of the test field $V_{t,2}^o$
\begin{equation}
    V_{t,2}^o= c\partial ce^{-2\phi} \partial\xi\,,
\end{equation}
to evaluate $\mathcal{A}_{2,2}^B,$ two $e^\phi T_F$ and one $-\partial \eta be^{2\phi}-\partial(\eta b e^{2\phi})$ from the PCOs must be contracted, and to evaluate $\mathcal{A}_{2,2}^F,$ one $e^\phi T_F$ and one$-\partial \eta be^{2\phi}-\partial(\eta be^{2\phi})$ must be contracted. Note that for $\mathcal{A}_{2,2}^B,$ one can also attempt to contract two $-\partial \eta be^{2\phi}-\partial(\eta b e^{2\phi})$ and one $c\partial\xi$ from the PCOs, however, this will vanish because the B-field vanishes on the anti-D3-brane. However, we find that there is exactly one more worldsheet boson than we need, and they cannot be contracted without making the correlator vanish. Therefore, we conclude $\mathcal{A}_{2,2}^B=\mathcal{A}_{2,2}^F=0.$

By using the conformal killing group, we shall fix the location of the closed string puncture in the upper half plane at $i$ and one open string puncture at $0.$ Let us denote the fundamental vertex region for a disk with C-O-O-O by $\mathcal{V}_{1,3}.$ Then, we can write, for the contributions from the interior of the moduli space,
\begin{align}
\mathcal{A}_{2,1}^{B,i}&=-\frac{i}{2!} \text{Tr}(f_i f_j g_k) \int_{\mathcal{V}_{1,3}} d t_1\wedge dt_2 \biggr\langle\oint dw_1 \mathfrak{B}_1\oint dw_2\mathfrak{B}_2\otimes\mathcal{X}(p_1)\otimes\mathcal{X}(p_2)\otimes\mathcal{X}(p_3)\nonumber\\
&\otimes  \tilde{V}_{NSNS}(z_0)\otimes c e^{-\phi}\tilde{\psi}^i(z_1)\otimes ce^{-\phi}\tilde{\psi}^j(z_2)\otimes ce^{-\phi}\tilde{\psi}^k (z_3) \biggr\rangle\,,
\end{align}
\begin{align}
\mathcal{A}_{2,1}^{F,i}=&-\frac{i}{2!} \text{Tr}(f_i f_j g_k) \int_{\mathcal{V}_{1,3}} d t_1\wedge dt_2 \biggr\langle\oint dw_1 \mathfrak{B}_1\oint dw_2\mathfrak{B}_2\otimes\mathcal{X}(p_1)\otimes\mathcal{X}(p_2)\nonumber\\
&\otimes  \tilde{V}_{RR}(z_0)\otimes c e^{-\phi}\tilde{\psi}^i(z_1)\otimes ce^{-\phi}\tilde{\psi}^j(z_2)\otimes ce^{-\phi}\tilde{\psi}^k (z_3) \biggr\rangle\,,
\end{align}
where
\begin{equation}
\mathfrak{B}_j:=\left( \sum_{i} \frac{\partial F_i}{\partial t_j} b(w_j)-\sum_i \frac{1}{\mathcal{X}(p_i)} \frac{\partial p_i}{\partial t_j}\partial\xi(p_i)\right)\,.
\end{equation}
Note that we assumed the ordering $z_1<z_2<z_3.$ We shall need to sum over the other ordering as well to obtain the full answer. Also, we included an additional factor of $i$ found in \cite{Alexandrov:2021shf}.

Let us study $\mathcal{A}_{2,1}^{B,i}.$ As we previously mentioned, we shall fix $z_2=0,$ $z_0=i,$ and treat $z_1$ and $z_3$ as moduli. Furthermore, we shall place PCOs at $p_1=z_1,~p_2=z_2,$ and $p_3=z_3.$ The resulting amplitude is then given as
\begin{align}
\mathcal{A}_{2,1}^{B,i}=-\frac{i}{2!}\text{Tr}(f_if_jg_k) \int_{\mathcal{V}_{1,3}}dz_1\wedge dz_3 \biggr\langle &\tilde{V}_{NSNS}(i) \otimes  (-i\sqrt{2} \partial \tilde{X}^i)(z_1)\nonumber\\
&\otimes(-i\sqrt{2} c\partial\tilde{X}^j-\eta e^\phi\tilde{\psi}^j)(0)\otimes (-i\sqrt{2}\partial\tilde{X}^k)(z_3)\biggr\rangle\,. 
\end{align}
Because there is no $\xi$ ghost insertion, only the first term of the open string vertex at $0$ can contribute. However, because the only worldsheet fermions present in the correlator are that of $\tilde{V}_{NSNS}$ and $H_{abc}\eta^{bc}=0,$ we find that $\mathcal{A}_{2,1}^{B,i}=0.$

Let us now evaluate $\mathcal{A}_{2,1}^{F,i}.$ We shall fix $z_2=0,$ $z_0=i,$ $p_1=z_2,$ and $p_2=z_3.$ Then we have
\begin{align}
\mathcal{A}_{2,1}^{F,i}=-\frac{i}{2!} \text{Tr}(f_if_jg_k) \int_{\mathcal{V}_{1,3}} dz_1\wedge dz_3\biggr\langle &\tilde{V}_{RR}(i) \otimes e^{-\phi}\tilde{\psi}^i(z_1)\nonumber\\
&\otimes(-i\sqrt{2} c\partial\tilde{X}^j-\eta e^\phi\tilde{\psi}^j)(0)\otimes (-i\sqrt{2}\partial\tilde{X}^k)(z_3)\biggr\rangle\,,
\end{align}
We can now use the doubling trick
\begin{equation}
e^{-\bar{\phi}/2}\overline{\Sigma}_\beta(i)= - e^{-\phi/2} (\Gamma_6)_{\beta}^{~\gamma} \Sigma_\gamma(-i)\,,
\end{equation}
and an identity
\begin{equation}
e^{-\phi}\tilde{\psi}^i(x_1) e^{-\phi/2}\Sigma_\alpha (x_2)e^{-\phi/2} \Sigma_\gamma(x_3)=-\frac{(\Gamma^i)_{\alpha\gamma}}{\sqrt{2}x_{12}x_{13}x_{23}}\,,
\end{equation}
to show that the matter CFT correlator is proportional to
\begin{equation}
f_i F_{abc}\text{Tr}\left( \Gamma^{abc}\Gamma^i \Gamma_6\right)\,.
\end{equation}
As $\text{Tr}(\Gamma^{abc} \Gamma^i\Gamma_6)=0,$ we again conclude that $\mathcal{A}_{2,1}^{F,i}=0.$

We write vertical integrations
\begin{align}
\mathcal{B}_{2,1}^B=&-\frac{i}{2}\text{Tr}(f_if_jg_k) \int_{\partial\mathcal{V}_{1,3}} Jac(t_{\|}) dt_{\|} \oint dw\biggr\langle\mathfrak{B}\otimes  \tilde{V}_{NSNS}(i)\otimes c e^{-\phi}\tilde{\psi}^i(z_1)\otimes ce^{-\phi}\tilde{\psi}^j(z_2)\nonumber\\
&\otimes  ce^{-\phi}\tilde{\psi}^k(z_3) \otimes \biggr[ (\xi(p_1)-\xi(W_1))\mathcal{X}(p_2)\mathcal{X}(p_3)+\mathcal{X}(W_1)(\xi(p_2)-\xi(W_2))\mathcal{X}(p_3)\nonumber\\&+\mathcal{X}(W_1)\mathcal{X}(W_2)(\xi(p_3)-\xi(W_3))\biggr]\biggr\rangle+\dots\,,
\end{align}
\begin{align}
\mathcal{B}_{2,1}^F=&-\frac{i}{2}\text{Tr}(f_if_jg_k) \int_{\partial\mathcal{V}_{1,3}} Jac(t_{\|}) dt_{\|} \oint dw\biggr\langle\mathfrak{B}\otimes  \tilde{V}_{RR}(i)\otimes c e^{-\phi}\tilde{\psi}^i(z_1)\otimes ce^{-\phi}\tilde{\psi}^j(z_2)\nonumber\\
&\otimes  ce^{-\phi}\tilde{\psi}^k(z_3) \otimes \biggr[ (\xi(p_1)-\xi(W_1))\mathcal{X}(p_2)+\mathcal{X}(W_1)(\xi(p_2)-\xi(W_2))\biggr]\biggr\rangle+\dots\,,
\end{align}
where $\dots$ denotes the terms that take into account the moduli dependence of the PCO jump. As $\dots$ does not contribute to the answer, we omitted them. Because $\mathcal{B}_{2,1}^F=\mathcal{B}_{2,1}^B,$ due to the closed-string spacetime supersymmetry, we can just compute one of them. We choose to explicitly compute $\mathcal{B}_{2,1}^F,$ as this amplitude involves less PCO jumps.

We use the following identity 
\begin{equation}
\langle \Sigma_\alpha(z_1)  \Sigma_\beta(z_2) \tilde{\psi}^i(z_3) \tilde{\psi}^j(z_4) \tilde{\psi}^k(z_5)\rangle\supset\frac{1}{2\sqrt{2}} (\Gamma^{ijk})_{\alpha\beta} z_{12}^{1/4}z_{13}^{-1/2}z_{14}^{-1/2}z_{15}^{-1/2} z_{23}^{-1/2}z_{24}^{-1/2}z_{25}^{-1/2}\,,
\end{equation}
and
\begin{equation}
e^{-\bar{\phi}/2}\overline{\Sigma}_\beta(\bar{z})= \Gamma_6 e^{-\phi/2}\Sigma_\beta (z)\,,
\end{equation}
to evaluate $\mathfrak{B}_{2,1}^F.$ Let us define 
\begin{align}
\mathfrak{B}_{2,1,0}^{F}&(p,q,W)=-\frac{g_s}{32\pi}\frac{1}{3!}  \text{Tr}(f_if_jg_k)\tilde{F}_{abc} \int_{\partial\mathcal{V}_{1,3}}Jac(t_{\|})dt_{\|} \oint dw \left(\sum_i \frac{\partial F_i}{\partial t_{\|}}\right)\nonumber\\
& \times\left[ \left(\frac{1}{(p-q)^2}-\frac{1}{(W-q)^2}\right) +\partial_q \left(\frac{1}{p-q}-\frac{1}{W-q} \right)\right] \mathfrak{C}_{bc}\mathfrak{C}_{ferm}^{ijk,abc}\mathfrak{C}_\phi\,,
\end{align}
where
\begin{equation}
\mathfrak{C}_{ferm}^{ijk,abc}= \frac{1}{2\sqrt{2}}\text{Tr}\left( \Gamma^{abc}\Gamma^{ijk}\Gamma_6\right) z_{12}^{1/4}z_{13}^{-1/2}z_{14}^{-1/2}z_{15}^{-1/2} z_{23}^{-1/2}z_{24}^{-1/2}z_{25}^{-1/2}z_{34}z_{35}z_{45}\,,
\end{equation}
\begin{equation}
\mathfrak{C}_\phi= z_{12}^{-1/4}z_{13}^{-1/2}z_{14}^{-1/2}z_{15}^{-1/2}z_{1q}z_{23}^{-1/2}z_{24}^{-1/2}z_{25}^{-1/2}z_{2q}z_{34}^{-1}z_{35}^{-1}z_{3q}^2z_{45}^{-1}z_{4q}^2z_{5q}^2\,,
\end{equation}
\begin{align}
\mathfrak{C}_{bc}=&\frac{C_{D^2}}{w-z_4} \biggr( \frac{1}{q-z_1} z_{23}z_{25}z_{35} -\frac{1}{q-z_2}z_{13}z_{15}z_{35} +\frac{1}{q-z_3}z_{12}z_{15}z_{25}-\frac{1}{q-z_5}z_{12}z_{13}z_{23}\biggr)\nonumber\\
&- (z_4\leftrightarrow z_5)\,.
\end{align}
Note that we denoted $z_2:= \bar{z}_1.$ Also, as $z_4$ and $z_5$ are the only movable vertices, we neglected the contraction between $b(w)$ and $c(z_i)$ for $i\neq 4,~5.$ Then, we compute
\begin{equation}
\mathcal{B}_{2,1}^F= \mathfrak{B}_{2,1,0}^F(p_1,p_2,W_1)+\mathfrak{B}_{2,1,0}^F(p_2,W_1,W_2)\,.
\end{equation}

To evaluate the vertical integration $\mathcal{B}_{2,1}^F,$, we need to choose the boundary regions of the moduli space and the PCO locations consistently. The boundary region of the moduli space is summarized in \S\ref{sec:loc cord COOO}. We did not fully specify the choice of the PCOs on the boundary regions in \S\ref{sec:loc cord COOO}, so we shall do so here. We need to choose the PCO configuration on the boundary so that PCO locations continuously vary without a gap. In principle, PCO locations must be chosen such that PCOs are located a finite distance away from the punctures. This is needed to avoid singularities. However, for the vertex operators we chose, there is no singularity as one PCO approaches an open string vertex. This was also manifest from the computation of the amplitude in the interior region of the moduli space. Hence, we can make a simpler choice of the PCO, allowing a collision of a PCO with an open string puncture. Also, note that the vertical integration due to the degeneration channel into (C-O)-(O-O-O-O) vanishes because, in the large stub limit, the vertical integration effectively moves PCOs within the (O-O-O-O) diagram. Since the states involved are on-shell primaries, moving PCOs within an on-shell amplitude does not alter the result. 

There are, in total, six boundaries we shall consider for the Feynman diagrams made of gluing a disk with C-O-O to a disk diagram with O-O-O. We shall parameterize the coordinates of the open string punctures and fix the locations of the PCOs on the boundaries. We shall choose the endpoints of the PCO jumps such that they agree with the PCO locations of the boundary of the Feynman regions due to (C-O)-(O-O-O)-(O-O-O) degenerations.

\begin{enumerate}[(a)]
    \item $z_3<z_1<0$ and $z_1\simeq 0.$

    We have
    \begin{equation}
        z_3=\frac{2t}{t^2-1}-\frac{(t^2+1)^2}{2\mu^4(t^2-1)^2}+\dots\,,\quad z_1= -\mu^{-4}+\dots\,,
    \end{equation}
    \begin{equation}
        p_1=0\,,\quad p_2=z_3\,,
    \end{equation}
    \begin{equation}
        W_1=-p_\pm \mu^{-4}+\dots\,,\quad W_2=-\frac{(\pm3i+\sqrt{3})t}{\mp3i+\sqrt{3}t^2}\mp\frac{3i\sqrt{3}(1+t^2)}{(\mp3i+\sqrt{3}t^2)^2\mu^2}+\dots\,,
    \end{equation}
    for $(2\mu^2)^{-1}\leq t\leq 1.$
    
    \item $z_3<z_1<0$ and $z_1\simeq z_3.$

    We have
    \begin{equation}
        z_3=\frac{2t}{t^2-1}-\frac{(t^2+1)^2}{2\mu^4(t^2-1)^2}+\dots\,,\quad z_1=\frac{2t}{t^2-1}+\frac{(t^2+1)^2}{2\mu^4(t^2-1)^2}+\dots\,,
    \end{equation}
    \begin{equation}
        p_1=0\,,\quad p_2=z_3\,,
    \end{equation}
    \begin{equation}
        W_1=-\frac{(\mp3i+\sqrt{3})t^{-1}}{\pm3i+\sqrt{3}t^{-2}}\mp \frac{3i\sqrt{3}(1+t^{-2})}{(\pm3i+\sqrt{3}t^{-2})^2\mu^2}+\dots\,,
    \end{equation}
    \begin{equation}
        W_2=\frac{2t}{t^2-1}\pm\frac{i\sqrt{3}(1+t^2)^2}{2(-1+t^2)^2\mu^4}+\dots\,,
    \end{equation}
    for $(2\mu^2)^{-1}\leq t\leq1.$
    
    \item $z_1<0<z_3$ and $z_1\simeq0.$

    We have
    \begin{equation}
        z_3=-\frac{2t}{t^2-1}-\frac{(1+t^2)^2}{2\mu^4(t^2-1)^2}+\dots\,,\quad z_1=-\mu^{-4}+\dots\,,
    \end{equation}
    \begin{equation}
        p_1=0\,,\quad p_2=z_3\,,
    \end{equation}
    \begin{equation}
        W_1=-p_\pm\mu^{-4}+\dots\,,\quad  W_2=\frac{(\pm3i+\sqrt{3})t}{\mp3i+\sqrt{3}t^2}\mp\frac{3i\sqrt{3}(1+t^2)}{(\mp3i+\sqrt{3}t^2)^2\mu^2}+\dots\,,
    \end{equation}
    for $(2\mu^2)^{-1}\leq t\leq1.$
    
    \item $z_1<0<z_3$ and $z_3\simeq0.$

    We have 
    \begin{equation}
        z_3=\mu^{-4}+\dots\,,\quad z_1=\frac{2t}{t^2-1}+\frac{(t^2+1)^2}{2\mu^4(t^2-1)^2}+\dots\,,
    \end{equation}
    \begin{equation}
        p_1=0\,,\quad p_2=z_3\,,
    \end{equation}
    \begin{equation}
        W_1=\frac{(\pm3i+\sqrt{3})t}{\mp3i+\sqrt{3}t^2}\mp\frac{3i\sqrt{3}(1+t^2)}{(\mp3i+\sqrt{3}t^2)^2\mu^2}+\dots\,,\quad W_2=p_\pm\mu^{-4}+\dots\,,
    \end{equation}
    for $(2\mu^2)^{-1}\leq t\leq1.$
    \item $0<z_3<z_1$ and $z_3\simeq0.$

    We have
    \begin{equation}
        z_1=-\frac{2t}{t^2-1}+\frac{(t^2+1)^2}{2\mu^4(t^2-1)^2}+\dots\,,\quad z_3=\mu^{-4}+\dots\,,
    \end{equation}
    \begin{equation}
        p_1=0\,,\quad p_2=z_3\,,
    \end{equation}
    \begin{equation}
        W_1=p_\pm\mu^{-4}+\dots\,,\quad W_2=\frac{(\pm3i+\sqrt{3})t}{\mp3i+\sqrt{3}t^2}\mp\frac{3i\sqrt{3}(1+t^2)}{(\mp3i+\sqrt{3}t^2)^2\mu^2}+\dots\,,
    \end{equation}
    for $(2\mu^2)^{-1}\leq t\leq1.$
    
    \item $0<z_3<z_1$ and $z_1\simeq z_3.$
    
    We have
    \begin{equation}
        z_1=-\frac{2t}{t^2-1}+\frac{(t^2+1)^2}{2\mu^4(t^2-1)^2}+\dots\,,\quad z_3=-\frac{2t}{t^2-1}-\frac{(t^2+1)^2}{2\mu^4(t^2-1)^2}+\dots\,,
    \end{equation}
    \begin{equation}
        p_1=0\,,\quad p_2=z_3\,,
    \end{equation}
    \begin{equation}
        W_1=-\frac{(\mp3i+\sqrt{3})t}{\pm3i+\sqrt{3}t^2}\mp \frac{3i\sqrt{3}(1+t^2)}{(\pm3i+\sqrt{3}t^2)^2\mu^2}+\dots\,,
    \end{equation}
    \begin{equation}
        W_2=-\frac{2t}{t^2-1}\pm\frac{i\sqrt{3}(1+t^2)^2}{2(-1+t^2)^2\mu^4}+\dots\,,
    \end{equation}
    for $(2\mu^2)^{-1}\leq t\leq1.$

\end{enumerate}

We found that the vertical integrations due to (a) and (c) vanish. Because we are taking the large stub limit, and all the vertex operators involved in the computations are on-shell, moving PCOs within a single disk diagram of the degeneration does not alter the result. In case (a), the disk diagram with C-O-O contains the closed string puncture with an open string puncture at $z_3,$ and the disk diagram with O-O-O contains the open string puncture at $0.$ As the PCOs are correctly distributed, one can conclude that the vertical integration due to (a) vanishes. Similarly, we find that PCOs are correctly distributed for (c), although the precise locations of the initial choice of the PCOs are a little different from the boundary conditions. 

On the other hand, individual components due to (b) and (f) are non-trivial. However sum of (b) and (f) vanish in the large stub limit. This can be understood as follows. In fact, the moduli spaces of (b) and (f) can be combined into one connected piece, as the cyclic ordering for (b) and (f) are equivalent. What this implies is that (b) and (f) individually may not give unambiguous answers, but they do only when they are combined. Since, for (b) and (f), the PCOs are again correctly distributed, we expect the vertical integration to vanish in the large stub limit as well. It is interesting to note that even (a) and (c) are, in fact, connected; in those cases, individual answers seem to vanish. 

The vertical integrations due to (d) and (e) are, however, non-trivial. This is because the degeneration limit forces wrong PCO distributions. For example, in the case of (d), we find that the disk diagram with C-O-O has no PCO, whereas the disk diagram with O-O-O has two PCO insertions. Therefore, we expect the vertical integration to give a non-trivial contribution. After taking into account the relative orientation of these two distinct contributions, we find
\begin{equation}
    \mathfrak{B}_{2,1}^F=i\frac{g_s}{32\sqrt{2}} \text{Tr}(f_if_jg_k)\tilde{F}_{abc}\text{Tr}(\Gamma^{abc}\Gamma^{ijk}\Gamma_6) \,.
\end{equation}
By using the identity, 
\begin{equation}
\text{Tr}(\Gamma^{abc}\Gamma^{ijk}\Gamma_6)=-16\epsilon^{abcijk}\,,
\end{equation}
we reduce $\mathfrak{B}_{2,1}^F$ to
\begin{equation}
\mathfrak{B}_{2,1}^F=-\frac{ig_s}{2\sqrt{2}}C_{D^2}\frac{1}{3!} \text{Tr}(f_if_jg_k) \tilde{F}_{abc}\epsilon^{abcijk}\,,
\end{equation}
By taking into account the contribution from the B-field, and summing over different orderings, we find
\begin{equation}
\mathcal{S}_{1,2}=-i \frac{\sqrt{2}}{2}g_s C_{D^2}\frac{1}{3!} [f_i,f_j] \tilde{F}_{abc}\epsilon^{abcijk} c\partial ce^{-\phi}\tilde{\psi}_k\,.
\end{equation}

\section{Vanising source terms}\label{app:diag}
We shall now argue that $\mathcal{S}_2,$ $\mathcal{S}_3,$ $\mathcal{S}_4,$ $\mathcal{S}_5,$ and $\mathcal{S}_6$ vanish. Same as before, we shall take the large stub limit. Since all the operators we are inserting are $L_0^+$ nilpotent, and there is no tachyon in the spectrum, in the large stub limit $(1-\Bbb{P})$ projected diagrams all vanish. Therefore, we can only consider the fundamental string vertices. 

\subsection{$\mathcal{S}_2$}
We shall start with $\mathcal{S}_2.$ $\mathcal{S}_2$ contains three distinct contributions
\begin{equation}
\mathcal{S}_{2,1}= -\tilde{R}_{S^3}^{-1}\mathcal{G}\Bbb{P}\left[ \frac{1}{2}\tilde{V}_{NSNS}^2\otimes\Bbb{P}\Psi_{1,0}^o\right]_{D^2}^o\,,\quad \mathcal{S}_{2,2}=-\tilde{R}_{S^3}^{-1}\mathcal{G}\Bbb{P}\left[\tilde{V}_{NSNS}\otimes\tilde{V}_{RR}\otimes\Bbb{P}\Psi_{1,0}^o\right]_{D^2}^o\,,
\end{equation}
\begin{equation}
\mathcal{S}_{2,3}=-\tilde{R}_{S^3}^{-1}\mathcal{G}\Bbb{P}\left[ \frac{1}{2}\tilde{V}_{RR}^2\otimes \otimes\Bbb{P}\Psi_{1,0}^o\right]_{D^2}^o\,.
\end{equation}

Let us study $\mathcal{S}_{2,1}.$ We shall argue that the following amplitude vanishes
\begin{equation}
-\tilde{R}_{S^3}^{-1} \left\{ \frac{1}{2}\tilde{V}_{NSNS}^2\otimes\Bbb{P}\Psi_{1,0}^o\otimes V_{t,1}^o\right\}_{D^2}\,.
\end{equation} 
As the total picture number, before inserting PCOs, is $-6,$ we shall insert a total of 4 PCOs. The non-vanishing contributions come from two different combinations of the PCO contractions. First, $e^\phi T_F$ is contracted from every PCO. Second, $e^\phi T_F$ should be chosen from two PCOs, $c\partial\xi$ from one PCO, and $-\partial\eta be^{2\phi}-\partial(\eta b e^{2\phi})$ from the remaining PCO. The tensor structure of the first contraction type is given as
\begin{equation}
H_{abc} H_{def} f_{[i}v_{j]}\langle \psi^{[a}\psi^b\psi^{c]}\psi^{[d}\psi^e\psi^{f]}  \psi^{[i}\psi^{j]} \psi^k \psi_k \rangle\,.
\end{equation}
Note that we contracted the worldsheet bosons of two PCOs against each other. If $\psi^k$ and $\psi_k$ are contracted against worldsheet fermions coupled to a single $H$ or $f_{[i}v_{j]},$ we get zero. Hence, we need to contract $\psi^k$ and $\psi_k$ against the worldsheet fermions coupled to different spacetime fields. If $\psi^k$ and $\psi_k$ are contracted against worldsheet fermions of two different $H$ fluxes, we get zero again due to the tensor structure. Similarly, the remaining case in which $\psi^k$ and $\psi_k$ are contracted against a worldsheet fermion coupled to $H$ and a worldsheet fermion coupled to $f_{[i}v_{j]}$ also vanishes.

The tensor structure of the second contraction of the matter fields is given as
\begin{equation}\label{eqn:S21 tens}
H_{abc} H_{def} f_{[i} v_{j]} \langle \psi^{[a}\psi^b\psi^{c]}\psi^{[d}\psi^e\psi^{f]}  \psi^{[i}\psi^{j]}  \rangle\,.
\end{equation}
There are, in total, three different classes of contractions. $\psi^i$ is contracted against $\psi^j,$ which vanishes. $\psi^i$ and $\psi^j$ are contracted against $\psi$ coupled to one $H.$ This also vanishes, as inevitably two of fermions coupled to the other $H$ flux should be contracted which vanishes. The remaining choice is to contract $\psi^i$ against $\psi$ coupled to one $H$ flux and contract $\psi^j$ against a fermion coupled to the other $H$ flux. Because replacing the indices $abc$ with $def$ is a symmetry, this contraction also vanishes. Note that the vertical integration has the same index structure. Therefore we conclude
\begin{equation}
-\tilde{R}_{S^3}^{-1} \left\{ \frac{1}{2}\tilde{V}_{NSNS}^2\otimes\Bbb{P}\Psi_{1,0}^o\otimes V_{t,1}^o\right\}_{D^2}=0\,.
\end{equation} 
Now we shall compute
\begin{equation}
-\tilde{R}_{S^3}^{-1}\left\{\frac{1}{2}\tilde{V}_{NSNS}^2\otimes\Bbb{P}\Psi_{1,0}^o\otimes V_{t,2}^o\right\}_{D^2}\,.
\end{equation}
To saturate the background $\phi$ charge and c-ghost number, there are only two possible choices of PCO contractions. First, one $-\partial\eta b e^{2\phi}-\partial(\eta be^{2\phi})$ must be picked from one PCO, and we contract $e^\phi T_F$ from the rest of the PCOs. Second, one $c\partial\xi$ shall be contracted in one of the PCOs, two $-\partial\eta b e^{2\phi}-\partial(\eta be^{2\phi})$ from the PCOs, and $e^\phi T_F$ from the rest. Because the second option cannot provide two derivatives of B-fields, the second option vanishes. The first option involves three bosons $\partial X$ in the correlator. Therefore, we also conclude that this contribution vanishes. Therefore, we conclude
\begin{equation}
\mathcal{S}_{2,1}=0\,.
\end{equation}

Let us compute $\mathcal{S}_{2,2}.$ We shall start by studying
\begin{equation}
-\tilde{R}_{S^3}^{-1}\left\{ \tilde{V}_{NSNS}\otimes\tilde{V}_{RR}\otimes\Bbb{P}\Psi_{1,0}^o\otimes V_{t,1}^o\right\}_{D^2}\,.
\end{equation}
To evaluate the above diagram, we shall insert 3 PCOs. There are two non-trivial contractions of the PCOs. The first one contracts $e^\phi T_F$ from all of the PCOs. Two of the worldsheet bosons in the PCOs must be contracted against each other to have a non-trivial contribution. The resulting tensor structure of the correlation function is then
\begin{equation}\label{eqn:S22 c1}
F_{abc}H_{def} f_{[i}v_{j]} \langle \psi^{[d} \psi^e\psi^{f]} \Sigma_\alpha (\Gamma^{abc}\Gamma_6)^{\alpha\beta}\Sigma_\beta \psi^{[i}\psi^{j]} \psi^k\psi_k \rangle\,.
\end{equation}
Using the ISD condition, we can rewrite the correlator as
\begin{equation}
H_{abc}H_{def} f_{[i}v_{j]} \langle \psi^{[d} \psi^e\psi^{f]} \Sigma_\alpha (\Gamma^{abc})^{\alpha\beta}\Sigma_\beta \psi^{[i}\psi^{j]} \psi^k\psi_k \rangle\,. 
\end{equation}
By contracting spin fields, we will find that the correlation shall take the following form
\begin{equation}
H_{abc} H_{def} f_{[i}v_{j]} \sum_{I,J} A_I \text{Tr}\left( \Gamma^{abc} \Gamma^{I_1\dots I_n}\right) (\eta^{J_1 J_2}\dots \eta^{J_{m-1}J_m}+\text{permutations})\,.
\end{equation}
Because the trace of Gamma matrices yields
\begin{equation}
\text{Tr}(\Gamma^{N_1}\dots \Gamma^{N_n})=\eta^{N_1N_2}\dots \eta^{N_{n-1}N_n}+\text{permutations}\,,
\end{equation}
we find that essentially the tensor structure of the correlation function is that of $\mathcal{S}_{2,1}.$ Therefore we conclude that \eqref{eqn:S22 c1} vanishes. The second option involves one $e^\phi T_F,$ one $c\partial\xi,$ and $-\partial\eta be^{2\phi}-\partial(\eta be^{2\phi}).$ The worldsheet boson $\partial X$ in $T_F$ must be contracted against the $B$ field. Then, the tensor structure of the correlation function is given as
\begin{equation}
F_{abc}H_{def} f_{[i}v_{j]} \langle \psi^{[d} \psi^e\psi^{f]} \Sigma_\alpha (\Gamma^{abc}\Gamma_6)^{\alpha\beta}\Sigma_\beta \psi^{[i}\psi^{j]} \rangle\,.
\end{equation}
One notices that using the ISD condition, we can again rewrite $F_{abc} \Gamma^{abc}\Gamma_6$ as $ H_{abc}\Gamma^{abc}$ with an overall normalization. This again leads to the same tensor structure as that we found in \eqref{eqn:S21 tens}, which vanishes. One can argue that $\langle V_{t,2}^o|\mathcal{S}_{2,2}\rangle$ using the ISD condition. Therefore, we conclude
\begin{equation}
\mathcal{S}_{2,2}=0\,.
\end{equation}

To study $\mathcal{S}_{2,3},$ we can again use the ISD condition for both of the $F_3$ fluxes. This leads to the conclusion
\begin{equation}
\mathcal{S}_{2,3}=0\,,
\end{equation}
and hence we find
\begin{equation}
\mathcal{S}_2=0\,.
\end{equation}

\subsection{$\mathcal{S}_3$}

Now we shall study $\mathcal{S}_3,$ $\mathcal{S}_4,$ $\mathcal{S}_6.$ As we discuss in \S\ref{app:S3+S4+S6}, there is a trick to show 
\begin{equation}
\langle V_{t,2}^o|\mathcal{S}_3+\mathcal{S}_4+\mathcal{S}_6\rangle=0\,.
\end{equation}
The idea is to notice that a linear combination of $V_{t,2}^o$ and $A_\mu ce^{-\phi}\tilde{\psi}^\mu$ is a $Q_B$ exact operator in the extended BRST complex defined in \cite{Belopolsky:1995vi}. As it is straightforward to show that $A_\mu ce^{-\phi}\tilde{\psi}^\mu$ has trivial tadpole, one can then use the main identities to show that $\langle V_{t,2}^o|\mathcal{S}_3+\mathcal{S}_4+\mathcal{S}_6\rangle=0.$ If, the extended complex of open string states can include string fields that contain non-trivial functions of $\tilde{X},$ this idea can be extended to show that even $\langle V_{t,1}^o|\mathcal{S}_3+\mathcal{S}_4+\mathcal{S}_6\rangle=0$ is true. But, at least to the author, it is not clear if such a definition makes sense. Therefore, we shall compute the overlap between $V_{t,1}^o$ and $\mathcal{S}_3,$ $\mathcal{S}_4,$ and $\mathcal{S}_6$ manually. 

Let us study $\langle V_{t,1}^o|\mathcal{S}_3\rangle.$ We find 4 distinct contributions
\begin{equation}
\mathcal{A}_{3,1}:=-\left\{ V_{t,1}^o\otimes \frac{1}{3!} \tilde{V}_{NSNS}^3\right\}_{D^2}\,,\quad \mathcal{A}_{3,2}:=-\left\{ V_{t,1}^o\otimes \frac{1}{2} \tilde{V}_{NSNS}^2\otimes\tilde{V}_{RR}\right\}_{D^2}\,,
\end{equation}
\begin{equation}
\mathcal{A}_{3,3}=-\left\{V_{t,1}^o\otimes\frac{1}{2} \tilde{V}_{NSNS}\otimes\tilde{V}_{RR}^2\right\}_{D^2}\,,\quad \mathcal{A}_{3,4}=-\left\{ V_{t,1}^o\otimes\frac{1}{3!}\tilde{V}_{RR}^3\right\}_{D^2}\,.
\end{equation}

To evaluate $\mathcal{A}_{3,1},$ we need to insert 5 PCOs. The are two ways to get non-trivial contributions. The first contribution is obtained by contracting $e^\phi T_F$ from all of the PCOs. The second contribution is due to contracting $e^\phi T_F$ from 3 PCOs, $c\partial\xi$ from one PCO, and $-\partial\eta be^{2\phi}-\partial(\eta be^{2\phi})$ from the rest. The tensor structure of the matter CFT correlator of the first type of PCO contractions is given as
\begin{equation}
H_{abc} H_{def} H_{ghi} f_j\eta_{kl} \langle \tilde{\psi}^a\tilde{\psi}^b\tilde{\psi}^c\tilde{\psi}^d\tilde{\psi}^e\tilde{\psi}^f\tilde{\psi}^g\tilde{\psi}^h\tilde{\psi}^i\tilde{\psi}^j\tilde{\psi}^k\tilde{\psi}^l\rangle\,.
\end{equation}
The structure of the matter CFT correlator for the second type of PCO contractions is given as
\begin{equation}
H_{abc} H_{def} H_{ghi} f_j \langle \tilde{\psi}^a\tilde{\psi}^b\tilde{\psi}^c\tilde{\psi}^d\tilde{\psi}^e\tilde{\psi}^f\tilde{\psi}^g\tilde{\psi}^h\tilde{\psi}^i\tilde{\psi}^j\rangle\,.
\end{equation}
Because in the first case, $\tilde{\psi}^k$ and $\tilde{\psi}^l$ are coupled to $\eta_{kl},$ the tensor structure of the correlator after contracting all the worldsheet fermions, the tensor structures are identical to that of the second case. Therefore, we shall focus on the second type of PCO contraction. Without loss of generality, we can contract $\tilde{\psi}^a $ against $\tilde{\psi}^j,$ then we are left with the tensor structure of
\begin{equation}
H_{abc} H_{def} H_{ghi} f^a \langle \tilde{\psi}^b\tilde{\psi}^c\tilde{\psi}^d\tilde{\psi}^e\tilde{\psi}^f\tilde{\psi}^g\tilde{\psi}^h\tilde{\psi}^i\rangle\,.
\end{equation}
As the contraction between worldsheet fermions is symmetric in indices, and the indices of $H_{abc}$ are fully anti-symmetric, every combination of the contractions vanishes. As the same tensor structure still persists in the vertical integration, the vertical integration also vanishes. Therefore, we conclude
\begin{equation}
\mathcal{A}_{3,1}=0\,.
\end{equation}

We shall now comment on the rest of the terms $\mathcal{A}_{3,i}$ for $i\neq1.$ Essentially, the reason why $\mathcal{A}_{3,1}$ vanishes is because there is no way to write down a term in the action involving three $H_3$ and a vector that respects the spacetime covariance without introducing derivatives. This implies that $\mathcal{A}_{3,i}$ for $i\neq1$ vanishes as well. The reasoning proceeds as follows. The vertex operator for the RR threeform flux
\begin{equation}
    g_sF_{ijk}c\bar{c} e^{-\phi/2}\Sigma_\alpha (\Gamma^{ijk})^{\alpha\beta}e^{-\bar{\phi}/2} \overline{\Sigma}_\beta\,,
\end{equation}
inserted in a disk diagram can be rewritten, using the doubling trick, as
\begin{equation}
    g_sF_{ijk} c\bar{c} e^{-\phi/2} \Sigma_\alpha (\Gamma^{ijk} \Gamma_6)^{\alpha\beta}e^{-\phi/2}\Sigma_\beta\,.
\end{equation}
Using the ISD condition,
\begin{equation}
    g_s F_{ijk}\Gamma^{ijk}\Gamma_6=-H_{ijk}\Gamma^{ijk}\,,
\end{equation}
the Ramond-Ramond vertex operator is then finally written as
\begin{equation}
    -H_{ijk}c\bar{c}e^{-\phi/2}\Sigma_\alpha(\Gamma^{ijk})^{\alpha\beta} e^{-\phi/2}\Sigma_\beta\,.
\end{equation}
Therefore, the tensor structures of the matter CFT correlator of $\mathcal{A}_{3,i}$ are identical, modulo relative numerical factors. As a result, we find that $\mathcal{A}_{3,i}=0$ for all $i.$

\subsection{$\mathcal{S}_4$}
Let us now study $\langle V_{t,1}^o|\mathcal{S}_4\rangle.$ We find four different contributions
\begin{equation}
\mathcal{A}_{4,1}:=-\tilde{R}_{S^3}^{-2}\left\{ V_{t,1}^o \otimes \tilde{V}_{NSNS} \otimes (\Bbb{P}\Psi_{2,0}^c)^{-1,-1}\right\}_{D^2}\,,\quad \mathcal{A}_{4,2}:=-\tilde{R}_{S^3}^{-2}\left\{ V_{t,1}^o \otimes \tilde{V}_{NSNS} \otimes (\Bbb{P}\Psi_{2,0}^c)^{-\frac{1}{2},-\frac{1}{2}}\right\}_{D^2}\,,
\end{equation}
\begin{equation}
\mathcal{A}_{4,3}:=-\tilde{R}_{S^3}^{-2}\left\{ V_{t,1}^o \otimes \tilde{V}_{RR} \otimes (\Bbb{P}\Psi_{2,0}^c)^{-1,-1}\right\}_{D^2}\,,\quad \mathcal{A}_{4,4}:=-\tilde{R}_{S^3}^{-2}\left\{ V_{t,1}^o \otimes \tilde{V}_{RR} \otimes (\Bbb{P}\Psi_{2,0}^c)^{-\frac{1}{2},-\frac{1}{2}}\right\}_{D^2}\,.
\end{equation}

Let us first study $\mathcal{A}_{4,1}.$ Let us recall the form of $\Bbb{P}(\Psi_{2,0}^c)^{-1,-1}$
\begin{align}
\tilde{R}_{S^3}^{-2}\Bbb{P}\Psi_2^{-1,-1}=&\mathcal{G}_{AB} c\bar{c} e^{-\phi}\tilde{\psi}^Ae^{-\bar{\phi}}\bar{\tilde{\psi}}^B+\mathcal{D} c\bar{c}(\eta \bar{\partial}\bar{\xi}e^{-2\bar{\phi}}-\partial\xi e^{-2\phi}\bar{\eta})\nonumber\\
&+\frac{i}{2\sqrt{2}}\mathcal{F}_A(\partial c+\bar{\partial}\bar{c}) c\bar{c} (e^{-\phi}\tilde{\psi}^Ae^{-2\bar{\phi}}\bar{\partial}\bar{\xi}+e^{-2\phi}\partial\xi e^{-\bar{\phi}}\bar{\tilde{\psi}}^A)\,,
\end{align}
where $\mathcal{G}_{AB},$ and $\mathcal{D},$ and $\mathcal{F}_A$ were determined in \S\ref{sec:KS sec order}
\begin{equation}
    \mathcal{G}_{AB}=-2\pi g_c^2 \left( T_{AB}-\frac{1}{8}\eta_{AB}T_{CD}\eta^{CD}\right) (\tilde{X}^4)^2-\frac{1}{12\pi}R_{AcBd}\tilde{X}^c\tilde{X}^d\,,
\end{equation}
\begin{equation}
    \mathcal{D}=-\frac{1}{2}\mathcal{G}_{AB}\eta^{AB}\,,\quad    \mathcal{F}_A=-\partial^B\mathcal{G}_{AB}\,.
\end{equation}
Therefore, $\mathcal{A}_{4,1}$ contains three distinct contributions
\begin{equation}
    \mathcal{A}_{4,1,1}:=-\left\{ V_{t,1}^o\otimes \tilde{V}_{NSNS} \otimes \mathcal{G}_{AB}c\bar{c} e^{-\phi}\tilde{\psi}^Ae^{-\bar{\phi}}\bar{\tilde{\psi}}^B\right\}_{D^2}\,,
\end{equation}
\begin{equation}
    \mathcal{A}_{4,1,2}:=-\left\{ V_{t,1}^o\otimes \tilde{V}_{NSNS} \otimes \mathcal{D}c\bar{c} (\eta \bar{\partial}\bar{\xi} e^{-2\bar{\phi}}-\partial \xi e^{-\phi}\bar{\eta})\right\}_{D^2}\,,
\end{equation}
\begin{equation}
    \mathcal{A}_{4,1,3}:=-\frac{i}{2\sqrt{2}}\left\{ V_{t,1}^o\otimes \tilde{V}_{NSNS} \otimes \mathcal{F}_A (\partial c+\bar{\partial} \bar{c}) c\bar{c} (e^{-\phi}\tilde{\psi}^A e^{-2\bar{\phi}}\bar{\partial}\bar{\xi}+e^{-2\phi}\partial\xi e^{-\bar{\phi}}\bar{\tilde{\psi}}^A)\right\}_{D^2}\,.
\end{equation}

To evaluate $\mathcal{A}_{4,1},$ we need to insert 3 PCOs. The non-trivial contributions due to $\mathcal{A}_{4,1,1}$ and $\mathcal{A}_{4,1,2}$ are obtained by contracting $e^\phi T_F$ from all of the PCOs. On the other hand, the non-trivial contribution due to $\mathcal{A}_{4,1,3}$ is obtained by contracting $e^\phi T_F$ from two PCOs and $-\partial \eta be^{2\phi}-\partial(\eta b e^{2\phi})$ from the remaining PCO. Each amplitude contains the following matter CFT correlator
\begin{equation}
    \mathcal{A}_{4,1,1}\propto H_{ijk} v_l \biggr\langle(c_1 H^2_{AB}\tilde{\psi}^4\tilde{\psi}^4+c_2\eta_{AB} |H|^2\tilde{\psi}^4\tilde{\psi}^4+c_3 R_{AcBd}\tilde{\psi}^c\tilde{\psi}^d)  \tilde{\psi}^A\tilde{\psi}^B \tilde{\psi}^i\tilde{\psi}^j\tilde{\psi}^k\tilde{\psi}^l\biggr\rangle\,,
\end{equation}
\begin{equation}
    \mathcal{A}_{4,1,2}\propto H_{ijk}v_l\biggr\langle c_4 \tilde{\psi}^4\tilde{\psi}^4\tilde{\psi}^i\tilde{\psi}^j\tilde{\psi}^k\tilde{\psi}^l  \biggr\rangle\,,
\end{equation}
\begin{equation}
    \mathcal{A}_{4,1,3}\propto H_{ijk}v_l\biggr\langle c_4 \tilde{\psi}^4\tilde{\psi}^4\tilde{\psi}^i\tilde{\psi}^j\tilde{\psi}^k\tilde{\psi}^l  \biggr\rangle\,.
\end{equation}
One can check $\mathcal{A}_{4,1,2}=\mathcal{A}_{4,1,3}=0$ because $H_{ijk}$ is an anti-symmetric rank 3 tensor. Similarly, one can also conclude that
\begin{equation}
    H_{ijk} v_l \biggr\langle (c_2\eta_{AB} |H|^2\tilde{\psi}^4\tilde{\psi}^4+c_3 R_{AcBd}\tilde{\psi}^c\tilde{\psi}^d)  \tilde{\psi}^A\tilde{\psi}^B \tilde{\psi}^i\tilde{\psi}^j\tilde{\psi}^k\tilde{\psi}^l\biggr\rangle=0\,,
\end{equation}
because $\eta_{AB}$ is a symmetric tensor, and $R_{AcBd}$ is a Weyl tensor. To show that the Weyl tensor contribution vanishes, one can use
\begin{equation}
R_{abcd}H^{abc}v^d= \frac{1}{3} R_{abcd}v^d(H^{abc}+H^{bca}+H^{cab})=\frac{1}{3}H^{abc}v^d (R_{abcd}+R_{bcad}+R_{cabd})=0\,,
\end{equation}
and a contraction of any two of the indices of the Weyl tensor vanishes. In the remaining term, the only contraction that does not obviously vanish is obtained by contracting $l$ with $k,$ $i$ with $4,$ and $j$ with $A,$ $B$ with $4,$ and their permutations. However, as one can check from \eqref{eqn:H in large radius}, $H^2_{A4}$ is non-trivial only for $A=4.$ Therefore, we conclude that 
\begin{equation}
    \mathcal{A}_{4,1}=0\,.
\end{equation}
We can again use the ISD condition and the doubling trick to also conclude that
\begin{equation}
    \mathcal{A}_{4,3}=0\,.
\end{equation}

Let us now evaluate $\mathcal{A}_{4,2}.$ To evaluate $\mathcal{A}_{4,2}$ we shall insert two PCOs. The only non-trivial PCO contraction is to contract $e^\phi T_F$ from all of the PCOs. The matter CFT correlator takes the following form
\begin{equation}
    H_{ijk} v_l H_{abc}F_{def} \biggr\langle \tilde{\psi}^i\tilde{\psi}^j\tilde{\psi}^k \tilde{\psi}^l \Sigma^\alpha (\Gamma^{abcdef})_{\alpha}^{\beta}\overline{\Sigma}_\beta+c.c\biggr\rangle\,.
\end{equation}
Using the doubling trick, we can rewrite the matter correlator as
\begin{equation}
        H_{ijk} v_l H_{abc}F_{def} \epsilon^{abcdef}\biggr\langle \tilde{\psi}^i\tilde{\psi}^j\tilde{\psi}^k \tilde{\psi}^l \Sigma^\alpha \Sigma_\alpha+c.c\biggr\rangle\,.
\end{equation}
The above matter correlator again vanishes, as the contraction necessarily involves contracting two indices of $H_{ijk}.$ Therefore, we conclude
\begin{equation}
    \mathcal{A}_{4,2}=0\,.
\end{equation}
By using the ISD condition, one can check that the matter CFT correlator of $\mathcal{A}_{4,4}$ takes the same tensor structure. Therefore, we find
\begin{equation}
    \mathcal{A}_{4,i}=0\,,
\end{equation}
for all $i.$

Note that because there is no way to contract $c\partial \xi$ and $-\partial \eta be^{2\phi}-\partial(\eta b e^{2\phi}),$ the vertical integration cannot yield a non-trivial result.

\subsection{$\mathcal{S}_5$}
In this section, we shall study $\langle V_{t,1}^o|\mathcal{S}_5\rangle$ and $\langle V_{t,2}^o|\mathcal{S}_5\rangle.$ There are two different contributions
\begin{equation}
    \mathcal{A}_{5,1}:= -\tilde{R}_{S^3}^{-3} \left\{ V_{t,i}^o\otimes \Bbb{P}\Psi_{1,0}^o\otimes \Bbb{P} (\Psi_{2,0}^c)^{-1,-1}\right\}_{D^2}\,,
\end{equation}
\begin{equation}
    \mathcal{A}_{5,2}:= -\tilde{R}_{S^3}^{-3} \left\{ V_{t,i}^o\otimes \Bbb{P}\Psi_{1,0}^o\otimes \Bbb{P} (\Psi_{2,0}^c)^{-\frac{1}{2},-\frac{1}{2}}\right\}_{D^2}\,.
\end{equation}

Let us first study $\mathcal{A}_{5,1}.$ $\mathcal{A}_{5,1}$ receives three distinct contributions
\begin{equation}
    \mathcal{A}_{5,1,1}:=-\tilde{R}_{S^3}^{-1}\left\{ V_{t,1}^o\otimes \Bbb{P}\Psi_{1,0}^o\otimes \mathcal{G}_{AB}c\bar{c} e^{-\phi}\tilde{\psi}^Ae^{-\bar{\phi}}\tilde{\psi}^B\right\}_{D^2}\,,
\end{equation}
\begin{equation}
    \mathcal{A}_{5,1,2}:=-\tilde{R}_{S^3}^{-1}\left\{ V_{t,1}^o\otimes \Bbb{P}\Psi_{1,0}^o\otimes \mathcal{D}c\bar{c} (\eta \bar{\partial}\bar{\xi}e^{-2\bar{\phi}}-\partial\xi e^{-2\phi}\bar{\eta})\right\}_{D^2}\,,
\end{equation}
\begin{equation}
    \mathcal{A}_{5,1,3}:=-\frac{i}{2\sqrt{2}}\tilde{R}_{S^3}^{-1}\left\{ V_{t,1}^o\otimes \Bbb{P}\Psi_{1,0}^o\otimes \mathcal{F}_{A}(\partial c+\bar{\partial}\bar{c})c\bar{c} (e^{-\phi}\tilde{\psi}^A e^{-2\bar{\phi}}\bar{\partial}\bar{\xi} +e^{-2\phi}\partial\xi e^{-\bar{\phi}}\bar{\tilde{\psi}}^A)\right\}_{D^2}\,.
\end{equation}
To evaluate $\mathcal{A}_{5,1},$ we shall insert two PCOs. The non-trivial contribution from each correlator comes from the following PCO contractions. For $\mathcal{A}_{5,1,1}$ and $\mathcal{A}_{5,1,2},$ $e^\phi T_F$ from all of the PCOs must be contracted. For $\mathcal{A}_{5,1,3},$ $e^\phi T_F$ from one PCO and $-\partial \eta be^{2\phi}-\partial(\eta be^{2\phi})$ from the remaining PCO must be contracted. Each amplitude contains the following matter CFT correlator
\begin{equation}
    \mathcal{A}_{5,1,1}\propto f_{[i} v_{j]} \biggr\langle(c_1 H^2_{AB}\tilde{\psi}^4\tilde{\psi}^4+c_2\eta_{AB} |H|^2\tilde{\psi}^4\tilde{\psi}^4+c_3 R_{AcBd}\tilde{\psi}^c\tilde{\psi}^d)  \tilde{\psi}^A\tilde{\psi}^B \tilde{\psi}^i\tilde{\psi}^j\biggr\rangle\,,
\end{equation}
\begin{equation}
    \mathcal{A}_{5,1,2}\propto f_{[i} v_{j]}\biggr\langle (c_4 \tilde{\psi}^4\tilde{\psi}^4+ c_5 \tilde{\psi}^a\tilde{\psi}_a)\tilde{\psi}^i\tilde{\psi}^j \biggr\rangle\,,
\end{equation}
\begin{equation}
    \mathcal{A}_{5,1,3}\propto f_{[i} v_{j]}\biggr\langle c_6 \tilde{\psi}^4\tilde{\psi}^4\tilde{\psi}^i\tilde{\psi}^j \biggr\rangle\,.
\end{equation}
As one can check, upon contracting the worldsheet fermions, all of the correlators vanish. We haven't yet computed the overlap between $\mathcal{S}_{5,1}$ and $V_{t,2}^o.$ One can easily see that such an overlap must vanish. To saturate the background $\phi$ charge, one must contract one more $-\partial\eta be^{2\phi}-\partial(\eta be^{2\phi})$ from the PCOs. This, in turn, implies that we have one less derivative to take. As first derivative of $\mathcal{G}_{AB},$ $\mathcal{D}$ vanish at the anti-D3-brane, and $\mathcal{F}_A$ vanishes at the anti-D3-brane, we therefore conclude
\begin{equation}
    \mathcal{A}_{5,1}=0\,.
\end{equation}

We shall now study $\mathcal{A}_{5,2}.$ To evaluate $\mathcal{A}_{5,2},$ we need to insert one PCO. Because only the first derivative $\Bbb{P}(\Psi_{2,0}^c)^{-\frac{1}{2},-\frac{1}{2}}$ along the radial direction is non-trivial evaluated at the anti-D3-brane, to have a non-trivial result $e^\phi T_F$ from the PCO must be contracted. This in turn implies that $\langle V_{t,2}^o|\mathcal{S}_{5,2}\rangle=0. $ The matter CFT correlator of the overlap against $V_{t,1}^o$ takes the following form
\begin{equation}
   f_{[i} v_{j]} H_{abc}F_{def} \biggr\langle \tilde{\psi}^i\tilde{\psi}^j\Sigma^\alpha (\Gamma^{abcdef})_{\alpha}^{\beta}\overline{\Sigma}_\beta+c.c\biggr\rangle\,.
\end{equation}
We can use the doubling trick to simplify the matter CFT correlator into
\begin{equation}
    f_{[i} v_{j]} H_{abc}F_{def}\epsilon^{abcdef} \biggr\langle \tilde{\psi}^i\tilde{\psi}^j\Sigma^\alpha \Sigma_\alpha+c.c\biggr\rangle\,,
\end{equation}
which again vanishes. Hence, we conclude
\begin{equation}
    \mathcal{A}_5=0\,.
\end{equation}

\subsection{$\mathcal{S}_6$}
Finally, we shall study $\mathcal{S}_6.$ Although we haven't fully specified the details of $\Psi_{3,0}^c,$ we know just enough to conclude that $\mathcal{S}_6$ vanishes. As we showed in \S\ref{sec:third order KS}, in the NSNS sector of the background solution there exists a modulus parametrized by
\begin{align}
    \Psi_{3,m}^{-1,-1}= &c\bar{c} \biggr[\mathcal{G}^{3,m}_{AB} e^{-\phi}\tilde{\psi}^A e^{-\bar{\phi}}\bar{\tilde{\psi}}^B +\mathcal{D}^{3,m} (\eta e^{-2\bar{\phi}} \bar{\partial}\xi-e^{-2\phi}\partial\xi \bar{\eta}) \nonumber\\
    &\quad+F_A^{3,m} (\partial c+\bar{\partial}\bar{c}) (e^{-\phi}\tilde{\psi}^A e^{-2\bar{\phi}}\bar{\partial}\bar{\xi}+ (\partial c+\bar{\partial}\bar{c}) e^{-2\phi}\partial\xi e^{-\bar{\phi}}\bar{\tilde{\psi}}^A)\biggr]\,.
\end{align}
where
\begin{equation}
    \partial^B\mathcal{G}_{AB}^{3,m}+\mathcal{F}_A^{3,m}+\partial_A\mathcal{D}^{3,m}=0\,,
\end{equation}
for linear functions $\mathcal{G}_{AB},$ $\mathcal{D}$ and a constant $\mathcal{F}_A,$ and the RR sector of the background solution is given by the Ramond-Ramond threeform flux. One can check that 
\begin{equation}
    \{ \Psi_{3,m}^{-1,-1}\otimes V_{t,1}^o\}_{D^2}\neq0.
\end{equation}
Therefore, we can adjust the modulus to make the NSNS contribution to $\mathcal{S}_{6}$ vanish. This is to adjust the background solution to ensure that the anti-D3-brane is properly located at the tip of the KS throat. Furthermore, in \S\ref{sec:first order KPV}, we already showed that the tensor structure of the $V_{t,1}^o$ coupling to a threeform flux forbids non-trivial source term. Therefore, we conclude
\begin{equation}
    \langle V_{t,1}^o|\mathcal{S}_6\rangle=0\,.
\end{equation}

\section{$\langle V_{t,2}^o|\mathcal{S}_3+\mathcal{S}_4+\mathcal{S}_6\rangle$}\label{app:S3+S4+S6}
We shall now study the overlap between $V_{t,2}^o$ and $\mathcal{S}_3,$ $\mathcal{S}_4,$ and $\mathcal{S}_6$ altogether. For the overlap 
\begin{equation}
    \langle V_{t,2}^o|\mathcal{S}_3+\mathcal{S}_4+\mathcal{S}_6\rangle\,,
\end{equation}
rather than tour-de-force computations, we shall use a hack based on the following observation. We find that the BRST operator acting on ghost number zero $L_0^+$ nilpotent state is 
\begin{equation}
Q_B(h ce^{-2\phi}\partial\xi)= \partial^2 h c\partial ce^{-2\phi}\partial \xi +i\sqrt{2} \partial_\mu h ce^{-\phi} \tilde{\psi}^\mu\,.
\end{equation}
Therefore, if we choose $h$ such that
\begin{equation}
h=\frac{1}{8}\tilde{X}^\mu\tilde{X}_\mu\,,
\end{equation}
then we can treat 
\begin{equation}
c\partial ce^{-2\phi}\partial\xi +i\frac{\sqrt{2}}{4}\tilde{X}_\mu ce^{-\phi}\tilde{\psi}^\mu\,,
\end{equation}
as a BRST exact operator. As we studied in the previous section, $ce^{-\phi}\tilde{\psi}^\mu$ has a zero overlap against the source terms. This can be simply seen by replacing $v_j$ with $v_\mu.$ Hence, we can effectively treat $V_{t,2}^o$ as a BRST operator when evaluating the source terms $\mathcal{S}_3+\mathcal{S}_4+\mathcal{S}_6.$  Using this BRST exactness, we shall use the main identities to show the following identity holds
\begin{equation}
\langle V_{t,2}^o| \mathcal{S}_3+\mathcal{S}_4+\mathcal{S}_6\rangle=0\,.
\end{equation}

We shall define a new vertex operator of zero ghost number
\begin{equation}
\tilde{V}^o_{t,2}=\frac{1}{8}\tilde{X}_\mu\tilde{X}^\mu ce^{-2\phi}\partial\xi\,. 
\end{equation}
We compute
\begin{align}
&\langle Q_B\tilde{V}_{t,2}^o| \mathcal{S}_3\rangle +\langle \tilde{V}_{t,2}^o|Q_B\mathcal{S}_3\rangle=\frac{g_c^2}{4} \tilde{R}_{S^3}^{-3}\biggr[\left\{ \tilde{V}_{t,2}^o\otimes \left[\frac{1}{3!}(\Psi_{1,0}^c)^3\right]_{S^2}^c\right\}_{D^2}\nonumber\\&+\left\{\tilde{V}_{t,2}^o \otimes\Psi_{1,0}^c \otimes \left[\frac{1}{2}(\Psi_{1,0}^c)^2 \right]_{S^2}^c\right\}_{D^2}\biggr]+g_o^2\tilde{R}_{S^3}^{-3}\biggr[ \left\{ \tilde{V}_{t,2}^o \otimes \frac{1}{2} (\Psi_{1,0}^c)^2 \otimes\left[ \Psi_{1,0}^c\right]_{D^2}^o\right\}_{D^2}\nonumber\\
&+\left\{ \tilde{V}_{t,2}^o\otimes\Psi_{1,0}^c\otimes\left[ \frac{1}{2}(\Psi_{1,0}^c)^2\right]_{D^2}^o\right\}_{D^2}\biggr]\,.
\end{align}
Note that the unusual factors of $g_c^2$ and $g_o^2$ are due to the different normalization we chose for the kinetic action of string fields. Note also 
\begin{equation}
\langle \tilde{V}_{t,2}^o|Q_B\mathcal{S}_3\rangle=0\,,
\end{equation}
because $Q_B\Psi_{1,0}^c=0,$ and we neglected terms that vanish in the large stub limit.\footnote{As was studied in \cite{Belopolsky:1995vi}, one needs to be careful when computing string vertices involving $Q_B.$ In particular, because the string vertices should be thought of as distributions, and $Q_B$ can introduce derivatives to $\delta(\sum_ip_i),$ just because an operator is BRST closed the string vertex involving $Q_B (V)$ does not always vanish. However, we don't expect such subtlety to arise in our case, as $\Psi_{n,0}^c$ only depend on the internal coordinates non-trivially, whereas $V_{t,2}^o$ contains a non-trivial position dependence along the four external directions. } The open string channels vanish, as we have confirmed that the tadpoles up to the second order in open string equations of motion are absent. Therefore, we can rewrite
\begin{align}\label{eqn:S3 t1}
&\langle Q_B\tilde{V}_{t,2}^o| \mathcal{S}_3\rangle =\frac{g_c^2}{4} \tilde{R}_{S^3}^{-3}\biggr[\left\{ \tilde{V}_{t,2}^o\otimes \left[\frac{1}{3!}(\Psi_{1,0}^c)^3\right]_{S^2}^c\right\}_{D^2}+\left\{\tilde{V}_{t,2}^o \otimes\Psi_{1,0}^c \otimes \left[\frac{1}{2}(\Psi_{1,0}^c)^2 \right]_{S^2}^c\right\}_{D^2}\biggr]\,.
\end{align}
Note again that we neglected terms that vanish in the large stub limit. We shall now compute
\begin{align}
&\langle Q_B\tilde{V}_{t,2}^o|\mathcal{S}_4\rangle+\langle\tilde{V}_{t,2}^o|Q_B\mathcal{S}_4\rangle=\frac{g_c^2}{4}\tilde{R}_{S^3}^{-3} \left\{\tilde{V}_{t,2}^o \otimes \left[ \Psi_{1,0}^c \otimes \Bbb{P} \Psi_{2,0}^c\right]_{S^2}^c\right\}_{D^2}\nonumber\\
&+g_o^2\tilde{R}_{S^3}^{-2}\biggr[\left\{\tilde{V}_{t,2}^o\otimes\Psi_{1,0}^c\otimes\left[ \Bbb{P}\Psi_{2,0}^c\right]_{D^2}^o \right\}_{D^2} +\left\{\tilde{V}_{t,2}^o\otimes\Bbb{P}\Psi_{2,0}^c\otimes\left[\Psi_{1,0}^c\right]_{D^2}^o \right\}_{D^2}\biggr]_{D^2}\,.
\end{align}
Because the open string tadpoles up to the second order were shown to be absent, we can rewrite the above equation as
\begin{align}\label{eqn:S4 t1}
\langle Q_B\tilde{V}_{t,2}^o|\mathcal{S}_4\rangle=&-\frac{g_c^2}{4} \left\{ \tilde{V}_{t,2}^o\otimes\Psi_{1,0}^c\otimes\left[\frac{1}{2}(\Psi_{1,0}^c)^2\right]_{S^2}^c \right\}_{D^2}+\frac{g_c^2}{4}\tilde{R}_{S^3}^{-3} \left\{\tilde{V}_{t,2}^o \otimes \left[ \Psi_{1,0}^c \otimes \Bbb{P} \Psi_{2,0}^c\right]_{S^2}^c\right\}_{D^2}\,.
\end{align}
We compute
\begin{align}
\langle Q_B\tilde{V}_{t,2}^o|\mathcal{S}_6\rangle+\langle \tilde{V}_{t,2}^o|Q_B\mathcal{S}_6\rangle=0\,.
\end{align}
Therefore, we find
\begin{equation}\label{eqn:S6 t1}
\langle Q_B\tilde{V}_{t,2}^o|\mathcal{S}_6\rangle= -\frac{g_c^2}{4}\tilde{R}_{S^3}^{-3} \biggr[\left\{ \tilde{V}_{t,2}^o\otimes \left[ \frac{1}{3!}(\Psi_{1,0}^c)^3\right]_{S^2}^c\right\}_{D^2} +\left\{ \tilde{V}_{t,2}^o\otimes\left[\Psi_{1,0}^c\otimes\Bbb{P}\Psi_{2,0}^c\right]_{S^2}^c\right\}_{D^2}\biggr]\,.
\end{equation}
By combining \eqref{eqn:S3 t1}, \eqref{eqn:S4 t1}, and \eqref{eqn:S6 t1}, we conclude
\begin{equation}
\langle V_{t,2}^o| \mathcal{S}_3+\mathcal{S}_4+\mathcal{S}_6\rangle=0\,,
\end{equation}
as promised.

\bibliographystyle{JHEP}
\bibliography{refs}
\end{document}